\documentclass[reprint,showpacs,preprintnumbers,amssymb,nofootinbib,superscriptaddress,aps]{revtex4-1}
\usepackage{graphicx,floatflt,amssymb,rotate,epsfig,url}
\usepackage[utf8]{inputenc}
\usepackage{mathrsfs}
\usepackage{amssymb}
\usepackage{amsmath}
\usepackage{color}
\usepackage{graphicx}
\usepackage{subfig}
\usepackage{multirow,bigdelim}
\usepackage{mathtools}
\usepackage{float}
\usepackage{pstricks}
\usepackage[toc,page]{appendix}
\usepackage{pifont}
\usepackage{braket}
\usepackage{bbold}
\usepackage[toc,page]{appendix}
\usepackage[colorlinks=true,
linkcolor=red,
urlcolor=blue,
citecolor=blue]{hyperref}

\usepackage{relsize}
\def\Babar{{\mbox{\slshape B\kern-0.1em{\smaller A}\kern-0.1em B\kern-0.1em{\smaller A\kern-0.2em R}}}}

\usepackage{enumitem}

\newcommand{\ba}{\begin{array}}
	\newcommand{\ea}{\end{array}}
\def\beq{\begin{equation}}
\def\eeq{\end{equation}}
\def\bea{\begin{eqnarray}}
\def\eea{\end{eqnarray}}
\def\nn{\nonumber}

\def\roughly#1{\mathrel{\raise.3ex\hbox
		{$#1$\kern-.75em\lower1ex\hbox{$\sim$}}}}

\def\sla#1{\raise.15ex\hbox{$/$}\kern-.57em #1}% Feynman slash

\def\bd{B_d^0}

\def\order{\lower 1.8ex \hbox{\LARGE\~{}}}

\def\rdast{{{\cal B}{(B\to D^{(\ast)}\tau\nu_{\tau})}}/{{\cal B}{(B\to D^{(\ast)}\ell\nu_{\ell})}}}

\def\bdell{B\to D^{(\ast)}\ell\nu_{\ell}}
\def\bd0tau{B\to D \tau\nu_{\tau}}

\def\be {\begin{equation}}
\def\ee {\end{equation}}

\usepackage[normalem]{ulem}

\definecolor{darkgreen}{cmyk}{1,0,1,0.4}

\definecolor{pink}{cmyk}{0.4,1,0.3,0}

\def\com2#1{\textcolor{red}{\it{#1}}}
\begin{document}
	
	%opening
	\title{`Deep' Dive into $b \to c$ Anomalies: Standardized and Future-proof Model Selection Using Self-normalizing Neural Networks}
	
	\author{Srimoy Bhattacharya}
	\email{bhattacharyasrimoy@gmail.com}
	\affiliation{Institute of Particle Physics and Key Laboratory of Quark and Lepton Physics (MOE), Central China Normal University, Wuhan, Hubei 430079, China}
	
	\author{Soumitra Nandi}
	\email{soumitra.nandi@iitg.ernet.in}
	\affiliation{Indian Institute of Technology, North Guwahati, Guwahati 781039, Assam, India }
	
	\author{Sunando Kumar Patra}
	\email{sunando.patra@gmail.com}
	\affiliation{Department of Physics, Bangabasi Evening College, 19 Rajkumar Chakraborty Sarani, Kolkata 700009, West Bengal, India }
	
	\author{Shantanu Sahoo}
	\email{shantanu\_sahoo@iitg.ac.in}
	\affiliation{Indian Institute of Technology, North Guwahati, Guwahati 781039, Assam, India }

	\begin{abstract}  
Noting the erroneous proclivity of information-theoretic approaches, like the Akaike information criterion (AIC), to select simpler models while performing model selection with a small sample-size, we address the problem of new physics model selection in $b\to c \tau \nu_{\tau}$ decays in this paper by employing a specific machine learning algorithm (self-normalizing neural networks, a.k.a. SNN) for supervised classification and regression, in a model-independent framework. While the outcomes of the classification with real data-set are compared with AIC, with the SNNs outperforming AIC$_c$ in all aspects of model selection, the regression-outcomes are compared with the results from Bayesian analyses; the obtained parameter spaces differ considerably, while keeping maximum posterior (MAP) estimates similar. A few of the two-operator scenarios with a tensor-type interaction are found to be the most probable solution for the data. We also test the effectiveness of our trained networks with the expected, more precise data in Belle-II. The trained networks and associated functionalities are supplied for the use of the community.
	\end{abstract}  
	
	\maketitle
	
%%%%%%%%%%%%%%%%%%%%%%%%%%%%%%%%%%%%%%%%%%%%%%%%%	
	\section{Introduction}\label{sec:intro}
%%%%%%%%%%%%%%%%%%%%%%%%%%%%%%%%%%%%%%%%%%%%%%%%%
%\subsubsection{Inadequacies of Standard Techniques}
%%%%%%%%%%%%%%%%%%%%%%%%%%%%%%%%%
Exclusive $\bdell$ (with $\ell = e, \mu, \tau$) decays have got a lot of attention in recent years both from the theoretical and experimental perspectives. Some of these modes are the useful probes of the Cabibbo-Kobyashi-Masakawa (CKM) matrix element $V_{cb}$ (for an update see \cite{Alberti:2014yda,Gambino:2016jkc,Grinstein:2017nlq,Bernlochner:2017jka,Jaiswal:2017rve,Amhis:2019ckw,hflav,Gambino:2019sif, Jaiswal:2020wer} and the references therein). At the same time, some of the observables in these modes are potentially sensitive to the physics beyond the standard model (BSM), (see \cite{Tanaka:2012nw,Bhattacharya:2015ida}). Measurements have been carried out with reasonable precision for the ratios $R(D^{(*)})=\rdast$ \cite{hflav,Aaij:2017tyk,Aaij:2017uff,Abdesselam:2019dgh}. While these are predicted in the standard model (SM) with uncertainties less than 3\% (for details, check out the refs. \cite{Bigi:2016mdz,Bigi:2017jbd,Jaiswal:2017rve,Gambino:2019sif, Bordone:2019vic, Bordone:2019guc,Jaiswal:2020wer}), there is some degree of discrepancy between the measured and predicted values of $R(D^{(*)})$. There is scope of improvement in both theory and experimental measurements. For that we have to wait for the lattice inputs on the respective form-factors at non-zero recoils and the results from high precision experiments like Belle-II and LHCb \cite{Albrecht:2017odf, Adamczyk:2019wyt}. 

It is not surprising that a multitude of new physics models can explain the observed discrepancies. However, a simple model-independent analysis with different new effective operators can already give us information about the size and the pattern of new contributions, which can further be utilized to build a new physics model or to constrain the parameters of an existing model. Among the plethora of works present in the literature, here we point out a few of such analyses which are based on the most updated results in both theory and experiment  \cite{Murgui:2019czp,Asadi:2019xrc,Shi:2019gxi,Jaiswal:2020wer,Iguro:2020cpg, Schacht:2020qot}. As all these studies suggest that many possible scenarios (one or two- operator) are capable of describing the observed data, one is confronted with the problem of model selection. The problem with a simple goodness-of-fit test approach after fitting a model suffers from the fact that a better fit does not necessarily imply superior predictive performance. A statistical analysis of a data-set with a model examines the capability of the model in question to adequately explain the important features of the data. An improper choice of model or method can lead to severely misleading conclusions, or disappointing predictive performances. Therefore, a crucial step in a typical data analysis is to consider a set of candidate models, and then select the most appropriate one that describes the data best.

Information criteria, such as Akaike Information Criterion (AIC) or the corrected version AIC$_c$, divergence measures like Kullback-Leibler divergence ($D_{KL}$), and cross-validation techniques like leave-one-out-cross-validation (LOOCV) have been used for model selection in various fields for quite a long period of time. In flavor physics, they have been used in the analysis of semileptonic $b\to c$ \cite{Bhattacharya:2016zcw,Bhattacharya:2018kig} and $b\to s$ LFUV anomalies \cite{Bhattacharya:2019dot,Biswas:2020uaq,Ciuchini:2019usw}. They have also been used to optimize the order of the $B\to D^{(*)}$ form-factor polynomials \cite{Jaiswal:2020wer,Iguro:2020cpg}. A comparative analysis of the selection results of AIC$_c$ and LOOCV has been carried out In ref.s \cite{Bhattacharya:2019dot,Biswas:2020uaq} and interesting differences have been found.
In the present context of the considered cases, using these criteria to settle the model selection question has the following limitations:
\paragraph{} For small, finite sample sizes, the results of AIC$_c$ as well as LOOCV, are unstable. In case of exclusive semileptonic $B\to D^{(*)}$, only four observables are measured till date, one of which ($P_{\tau}(D^*)$) is really imprecise. There are, in total, seven such measurements possible in those channels, except the branching fractions, most of which are not measured yet. The status of other observables, such as $\mathcal{R}_{J/\Psi}$, was ambiguous till very recently, as the SM predictions for those were extremely model-dependent and differed over a wide range. Lattice predictions of the form factors at zero and non-zero recoils have come out only during the course of the present work. Using these meager number of observables to do a stable statistical model-selection is problematic at best.
\paragraph{} As the application of any information criterion (IC) depends on the maximum-likelihood-estimator (MLE) estimate of the model parameters beforehand, bringing more complex models (with dimensions equal or more than the available data) in the mix remains out of question. This restricts the model space severely and finding the underlying structure of the data-distribution for model-separation becomes improbable. Also, as they depend only on the MLE estimate, uncertainty of the data-distribution are not taken into account. This leads to the selection of models that are too simple.
\paragraph{} The statistical methods, as mentioned above, can only perform a model selection for a given set of data and can not be generalized to future, more precise data. 

It is well known that a model with too few parameters can involve making unrealistically simple assumptions, which leads to high bias, poor predictions due to under-fitting, and consequently, missed opportunities for insight. On the other hand, models, with greater complexity than needed, may over-fit the data, and generally tend to have poor predictive performance. Any machine learning algorithm, optimized with a well-chosen objective function, can tune the desired complexity of the model and thus takes care of both over and under-fitting by minimizing the generalization error (by using unseen data-sets for validation). This fact can be used for not only regression in presence of a specific model, but also to assign inverse probabilities to models, given data; this enable one to do `classification'/model selection. For the stated classification/model-selection algorithm, as well as for performing regression for a specific model in the present analysis, we choose a special type of artificial neural network, called Self-normalizing Neural Networks (SNN) \cite{SNNMain}. This network is well suited for deep supervised learning and was introduced very recently to overcome the shortcomings of vanilla fully-connected-networks (FNN) in high-level abstract representation problems. Using the self-normalizing properties (like variance stabilization which in turn avoids exploding and vanishing gradients) of the unique activation function called scaled exponential linear units (SELU), the SNNs have stable outputs over longer chains in face of input-perturbations and their variances always tend to that of a unit normal (given unit-normalized inputs). This enables the SNNs to robustly train over many layers \cite{SNNMain}.

%%%%%%%%%%%%%%%%%%%%%%%%%%%%%%%%%%%%%%%%%%%%%%%%%
	\section{Background}
	\subsubsection{Theoretical Framework}
%%%%%%%%%%%%%%%%%%%%%%%%%%%%%%%%%%%%%%%%%%%%%%%%%

In an effective theory framework, the most general effective Hamiltonian describing the $b \to c \ell \nu_\ell$ transitions (where $\ell = e, \mu$  or $\tau$ ) can be written as 
\begin{align}\label{eq:effham}
	{\cal H}_{eff} &= \frac{4 G_F}{\sqrt{2}} V_{cb} \Big[( \delta_{\ell\tau} + C_{V_1}^{\ell}) {\cal O}_{V_1}^{\ell} + C_{V_2}^{\ell} {\cal O}_{V_2}^{\ell} + C_{S_1}^{\ell} {\cal O}_{S_1}^{\ell} \nn \\
	& + C_{S_2}^{\ell} {\cal O}_{S_2}^{\ell} + C_{T}^{\ell} {\cal O}_{T}^{\ell}\Big]\,,
\end{align}
where $G_F$ is the Fermi coupling constant, $V_{cb}$ is the Cabibbo-Kobayashi-Maskawa (CKM) matrix element. Here we have considered only the dimension-six operators, and the operator basis is given by 
\begin{align}\label{eq:opbasis}
	{\cal O}_{V_1}^{\ell} &= ({\bar c}_L \gamma^\mu b_L)({\bar \tau}_L \gamma_\mu \nu_{\ell L}) \nn, \\
	{\cal O}_{V_2}^{\ell} &= ({\bar c}_R \gamma^\mu b_R)({\bar \tau}_L \gamma_\mu \nu_{\ell L}) \nn, \\
	{\cal O}_{S_1}^{\ell} &= ({\bar c}_L  b_R)({\bar \tau}_R \nu_{\ell L}) \nn, \\
	{\cal O}_{S_2}^{\ell} &= ({\bar c}_R b_L)({\bar \tau}_R \nu_{\ell L}) \nn, \\
	{\cal O}_{T}^{\ell} &= ({\bar c}_R \sigma^{\mu\nu} b_L)({\bar \tau}_R \sigma_{\mu\nu} \nu_{\ell L})\,.
\end{align}
These operators are weighted by the corresponding Wilson coefficients $C^\ell_W (W = V_1, V_2. S_1, S_2, T)$. Note that only the left-handed neutrinos are considered here. We have neglected the possibility of any NP effects in transitions involving light leptons ($\ell = e$ and $\mu$) in this analysis. So far, the data is in good agreement with the respective SM estimates. The possible NP contributions are considered to be present only in the third generation of leptons ($\tau$).
%%%%%%%%%%%%%%%%%%%%%%%%%%%%%%%%%%%%%%%%%%%%%%%%%
\subsubsection{Observables}\label{sec:thobs}
%%%%%%%%%%%%%%%%%%%%%%%%%%%%%%%%%%%%%%%%%%%%%%%%%
A brief discussion on various observables related to $b\to c \ell\nu_{\ell}$ decays is given in this subsection. Measurements are available only for some of them, and in our analysis, we have used them as inputs for parameter-inference after model selection. We also provide the predictions of the Bayesian fits corresponding to the best selected models for many of the observables.
%We have created trained network with the rest of the observables in fold, which can be used for prediction, regression, and model selection with future measurements of those observables.

\paragraph{$B \to D^{(*)}\tau \nu_\tau$ :}
With the given hamiltonian in eq. (\ref{eq:effham}), one can derive the differential decay rates for $B \to D^{(*)}\tau \nu_\tau$. These rates can be expressed in terms of helicity amplitudes,  sensitive to the different new contributions in these decays. For details, see \cite{Bhattacharya:2016zcw, Sakaki:2013bfa} and the references therein. The significant sources of uncertainties in these decays are the form-factors, whose shapes in the full kinematically allowed range of $q^2$ are the things that we need to know. Different ways of parametrizing the form factors have been developed to this goal. We have used the Boyd-Grinstein-Lebed (BGL) parametrization \cite{Boyd_1997} in this analysis. All the relevant inputs associated with the form-factor parameters are taken from \cite{Jaiswal:2020wer}. In terms of the differential decay distributions, the ratios $\mathcal{R_{D^{(*)}}}$ can be defined as:
\begin{align}\label{eq:Rth}
\nn\mathcal{R}_{D^{(*)}} &= \left[\int^{q^2_{max}}_{m^2_{\tau}} \frac{d\Gamma\left(\overline{B} \rightarrow D^{(*)}
	\tau \overline{\nu}\right)}{d q^2} d q^2\right]\\
&\times \left[\int^{q^2_{max}}_{m^2_{\ell}} 
\frac{d\Gamma\left(\overline{B} \rightarrow D^{(*)} \ell \overline{\nu}\right)}{d q^2} d q^2\right]^{-1}\,,
\end{align}
with $q^2_{max}= (m_B - m_{D^{(*)}})^2$, and $\ell=e$ or $\mu$.
Along with these, there are other observables like the $\tau$-polarization asymmetry ($P_\tau(D^{(*)})$), $D^*$ -longitudinal polarization ($F_L^{D^*}$) and  the lepton forward-backward asymmetry ($\mathcal{A}_{FB}^{(*)}$) which are sensitive to NP. The definitions of these observables and the respective SM predictions are given in \cite{Bhattacharya:2018kig,Jaiswal:2020wer}. 

The current experimental status of the observables as mentioned above are given in table \ref{tab:expobs}. Here, the first uncertainty is statistical and the second one is systematic. The value of $\mathcal{R}_{D}$ and $\mathcal{R}_{D^*}$ exceed their SM predictions \cite{Amhis:2019ckw} by $1.4 \sigma$ and $2.5 \sigma$ respectively. If we consider the correlation of $\mathcal{R}_{D}-\mathcal{R}_{D^*}$, which is -0.38, the BSM-SM deviation increases to $\sim3.08\sigma$. The measurement of $\tau$ polarization asymmetry is done by Belle collaboration. Although the data is very imprecise, it is consistent with the SM prediction  \cite{Bhattacharya:2018kig}. Recently, the first measurement of $D^*$- meson polarization in the decay $B^0 \to D^{*-} \tau^+ \nu_\tau$ was reported by Belle collaboration \cite{Abdesselam:2019wbt}. This result agrees within about 1.7 standard deviations of the SM prediction. 

\paragraph{$\Lambda_b \to \Lambda_c \tau \nu_\tau$ :}
The primary quark level transition $b \to c \tau \nu_\tau$ can also be probed in $\Lambda_b$ decays. Similar to the ratios like $\mathcal{R}_{D^{(*)}}$, here we can define an observable:
\begin{align}\label{eq:Rlambda}
\mathcal{R}^\ell_{\Lambda}  =  \frac{\mathcal{B} \left(\Lambda_b \to \Lambda_c \tau \bar{\nu}_{\tau}\right)}{\mathcal{B}\left(\Lambda_b \to \Lambda_c \ell \bar{\nu}_{\ell}\right)}\,.
\end{align}
The other relevant observable associated with this mode is the forward-backward asymmetry $(\mathcal{A}_{FB}^\Lambda)$. Theoretical expressions for the differential $q^2$ distribution of these observables are given in \cite{Datta:2017aue}. For the form-factors, we have followed the inputs/method given in \cite{Detmold:2013nia}. No data is currently available for either of these observables. We, however, have provided the predictions of $\mathcal{R}^\ell_{\Lambda}$ from Bayesian fit results corresponding to the best selected models.

\paragraph{$B_c \to J/\psi \tau \nu_\tau$ :}
Similar to the ratios defined in eqs. \ref{eq:Rth} and \ref{eq:Rlambda}, $\mathcal{R}_{J/\psi}$ can be defined in $B_c \to J/\psi \tau \nu_\tau$ by replacing the respective mesons. In our previous analysis \cite{Bhattacharya:2018kig} we had studied the NP sensitivity of $\mathcal{R}_{J/\psi}$. The major source of contention about the theoretical estimate of the observable was the QCD modeling of the form-factors. Depending on that, the central value of the SM estimate of $\mathcal{R}_{J/\psi}$ had a range from 0.25 to 0.29. The theoretical uncertainties of the form factor parameters coming from different parametrizations makes this theoretical range even larger. However, in a very recent result from lattice \cite{Harrison:2020gvo}, the form-factors in $B_c \to J/\psi$ semileptonic decays are extracted in the full physical $q^2$ range. We have incorporated these new results from lattice as theory inputs in our analysis. Besides $\mathcal{R}_{J/\psi}$, there are a few other observables like the forward-backward asymmetry $\mathcal{A}^{J/\psi}_{FB}$, the $\tau$ polarization asymmetry $P^{J/\psi}_\tau$, and the longitudinal polarization fraction $F_L^{J/\psi}$ for which experimental results are not yet available. Detailed description of the form-factors and full analytical expressions for the considered observables are given in \cite{Harrison:2020nrv}. Table \ref{tab:expobs} also includes the recent measurements of $\mathcal{R}_{J/\psi}$ by LHCb. As the experimental uncertainty is large, the value of  $\mathcal{R}_{J/\psi}$ is still consistent with the SM prediction within $90\%$ C.L range. %\hlas{For the other observables, we have given predictions.} 
%%%%%%%%%%%%%%%%%%%%%%%%%%%%%%%%%%%%%%%%%%%%%%%%%
\begin{table}[!t]
	\begin{center}
		\begin{ruledtabular}
		\begin{tabular}{lc}
			Observables	 & Measurement   \\
			\hline
			$	\mathcal{R}_{D}$\cite{Amhis:2019ckw}  ~~~~~~~\rdelim\}{4}{4mm}[ 4-obs. data-set] & 0.340(27)(13)\\
			$	\mathcal{R}_{D^{*}}$ \cite{Amhis:2019ckw} & 0.295(11)(8) \\
			$P_\tau(D^{*})$ \cite{Hirose_2017}   &  $ -0.38(51)(~^{+0.21}_{-0.16})$\\
			$F_L^{D^*}$ \cite{Abdesselam:2019wbt}   & 0.60(8)(4)  \\
			$\mathcal{R}_{J/\psi}$ \cite{Aaij:2017tyk}  & 0.71(17)(18)	\\
		\end{tabular}
		\caption{Present experimental status of the observables. The two types of uncertainties are respectively statistical and systematic.} 
		\label{tab:expobs}
		\end{ruledtabular}
	\end{center}
\end{table}
%%%%%%%%%%%%%%%%%%%%%%%%%%%%%%%%%%%%%%%%%%%%%%%%%
%%%%%%%%%%%%%%%%%%%%%%%%%%%%%%%%%%%%%%%%%%%%%%%%%
\begin{figure*}[!t]
	\small
	\centering
	\subfloat[Single operator real WCs]{\includegraphics[width=0.47\textwidth]{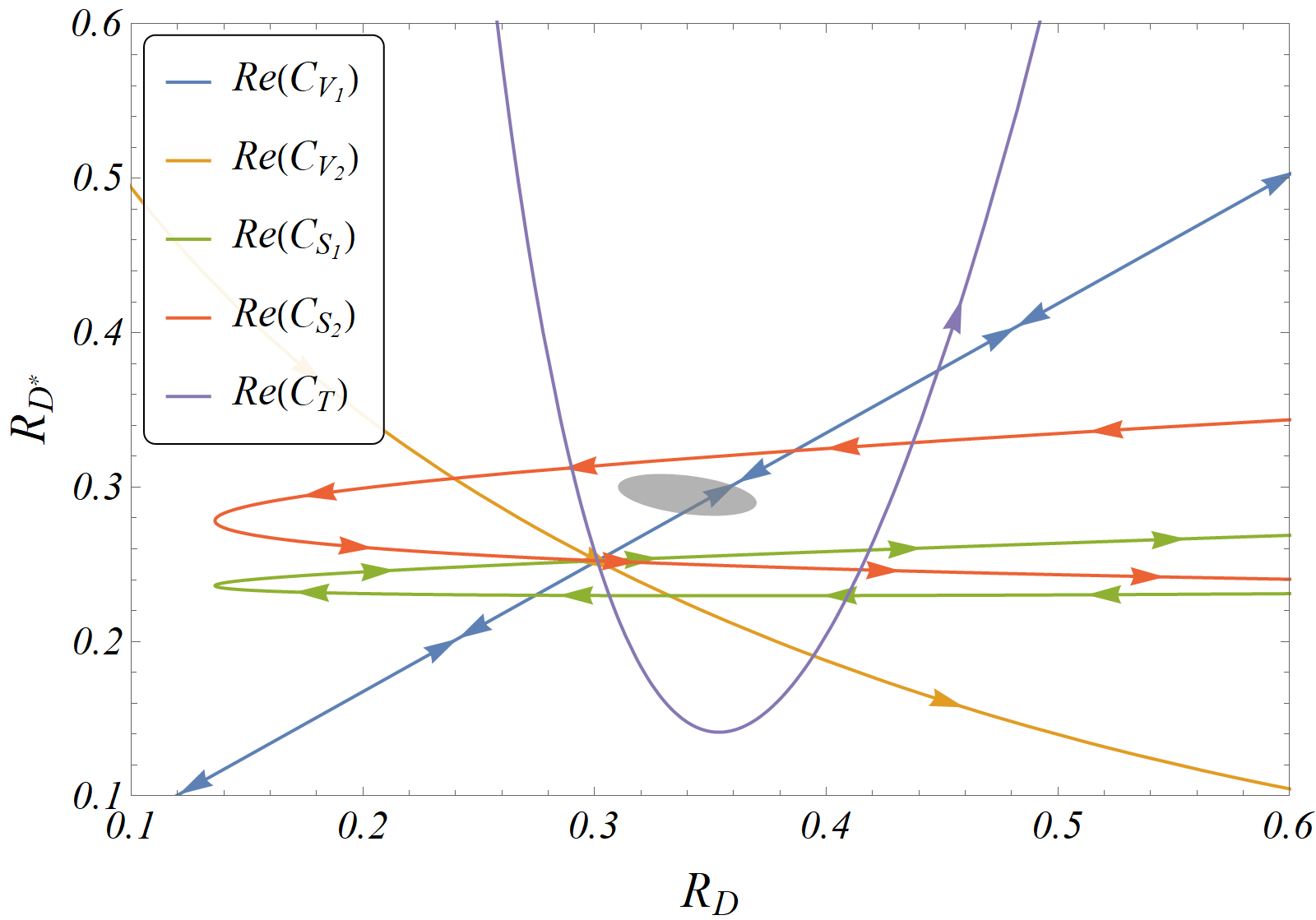}\label{fig:motivPlot1}}~
	\subfloat[Some single operator complex WCs]{\includegraphics[width=0.47\textwidth]{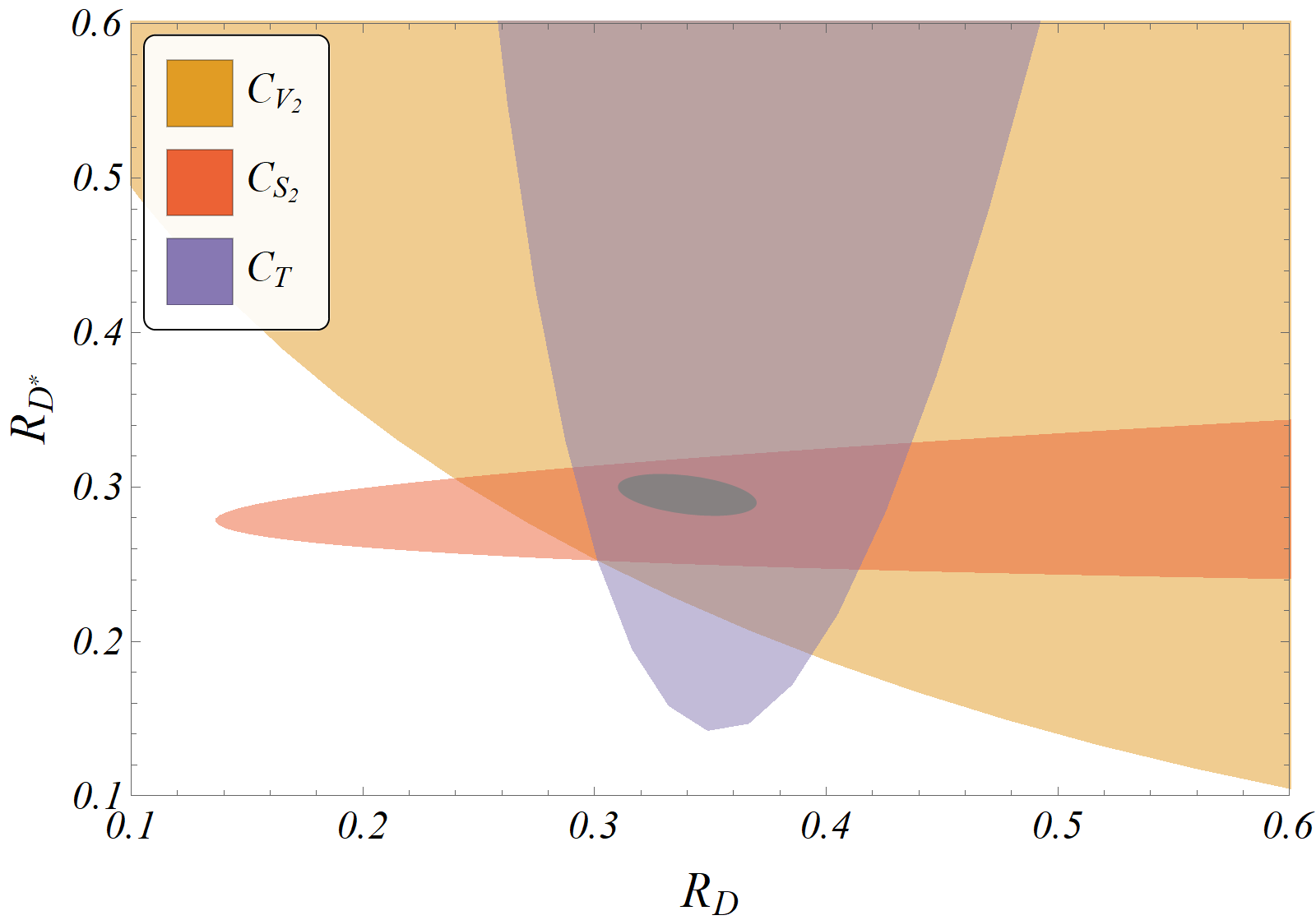}\label{fig:motivPlot2}}
	\caption{Illustrative example to show the complexity of an inverse problem in presence of competing models.}
	\label{fig:motivPlot}
\end{figure*}
%%%%%%%%%%%%%%%%%%%%%%%%%%%%%%%%%%%%%%%%%%%%%%%%%	
	\section{ Problem Statement}\label{sec:motiv}
		\subsection{Model Selection: A Recap}\label{sec:modsel}
%%%%%%%%%%%%%%%%%%%%%%%%%%%%%%%%%%%%%%%%%%%%%%%%%
\subsubsection{The Premise}\label{sec:premise}
As discussed above, presently the standard way of explaining the deviations of LFUV observables from SM is to consider NP in only $b\to c\tau\nu$ channels and construct the most general effective Hamiltonian with all possible four-Fermi operators. This gives rise to five, in general complex, NP Wilson coefficients (WCs). In principle, any high-scale theory can contribute to these decay modes. Since the new particles are expected to be rather heavy, respecting the SM gauge symmetries, their contributions can be parametrized to a good approximation in terms of dimension-six effective operators as given in eq.\ref{eq:effham}. Fixing such a theory beforehand and then trying to find the parameter space of that chosen theory, allowed by these observables, is the standard statistical inference problem. In that case, implementing constraints from other possible channels affected by that theory is the most non-trivial task at hand; as examples, see \cite{Barman:2018jhz,Borah:2020swo}. In practice, instead of constraining the parameter space of a specific theory, an attempt is made to constraint possible combinations of WCs, which will be affected by that theory. This gives us a chance of constraining classes of theories at a time. From hereon, any such combination of WCs will be called a model, i.e., `model' means a combination of NP WCs, not any underlying theory.

In principle, any number of theories could give rise to deviations in these observables. Given an observed data-set, selecting the best (optimal) candidate model to explain the data is a different type of problem, known as the `inverse problem'. The notion of an `optimal model' is very close to the concept of parsimony \cite{boxjenkins}, which in statistical terms is equivalent to the optimization of both bias and variance. Bias decreases and variance increases with varying model complexity. The goal thus becomes to find the `optimally complex' model, which corresponds to minimizing some combination of bias and variance \cite{breiman}. 

%%%%%%%%%%%%%%%%%%%%%%%%%%%%%%%%%%%%%%%%%%%%
\subsubsection{Cross-Validation}\label{sec:crossvalid}
In an ideal regression scenario, in the presence of a very high volume of data, finding the optimal model boils down to using `cross-validation'. If we break up the available data in two parts, one for the model to fit, other to then validate the fit results, then some estimation of the predictive power of the fitted model can be obtained by creating an objective function. If we then repeat this procedure randomly, ensuring to obtain the contribution of every data point in both fitting and validation set, an average measure of the objective function for the model in question can be obtained. In presence of a collection of such candidate models, this measure can then be used to separate the `best' or `optimal' models from the less applicable ones. The best, and computationally most expensive, practice regarding cross-validation is to take out one data-point at a time for validation and repeating the process for all data. This is called leave-one-out-cross-validation (LOOCV). 

%%%%%%%%%%%%%%%%%%%%%%%%%%%%%%%%%%%%%%%%%%%%
\subsubsection{Information Criteria}\label{sec:infocrit}
In realistic scenarios, however, the number of available data is small and often comparable to the size of the most complex model. Also, the pool of candidate models may be too large for the data to be capable of discerning the best ones. For small sample sizes, cross-validation results are unstable and have questionable applicability \cite{BELEITES,Varoquaux}. An alternative way of finding the optimal model comes from information theory. In the frequentist interpretation, given that there is a true model, which is the source of the data, all models used by a practitioner are approximations of that true one. There would be information loss due to this approximation and that loss would be minimum for the best approximating model. In Bayesian/subjective probability interpretation, where no `true' model exists, but the data-distribution is `true', information is lost when a prior distribution $Q$ is used to approximate a posterior $P$, which is representative of the `true' distribution of the data. This information loss can be quantified by some estimator measuring the `divergence' between $P$ and $Q$.

A popular choice of such a divergence is the $D_{KL}$, or `relative entropy', between target distribution $P$ and obtained distribution $Q$ \cite{kullback1951}:
\begin{align}\label{eq:KLdiv}
	D_{KL}(P||Q) = \int_{-\infty}^{\infty} p(\vec{x}) \log{\frac{p(\vec{x})}{q(\vec{x})}} d\vec{x}\,,
\end{align} 
for continuous probability distributions of random variable vectors $\vec{x}$, where $p$ and $q$ are probability density functions of $P$ and $Q$, respectively\footnote{For discrete distributions, the integral changes to summation over the \emph{probability space}.}. Unlike a `distance' measure, this does not need to be symmetric, i.e., $D_{KL}(P||Q) \neq D_{KL}(Q||P)$, in general. The best models approximating the data should thus have the smallest divergence estimates.

To estimate the amount of information loss, depending on these divergences, an IC for each model can be created:
\begin{align}
	IC = - 2 log(\mathcal{L}(\hat{\theta}|y)) + 2 p^*\,,
\end{align}
where $\mathcal{L}(\hat{\theta}|y)$ is the likelihood of the model estimated at the MLE $\hat{\theta}$ of the model parameter vector $\theta$, $y$ is the data-distribution and $p^*$ is the generalized dimension of the model: $p^* = Tr(J^{-1} K)$, where $J$ is the expectation of the information matrix and $K$ is the variance of the score vector. These are taken with respect to the unknown data-generating density $g(y)$. It is instructive to note that the log-likelihood at MLE is an estimate of the bias and $p^*$ is a penalty-function for model-complexity. Together, minimizing any IC over all possible models is thus equivalent to the model-selection problem stated before.

Under the assumption that some approximating model $f_i$ is correct, i.e., $g(y) = f_i(\hat{\theta}|y)$, $p^* \equiv length(\theta) = p$, where $p$ is the number of parameters in the model. In this case, the IC is called Akaike Information Criterion (AIC) \cite{akaike}. Though it has been shown that minimizing AIC is asymptotically equivalent to cross-validation \cite{shibata}, AIC is prone to select more complex models with increasing sample size. A second-order, corrected version of AIC, noted as AIC$_c$ \cite{sugiura78}, takes the sample-size $N$ into account in the following manner:
\begin{align}
	\text{AIC}_c = - 2 log(\mathcal{L}(\hat{\theta}|y)) + 2 p \frac{N}{N - p - 1}\,.
\end{align}
Whereas AIC$_c$ for all models under consideration are on a relative scale, $\Delta^{AIC}_i = {\rm AIC}^i_c - {\rm AIC}^{min}_c$ estimates the relative expected information loss for the $i$-th model, enabling us to rank them in increasing order of $\Delta^{AIC}_i$. It is possible to quantify the weight of evidence$/$probability in favor of a model by defining a set  of positive ``Akaike weights", $w^{\Delta\text{AIC}_c}_i = (e^{(-\Delta^{AIC}_i / 2)})/(\sum_{r = 1}^R e^{(-\Delta^{AIC}_r / 2)})$, adding up to $1$ \cite{Burnham}. This is equivalent to applying a Soft-max function, to find class probability, in multi-class classification problems.
%%%%%%%%%%%%%%%%%%%%%%%%%%%%%%%%%%%
\begin{figure*}[hbt]
	\small
	\centering
	\includegraphics[width=0.95\textwidth]{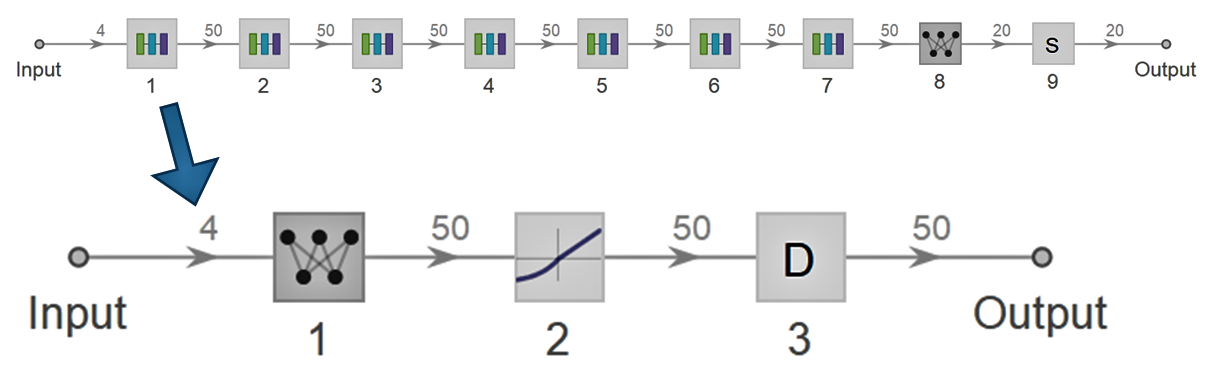}
	\caption{Structure of the SNN used in this analysis. The graph above shows the full network, and the one below shows the expanded form of one module containing one fully-connected (linear) layer of 50 neurons, a SELU activation layer, and one Alpha-Dropout layer. Layers labeled 8 and 9 are a linear layer of width same as the number of possible models and a Soft-max layer, respectively. The shown network is for the 4-observable data-set. For other data-sets, the input dimension would change accordingly. During training for classification, a cross-entropy loss layer is added to the output and the target, while for regression, it is replaced with a mean-squared loss layer.}
	\label{fig:netgraph}
\end{figure*}
%%%%%%%%%%%%%%%%%%%%%%%%%%%%%%%%%%
%%%%%%%%%%%%%%%%%%%%%%%%%%%%%%%%%%%
\begin{figure*}[hbt]
	\small
	\centering
	\subfloat[Loss (4 Obs.)]{\includegraphics[width=0.32\textwidth]{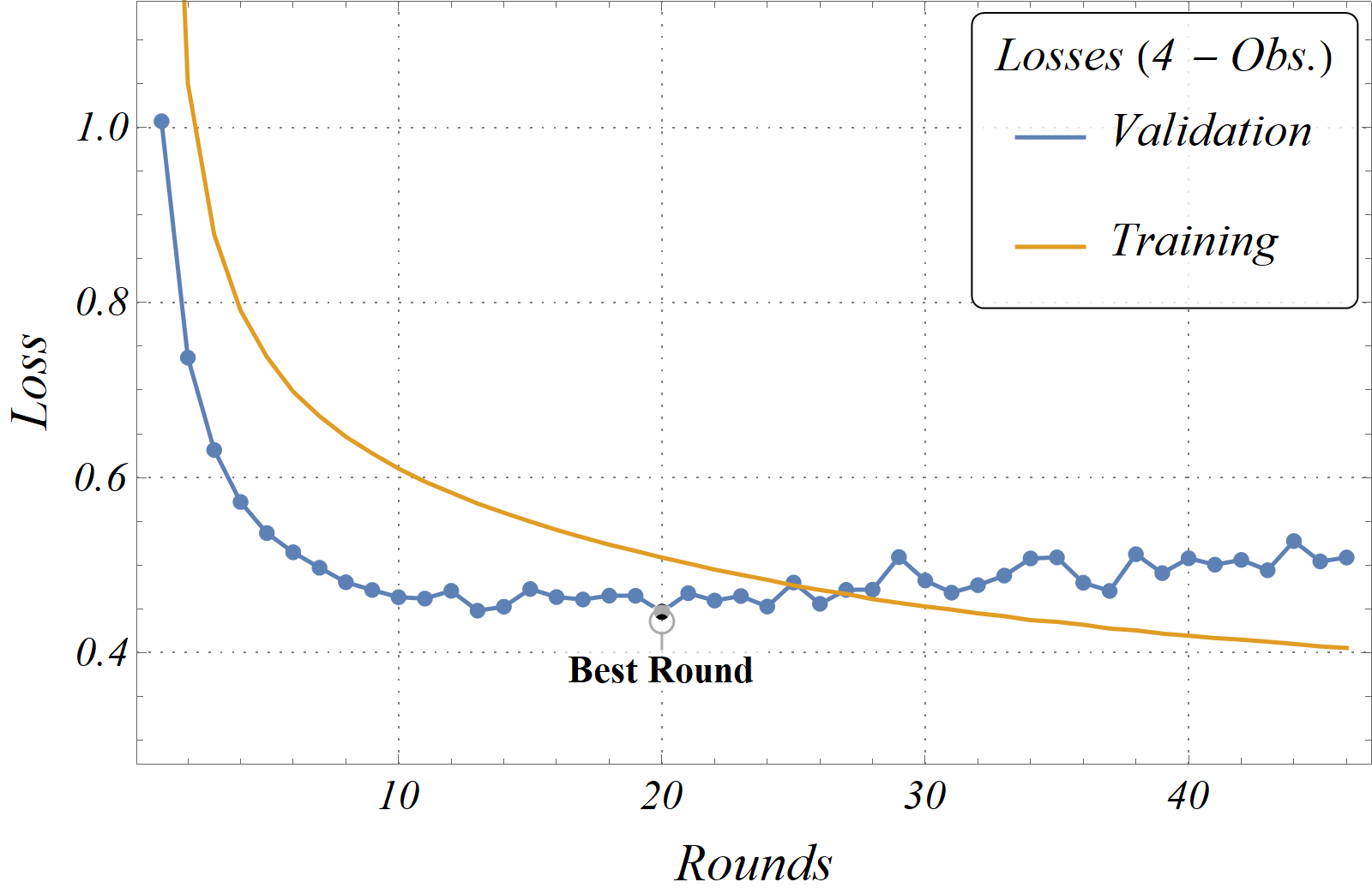}\label{fig:loss4}}~
	\subfloat[Loss (5 Obs.)]{\includegraphics[width=0.32\textwidth]{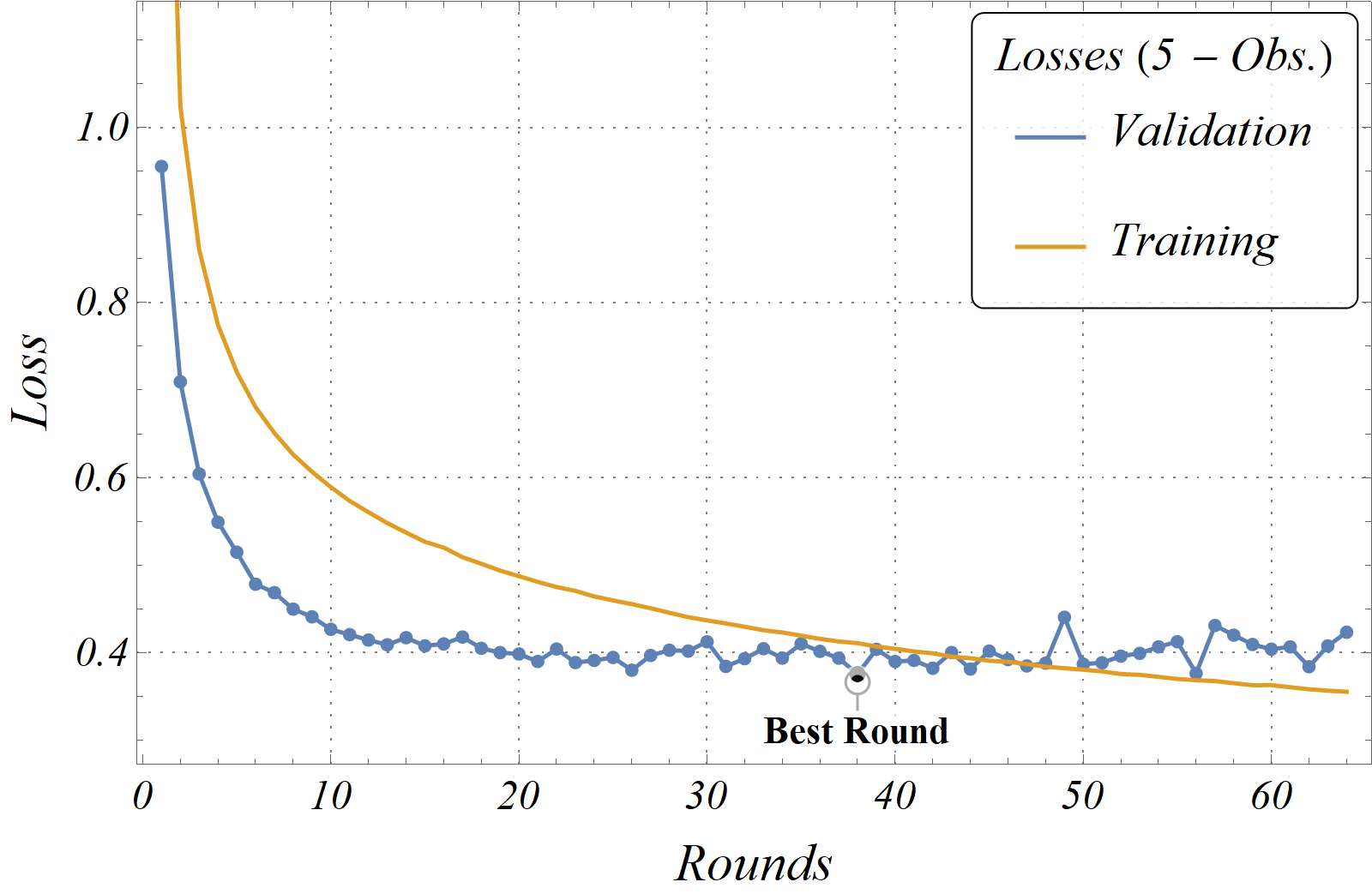}\label{fig:loss5}}~
	\subfloat[Loss (13 Obs.)]{\includegraphics[width=0.32\textwidth]{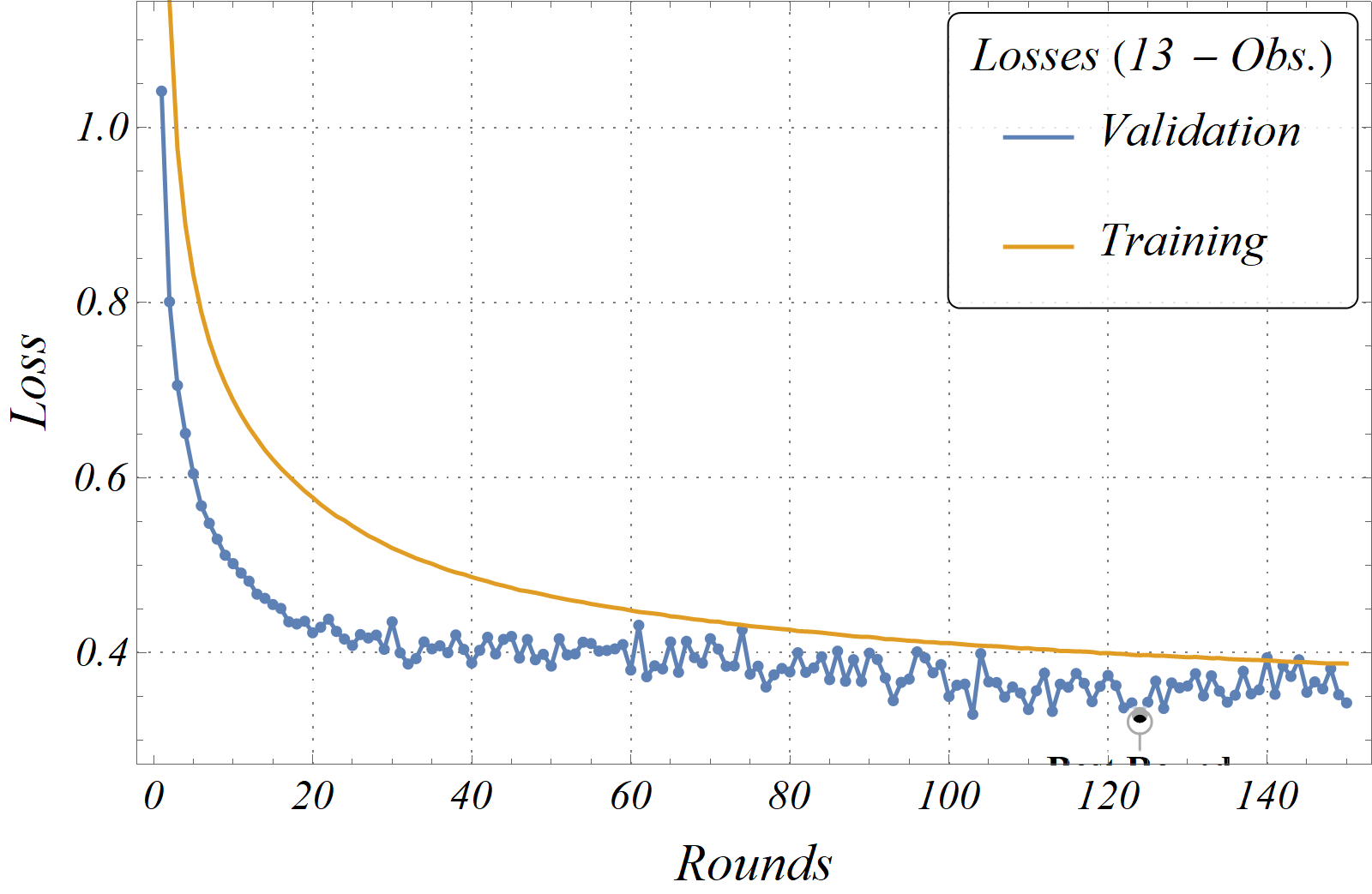}\label{fig:loss13}}\\
	\subfloat[Error (4 Obs.)]{\includegraphics[width=0.32\textwidth]{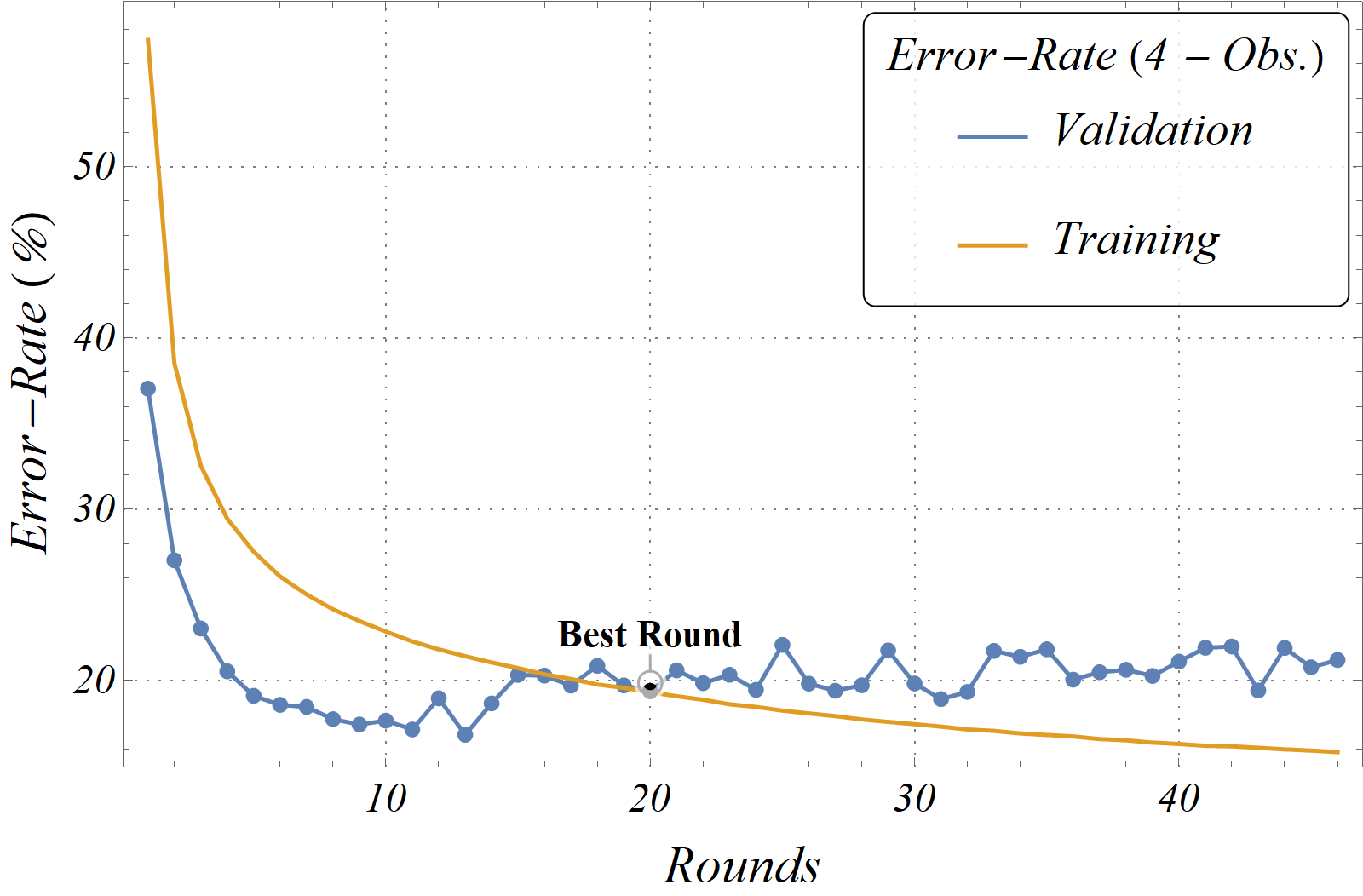}\label{fig:error4}}~
	\subfloat[Error (5 Obs.)]{\includegraphics[width=0.32\textwidth]{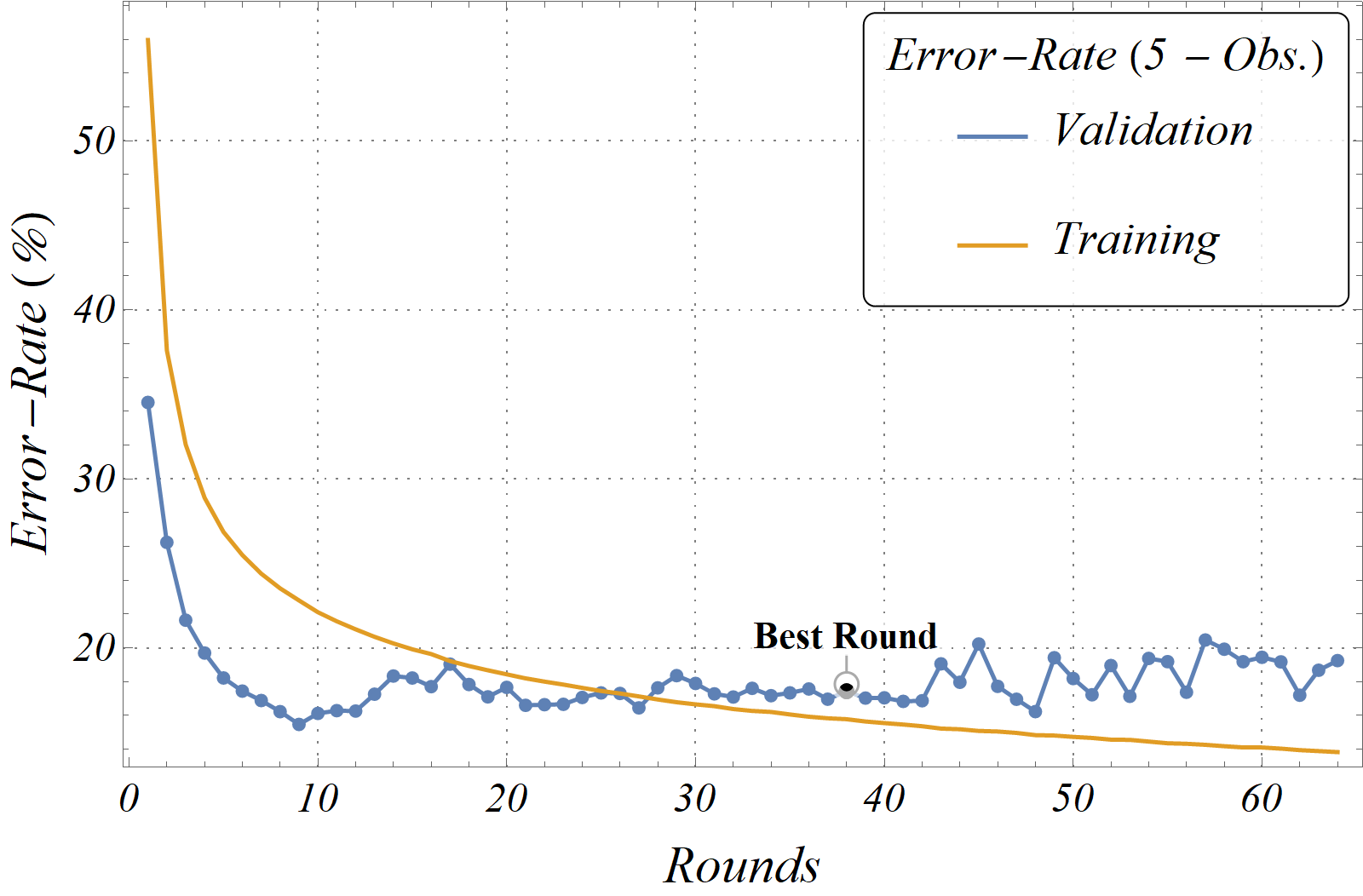}\label{fig:error5}}~
	\subfloat[Error (13 Obs.)]{\includegraphics[width=0.32\textwidth]{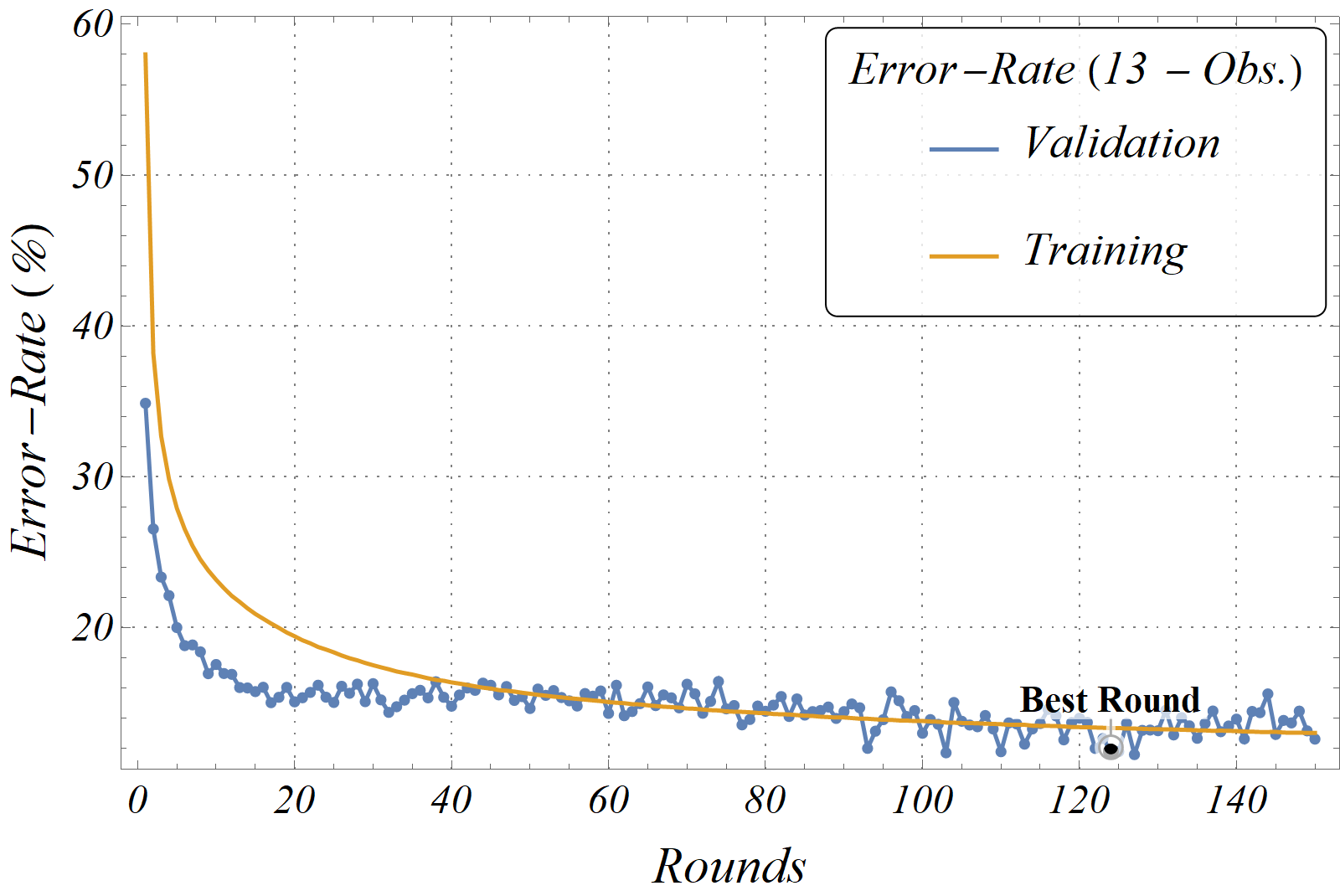}\label{fig:error13}}
	\caption{Plot of Loss and Error-rate for both training (orange) and validation (blue), w.r.t. number of rounds, for all data-sets. Best rounds are also shown. See section \ref{sec:methodclass} for details.}
	\label{fig:errloss}
\end{figure*}
%%%%%%%%%%%%%%%%%%%%%%%%%%%%%%%%%%
%%%%%%%%%%%%%%%%%%%%%%%%%%%%%%%%%%%
\begin{figure}[hbt]
	\small
	\centering
	\includegraphics[width=0.45\textwidth]{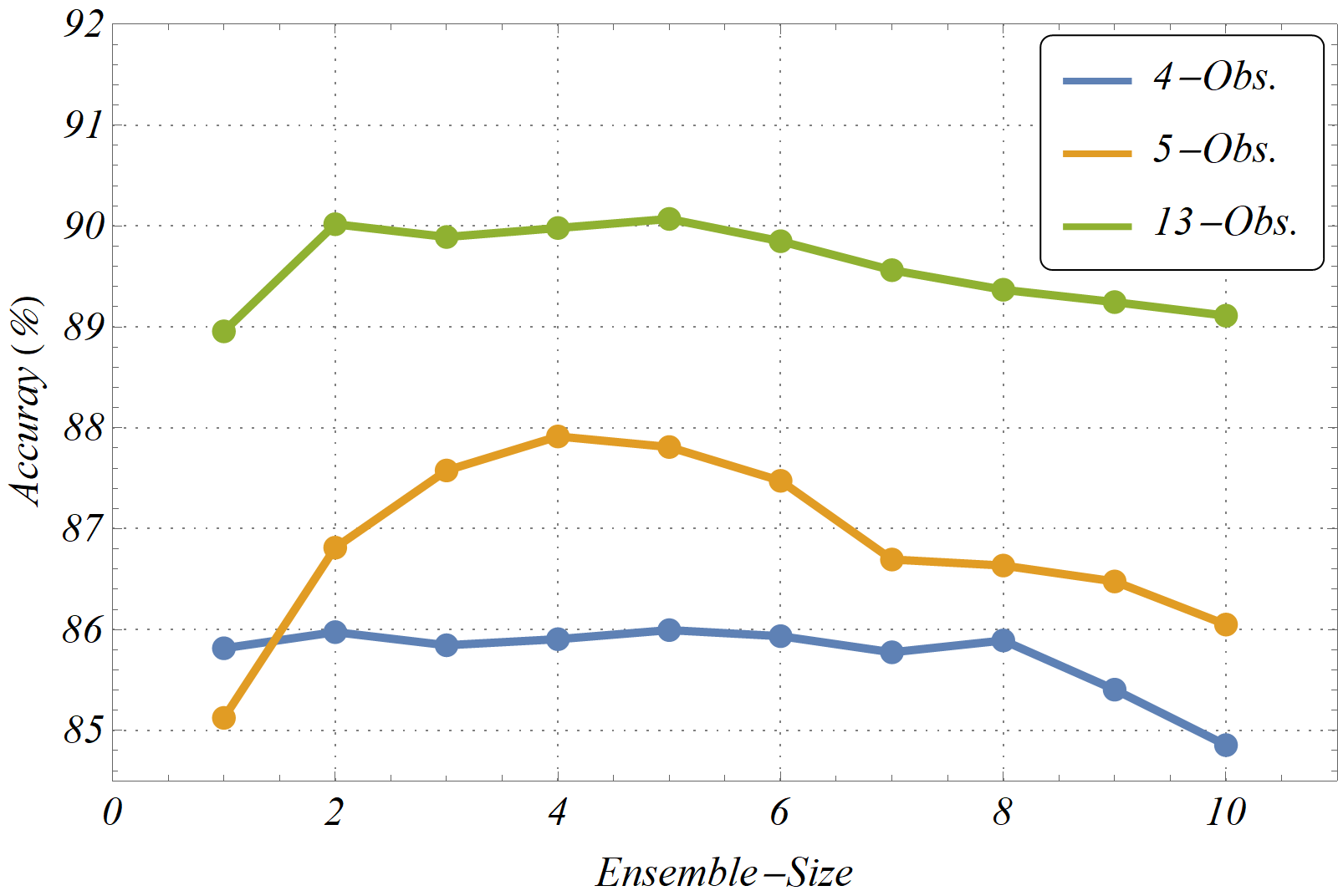}
	\caption{Variation of classification accuracy of `ensemble-net' with number of constituent nets. Each point corresponding to a specific $n$ represents the highest accuracy possible among all $n$-combinations.}
	\label{fig:accucompare}
\end{figure}
%%%%%%%%%%%%%%%%%%%%%%%%%%%%%%%%%%%
%%%%%%%%%%%%%%%%%%%%%%%%%%%%%%%%%%%
\begin{figure*}[hbt]
	\small
	\centering
	\subfloat[4-obs.]{\includegraphics[width=0.32\textwidth]{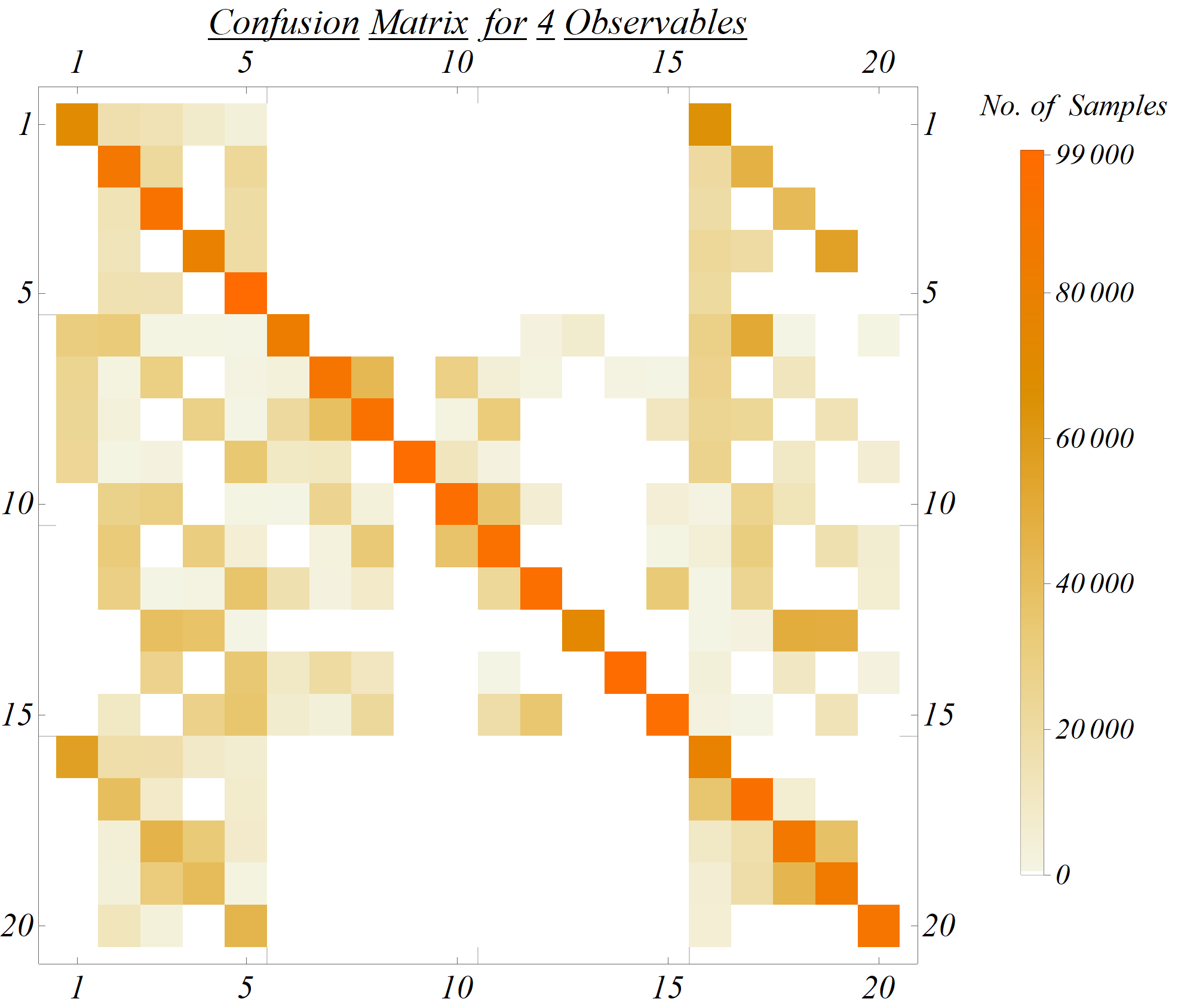}\label{fig:confusionob4}}~
	\subfloat[5-obs.]{\includegraphics[width=0.32\textwidth]{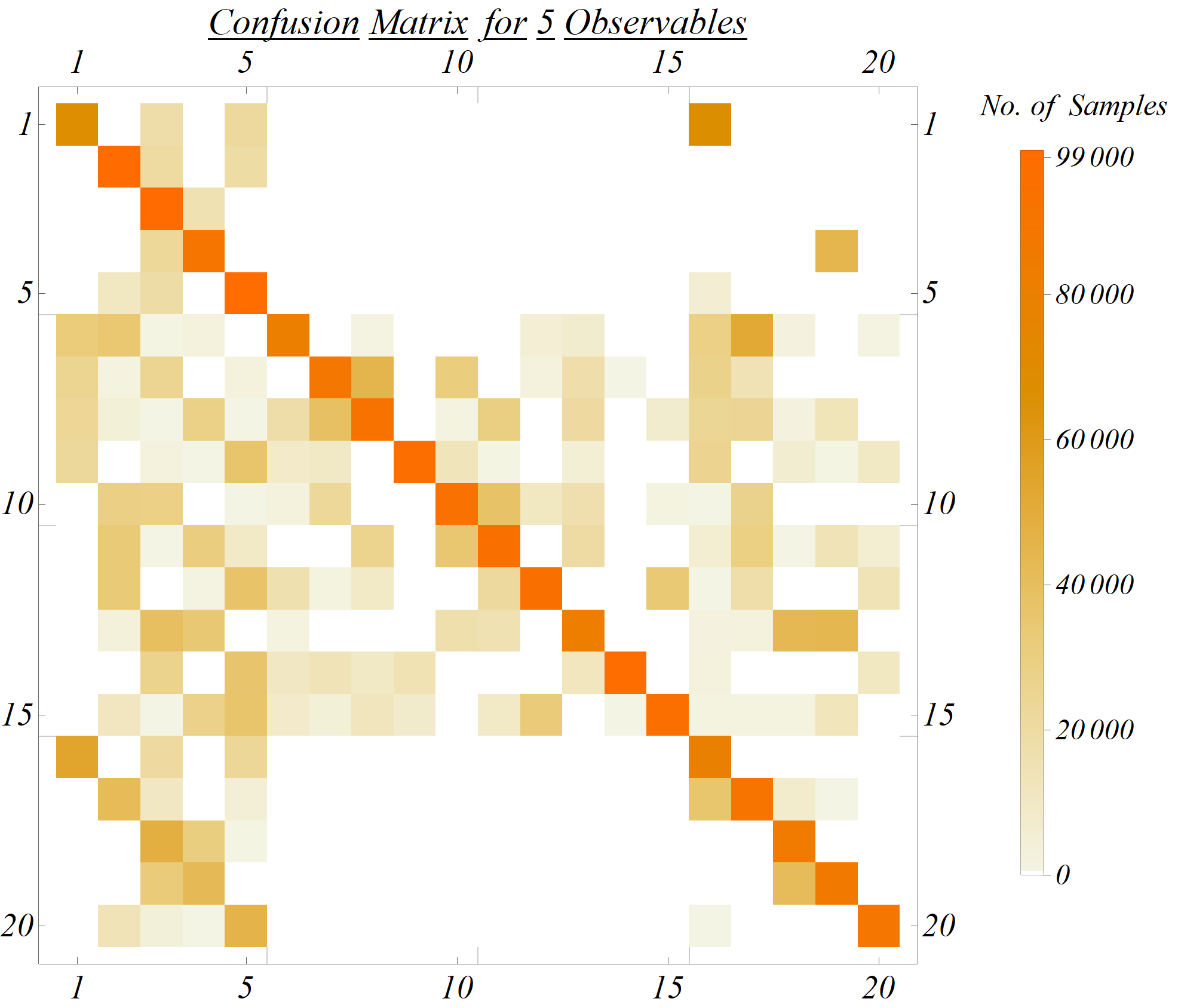}\label{fig:confusionob5}}~
	\subfloat[13-obs.]{\includegraphics[width=0.32\textwidth]{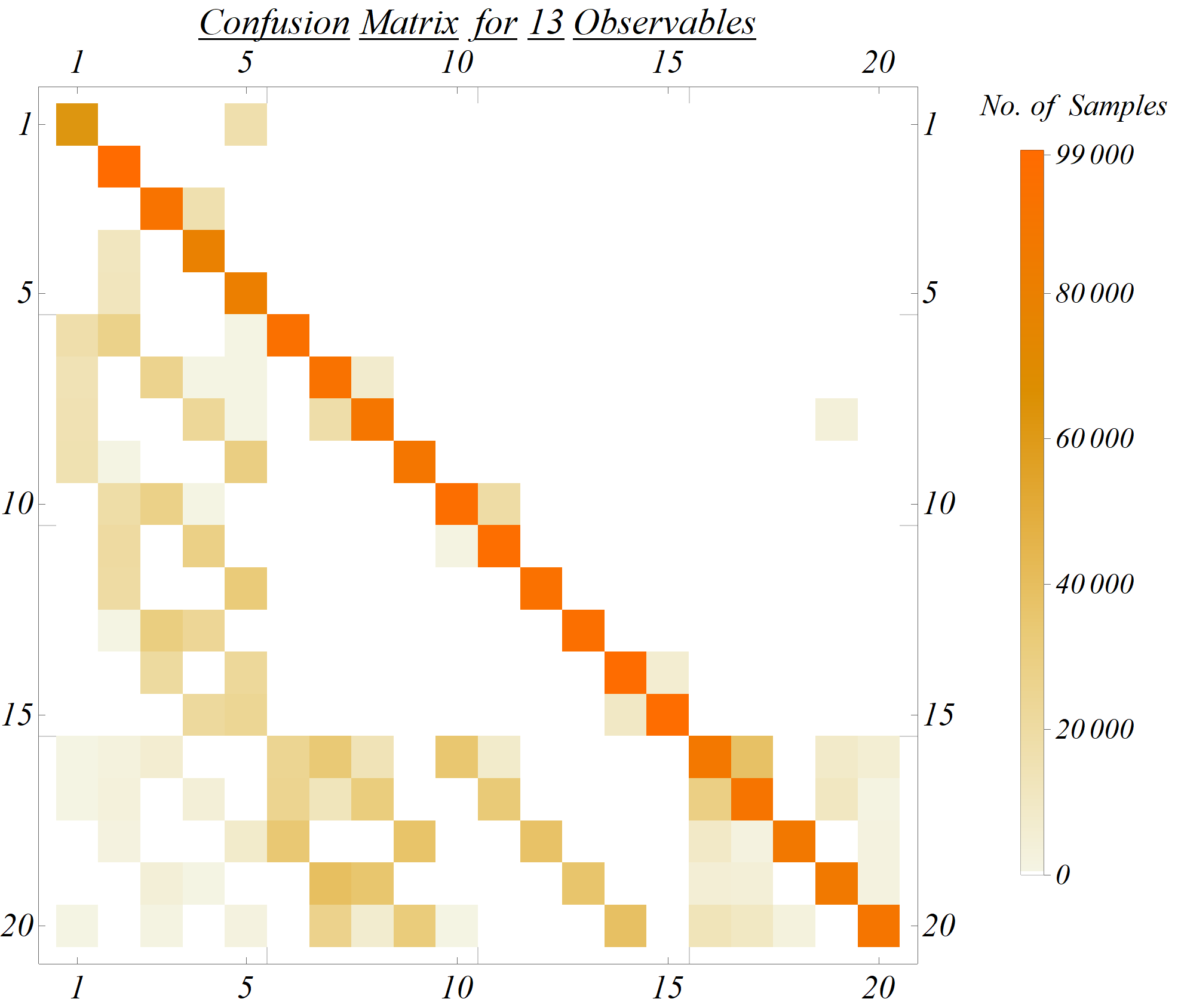}\label{fig:confusionob13}}
	\caption{Confusion matrices for ensemble-nets for all data-sets, after applied on test-data. See section \ref{sec:methodclass}.}
	\label{fig:confusion}
\end{figure*}
%%%%%%%%%%%%%%%%%%%%%%%%%%%%%%%%%%

%%%%%%%%%%%%%%%%%%%%%%%%%%%%%%%%%%
\subsection{Proposal: Supervised Classification}\label{sec:superclass}
%%%%%%%%%%%%%%%%%%%%%%%%%%%%%%%%%%
Unlike other real-world problems, where the data comes first and the model-space is then distributed to explain those, a global collection of all possible models is already available in the form of the combinations of WCs for the particular case at hand. This gives rise to an unique opportunity of using our theoretical knowledge of the observables for model selection. It is possible, using that knowledge, to know beforehand, which models best explain any precise data in the future. This is not possible using general statistical techniques, which require an {\em a priori} data-distribution for any inference, be it regression or model-selection.

%\subsubsection{Illustrative Example}\label{sec:motivexample}
As an instructive example of the last point, let us consider just the two LFUV observables $\mathcal{R}_{D^{(*)}}$ and five scenarios$/$models with only one type of real WC. Fig. \ref{fig:motivPlot1} shows the parametric variation of these observables with each WC. The arrows show the direction of change with the increase of the WCs\footnote{For the case of $Re(C_{V_1})$, the arrows switch direction for negative and positive values of the parameter.}. Obviously, they all intersect at the SM estimate of the observables. This is naturally an idealistic scenario. In presence of theoretical uncertainties (e.g. from form-factor estimates), these will be fuzzy lines, but the idea holds. For these simple models at least, position of the experimental results on this plane points us somewhat to possible model(s) responsible for that result. This inference becomes more probabilistic with the inclusion of experimental uncertainties, which are still considerable (as shown by the gray ellipse corresponding to the $68\%$ confidence region of present global average of $\mathcal{R}_{D^{(*)}}$ \cite{Amhis:2019ckw})  Overlap regions of the model-lines are positions of high entropy, with multiple models present as candidates for explaining the data situated at those points.

The situation quickly worsens after the inclusion of models with more than one parameter. The overlap regions become too big to reasonably isolate the best candidate models visually, even when experimental results are very precise. In fig. \ref{fig:motivPlot2} we show the possible regions for just three models with complex WCs. The boundaries of these regions are the corresponding cases from Fig. \ref{fig:motivPlot1}. It quickly becomes evident that discriminating between the models using just the experimental result becomes impossible (the global average is equivalently likely from all three models). 

The model-discriminating capability of the data increases if we consider more observables. However, we face a different kind of problem in that case. It becomes increasingly problematic to visualize the attributes of the higher-dimensional data-set that are responsible for discriminating between the models. This is the quintessential `curse of dimensionality' \cite{bellman2013dynamic}.

To solve the `inverse problem' stated in section \ref{sec:premise}, inspired by the arguments as mentioned earlier, we may be motivated to define an `inverse' classifier function which, when given an $N$-dimensional data-vector as input, churns out a specific model as the best candidate. More generally, we expect to construct a vector of probability content for each possible model. Though traditional statistical procedures will be of little help to us here, newly emerging machine learning techniques give us a clue for solving this using supervised learning. Unsupervised learning methods on the other hand can give us a handle in visualizing the underlying features of the data by reducing the feature-dimension but are not the concern for the present work.

%%%%%%%%%%%%%%%%%%%%%%%%%%%%%%%%%%%%%%%%%
\begin{table*}[!t]
	\begin{ruledtabular}
		\small
		\renewcommand*{\arraystretch}{1.2}
		\begin{tabular}{*{6}{c}}
			Index & Parameters & Index & Parameters & Index & Parameters \\
			\hline
			$1$(1)  &  $\left.\text{Re(}C_{V_1}\right)$  &  $16$  &  $\left.\text{Re(}C_{V_1}\text{), Re(}C_{V_2}\text{), Re(}C_{S_1}\right)$  &  $31$(16)  &  $C_{V_1}$  \\
			$2$(2)  &  $\left.\text{Re(}C_{V_2}\right)$  &  $17$  &  $\left.\text{Re(}C_{V_1}\text{), Re(}C_{V_2}\text{), Re(}C_{S_2}\right)$  &  $32$(17)  &  $C_{V_2}$  \\
			$3$(3)  &  $\left.\text{Re(}C_{S_1}\right)$  &  $18$  &  $\left.\text{Re(}C_{V_1}\text{), Re(}C_{V_2}\text{), Re(}C_T\right)$  &  $33$(18)  &  $C_{S_1}$  \\
			$4$(4)  &  $\left.\text{Re(}C_{S_2}\right)$  &  $19$  &  $\left.\text{Re(}C_{V_1}\text{), Re(}C_{S_1}\text{), Re(}C_{S_2}\right)$  &  $34$(19)  &  $C_{S_2}$  \\
			$5$(5)  &  $\left.\text{Re(}C_T\right)$  &  $20$  &  $\left.\text{Re(}C_{V_1}\text{), Re(}C_{S_1}\text{), Re(}C_T\right)$  &  $35$(20)  &  $C_{T}$  \\
			$6$(6)  &  $\left.\text{Re(}C_{V_1}\text{), Re(}C_{V_2}\right)$  &  $21$  &  $\left.\text{Re(}C_{V_1}\text{), Re(}C_{S_2}\text{), Re(}C_T\right)$  &  $36$  &  $C_{V_1}, ~C_{V_2}$  \\
			$7$(7)  &  $\left.\text{Re(}C_{V_1}\text{), Re(}C_{S_1}\right)$  &  $22$  &  $\left.\text{Re(}C_{V_2}\text{), Re(}C_{S_1}\text{), Re(}C_{S_2}\right)$  &  $37$  &  $C_{V_1}, ~C_{S_1}$  \\
			$8$(8)  &  $\left.\text{Re(}C_{V_1}\text{), Re(}C_{S_2}\right)$  &  $23$  &  $\left.\text{Re(}C_{V_2}\text{), Re(}C_{S_1}\text{), Re(}C_T\right)$  &  $38$  &  $C_{V_1}, ~C_{S_2}$  \\
			$9$(9)  &  $\left.\text{Re(}C_{V_1}\text{), Re(}C_T\right)$  &  $24$  &  $\left.\text{Re(}C_{V_2}\text{), Re(}C_{S_2}\text{), Re(}C_T\right)$  &  $39$  &  $C_{V_1}, ~C_{T}$  \\
			$10$(10)  &  $\left.\text{Re(}C_{V_2}\text{), Re(}C_{S_1}\right)$  &  $25$  &  $\left.\text{Re(}C_{S_1}\text{), Re(}C_{S_2}\text{), Re(}C_T\right)$  &  $40$  &  $C_{V_2}, ~C_{S_1}$  \\
			$11$(11)  &  $\left.\text{Re(}C_{V_2}\text{), Re(}C_{S_2}\right)$  &  $26$  &  $\left.\text{Re(}C_{V_1}\text{), Re(}C_{V_2}\text{), Re(}C_{S_1}\text{), Re(}C_{S_2}\right)$  &  $41$  &  $C_{V_2}, ~C_{S_2}$  \\
			$12$(12)  &  $\left.\text{Re(}C_{V_2}\text{), Re(}C_T\right)$  &  $27$  &  $\left.\text{Re(}C_{V_1}\text{), Re(}C_{V_2}\text{), Re(}C_{S_1}\text{), Re(}C_T\right)$  &  $42$  &  $C_{V_2}, ~C_{T}$  \\
			$13$(13)  &  $\left.\text{Re(}C_{S_1}\text{), Re(}C_{S_2}\right)$  &  $28$  &  $\left.\text{Re(}C_{V_1}\text{), Re(}C_{V_2}\text{), Re(}C_{S_2}\text{), Re(}C_T\right)$  &  $43$  &  $C_{S_1}, ~C_{S_2}$  \\
			$14$(14)  &  $\left.\text{Re(}C_{S_1}\text{), Re(}C_T\right)$  &  $29$  &  $\left.\text{Re(}C_{V_1}\text{), Re(}C_{S_1}\text{), Re(}C_{S_2}\text{), Re(}C_T\right)$  &  $44$  &  $C_{S_1}, ~C_{T}$  \\
			$15$(15)  &  $\left.\text{Re(}C_{S_2}\text{), Re(}C_T\right)$  &  $30$  &  $\left.\text{Re(}C_{V_2}\text{), Re(}C_{S_1}\text{), Re(}C_{S_2}\text{), Re(}C_T\right)$  &  $45$  &  $C_{S_2}, ~C_{T}$  \\
		\end{tabular}
	\end{ruledtabular}
	\caption{List of all models used in this analysis. The indices within parentheses are for the 4 and 5 observable data-set. Those without, are for the 13 - observable one. See section \ref{sec:methodclass}.}
	\label{tab:modlst}
\end{table*} 
%%%%%%%%%%%%%%%%%%%%%%%%%%%%%%%%%%%%%%%%%
%%%%%%%%%%%%%%%%%%%%%%%%%%%%%%%%%%%%%%%%%
\begin{table}[!t]
	\begin{ruledtabular}
		\renewcommand*{\arraystretch}{1.2}
		\begin{tabular}{*{2}{c}}
			Training Info. & Details\\
			\hline
			Batches/Round  &  $170$  \\
			Batch Size  &  $10000$  \\
			Best Valid. Round  &  $20$  \\
			Final $\ell$  &  $0.0009998$  \\
			Initial $\ell$  &  $0.001$  \\
			Avg. Batches/Second  &  $81.65$  \\
			Final Round Loss  &  $0.406$  \\
			Final Round Error  &  $14.48\%$  \\
			Total Batches  &  $7820$  \\
			Total Rounds  &  $46$  \\
			%			Total Training Time (Sec)  &  $188.093$  \\
			Validation Loss  &  $0.51$  \\
			Validation Error-Rate  &  $21\%$  \\
		\end{tabular}
	\end{ruledtabular}
	\caption{Details of training of a single SNN for the 4-obs. data-set. See section \ref{sec:methodclass} for details.}
	\label{tab:netdetails}
\end{table} 
%%%%%%%%%%%%%%%%%%%%%%%%%%%%%%%%%%%%%%%%%%%%
%%%%%%%%%%%%%%%%%%%%%%%%%%%%%%%%%%%%%%%%%
\begin{table}[hbt]
	\begin{ruledtabular}
		\renewcommand*{\arraystretch}{1.2}
		\begin{tabular}{*{4}{c}}
			Measure  				&  4 Obs.	  &  5 Obs.		&  13 Obs. \\
			\hline
			Accuracy		  		&  $85.88\%$  &  $88.67\%$  &  $89.68\%$  \\
			Cohen's $\kappa$  		&  $85.14\%$  &  $88.08\%$  &  $89.42\%$  \\
			Error     		  		&  $14.12\%$  &  $11.33\%$  &  $10.32\%$  \\
			Geometric Mean Prob.  	&  $69.54\%$  &  $71.97\%$  &  $71.75\%$  \\
			Mean Cross-Entropy	    &  $0.363$	  &  $0.329$    &  $0.332$  \\
			Mean Decision Utility   &  $85.88\%$  &  $88.67\%$  &  $89.68\%$  \\
			Perplexity			    &  $1.44$	  &  $1.39$ 	&  $1.39$  \\
			Scott's $\pi$ 			&  $85.13\%$  &  $88.07\%$  &  $89.42\%$  \\
		\end{tabular}
	\end{ruledtabular}
	\caption{Results of various performance measures for each ensemble-nets trained on the three observable-sets. See section \ref{sec:methodclass} for details.}
	\label{tab:netmeasuresClass}
\end{table} 
%%%%%%%%%%%%%%%%%%%%%%%%%%%%%%%%%%%%%%%%%
%%%%%%%%%%%%%%%%%%%%%%%%%%%%%%%%%%%%%%%%%
 \begin{table}[t]
 	\begin{ruledtabular}
 		\renewcommand*{\arraystretch}{1.2}
 		\begin{tabular}{*{2}{c}}
 			Classifier & Accuracy\\
 			\hline
 			\textbf{ensemble-SNN}   &  \textbf{89.08\%}  \\
 			Random Forest  &  $82.70\%$  \\
 			Decision Tree  &  $78.27\%$  \\
 			Nearest Neighbors &  $74.22\%$  \\
 			Gradient Boosted Trees  &  $74.10\%$  \\
 			Support Vector Machine  &  $63.82\%$  \\
 			Markov Process  &  $50.76\%$  \\
 			Naive Bayes  &  $46.64\%$  \\
 			Logistic Regression  &  $29.80\%$  \\
 		\end{tabular}
 	\end{ruledtabular}
 	\caption{Comparison of accuracy of ensemble-SNN with that of various shallow learning processes. See section \ref{sec:methodclass} for details.}
 	\label{tab:classcompare}
 \end{table} 
%%%%%%%%%%%%%%%%%%%%%%%%%%%%%%%%%%%%%%%%%%%%
%%%%%%%%%%%%%%%%%%%%%%%%%%%%%%%%%%%%%%%%%%%%
\begin{table*}[hbt]
	\begin{ruledtabular}
		\renewcommand*{\arraystretch}{1.2}
		\begin{tabular}{*{8}{c}}
			Data-Set & Models & Parameters & Aggregate  & $D_{KL}$ & SNN-Central & $\Delta\text{AIC}_c$ & $w^{\Delta\text{AIC}_c}$ \\
			& (SNN-Aggregate) &  & Prob. ($\%$) & Serial   & Serial   & Serial  & ($\%$) \\
			\hline
			& \textbf{12}  &  $Re(C_{V_2}),~Re(C_{T})$  	&  \textbf{38.48}  &  \textbf{2}  &  \textbf{1}  &  $18$  &  $0.07$  \\
			& \textbf{15}  &  $Re(C_{S_2}),~Re(C_{T})$  	&  \textbf{26.30}  &  \textbf{3}  &  \textbf{5}  &  \textbf{6}   &  \textbf{0.32}  \\
			& $13$  &  $Re(C_{S_1}),~Re(C_{S_2})$  	&  $11.53$  &  $9$  &  $7$  &  $9$   &  $0.29$  \\
			& \textbf{14}  &  $Re(C_{S_1}),~Re(C_{T})$  	&  $8.04$  	&  \textbf{1}  &  $15$ &  $17$  &  $0.11$  \\
			4-Obs.
			& $8$  	&  $Re(C_{V_1}),~Re(C_{S_2})$  	&  $5.01$  	&  $18$ &  \textbf{3}  &  \textbf{7}   &  \textbf{0.32}  \\
			& \textbf{10}  &  $Re(C_{V_2}),~Re(C_{S_1})$  	&  $2.80$  	&  \textbf{6}  &  \textbf{2}  &  $11$  &  $0.16$  \\
			& $6$  	&  $Re(C_{V_1}),~Re(C_{V_2})$  	&  $2.19$  	&  $14$ &  $9$  &  $12$  &  $0.12$  \\
			& \textbf{11}  &  $Re(C_{V_2}),~Re(C_{S_2})$  	&  $1.74$  	&  \textbf{4}  &  \textbf{4}  &  \textbf{8}   &  \textbf{0.32}  \\
			& $19$  &  $C_{S_2}$  					&  $1.18$  	&  $17$ &  $6$  &  $10$  &  $0.29$  \\
			%	$9$  &  $\{Re(C_{V_1}),Re(C_{T})\}$  &  $0.81$  &  $5$  &  $8$  &  $15$  &  $0.12$  \\
			%	$17$  &  $C_{V_2}$  &  $0.59$  &  $15$  &  $12$  &  $13$  &  $0.12$  \\
			%	$7$  &  $\{Re(C_{V_1}),Re(C_{S_1})\}$  &  $0.58$  &  $11$  &  $16$  &  $14$  &  $0.12$  \\
			%	$4$  &  $\{Re(C_{S_2})\}$  &  $0.41$  &  $10$  &  $13$  &  $2$  &  $38.92$  \\
			%	$20$  &  $C_{T}$  &  $0.19$  &  $7$  &  $11$  &  $19$  &  $0.03$  \\
			%	$5$  &  $\{Re(C_{T})\}$  &  $0.07$  &  $12$  &  $14$  &  $3$  &  $10.15$  \\
			%	$16$  &  $\{Re(C_{V_1}),\text{icv1}\}$  &  $0.02$  &  $13$  &  $20$  &  $16$  &  $0.12$  \\
			%	$2$  &  $\{Re(C_{V_2})\}$  &  $0.02$  &  $20$  &  $10$  &  $5$  &  $0.34$  \\
			%	$1$  &  $\{Re(C_{V_1})\}$  &  $0.01$  &  $16$  &  $19$  &  $1$  &  $46.95$  \\
			%	$18$  &  $\{Re(C_{S_1}),\text{ics1}\}$  &  $0.01$  &  $8$  &  $17$  &  $20$  &  $0.$  \\
			%	$3$  &  $\{Re(C_{S_1})\}$  &  $0.$  &  $19$  &  $18$  &  $4$  &  $1.14$  \\
			\hline
			&  \textbf{12}  &  $Re(C_{V_2}),~Re(C_{T})$  &  \textbf{42.62}  &  \textbf{1}  &  \textbf{2}  &  $17$  &  $0.69$  \\
			&  $13$  &  $Re(C_{S_1}),~Re(C_{S_2})$  &  $15.71$  &  $19$  &  \textbf{6}  &  \textbf{7}  &  \textbf{3.19}  \\
			&  \textbf{15}  &  $Re(C_{S_2}),~Re(C_{T})$  &  $8.56$  &  \textbf{4}  &  $10$  &  \textbf{4}  &  \textbf{3.37}  \\
			&  $6$  &  $Re(C_{V_1}),~Re(C_{V_2})$  &  $6.7$  &  $16$  &  $12$  &  $11$  &  $1.33$  \\
			5-Obs. &  $8$  &  $Re(C_{V_1}),~Re(C_{S_2})$  &  $6.54$  &  $18$  &  \textbf{4}  &  \textbf{5}  &  \textbf{3.31}  \\
			&  \textbf{14}  &  $Re(C_{S_1}),~Re(C_{T})$  &  $6.09$  &  \textbf{2}  &  $7$  &  $15$  &  $1.14$  \\
			&  $7$  &  $Re(C_{V_1}),~Re(C_{S_1})$  &  $4.63$  &  $8$  &  $11$  &  $12$  &  $1.29$  \\
			&  \textbf{11}  &  $Re(C_{V_2}),~Re(C_{S_2})$  &  $3.7$  &  \textbf{3}  &  \textbf{3}  &  \textbf{6}  &  \textbf{3.3}  \\
			&  \textbf{10}  &  $Re(C_{V_2}),~Re(C_{S_1})$  &  $2.66$  &  \textbf{6}  &  \textbf{1}  &  $9$  &  $1.67$  \\
			&  $17$  &  $C_{V_2}$  &  $1.45$  &  $15$  &  $9$  &  $10$  &  $1.33$  \\
			%	$9$  &  $\{Re(C_{V_1}),Re(C_{T})\}$  &  $0.96$  &  $5$  &  $5$  &  $13$  &  $1.27$  \\
			%	$20$  &  $\{Re(C_{T}),\text{ict}\}$  &  $0.1$  &  $20$  &  $17$  &  $18$  &  $0.29$  \\
			%	$2$  &  $\{Re(C_{V_2})\}$  &  $0.09$  &  $14$  &  $8$  &  $19$  &  $0.25$  \\
			%	$5$  &  $\{Re(C_{T})\}$  &  $0.07$  &  $7$  &  $16$  &  $3$  &  $7.64$  \\
			%	$19$  &  $\{Re(C_{S_2}),\text{ics2}\}$  &  $0.05$  &  $11$  &  $15$  &  $8$  &  $3.07$  \\
			%	$4$  &  $\{Re(C_{S_2})\}$  &  $0.04$  &  $12$  &  $18$  &  $2$  &  $30.$  \\
			%	$16$  &  $\{Re(C_{V_1}),\text{icv1}\}$  &  $0.01$  &  $17$  &  $19$  &  $14$  &  $1.24$  \\
			%	$3$  &  $\{Re(C_{S_1})\}$  &  $0.01$  &  $10$  &  $13$  &  $16$  &  $0.74$  \\
			%	$1$  &  $\{Re(C_{V_1})\}$  &  $0.01$  &  $13$  &  $20$  &  $1$  &  $34.84$  \\
			%	$18$  &  $\{Re(C_{S_1}),\text{ics1}\}$  &  $0.01$  &  $9$  &  $14$  &  $20$  &  $0.03$  \\
		\end{tabular}
	\end{ruledtabular}
	\caption{Main results of model selection using SNN classifier for the 4 and 5-obs cases. The $2^{nd}$ - $4^{th}$ columns list best model-indices, parameters, and classification probabilities ($>1\%$). Next 3 columns list their positions in lists sorted by $D_{KL}$, SNN applied on just central values, and $\Delta\text{AIC}_c$, respectively. Corresponding $w^{\Delta\text{AIC}_c}$ values are listed in the last column. See section \ref{sec:resclass} for details.}
	\label{tab:classnetrescompare}
\end{table*} 
%%%%%%%%%%%%%%%%%%%%%%%%%%%%%%%%%%%%%%%%%%%%
One possible methodology to do this would be to perform the following procedure:
\begin{itemize}
	\item Select a specific model. Varying the parameters of that model in a wide, theoretically allowed region, populate the $N$-dimensional data-space.
	\item Keep each combination of model-information for a specific parameter-value set and corresponding data-vector as set of rules or association sets.
	\item Repeat this for all possible models. Combine the data-set.
	\item Train a sufficiently complex machine-learning algorithm by providing a data-set of data-vectors as inputs and corresponding model-information as target output. Thus it becomes a classification problem of dimension equal to the number of allowed models.
\end{itemize}
If the training is properly done (by minimizing cross-entropy loss), given the real data-set, we expect the trained algorithm to find the best probable classes (models) for generating the data. The reason is that for a fixed reference distribution $P$, $D_{KL}$ is identical to cross-entropy up to an additive constant and minimizing one is equivalent to minimizing the other \cite{Goodfellow-et-al-2016}. Furthermore, as the classifier is trained on the whole parameter-space, this algorithm is future-proof given the same set of observables, i.e., if some future experiment performs more precise measurements on any number of the said observables, the algorithm will work with same accuracy on that updated data-set, and equally quickly, provided the underlying theoretical information (e.g. form-factors) remains unchanged.

Creating a fine-grained data-set for training is computation-intensive, but has further goodies to provide. If instead of saving just the model information for each data-point, we save the parameter-value vector in the `input-target' association as well, then using those, we can potentially create a `future-proof' predictor algorithm for each model as well\footnote{Because minimizing MSE, as happens in case of training of a predictor (when variance is independent of inputs), is equivalent to maximizing the log-likelihood \cite{Goodfellow-et-al-2016}.}. In other words, given any data-set, from past, future, or present, a well-trained regression algorithm should potentially find the parameter space with equal ease. In the best case, this can be used as an extremely fast automated statistical inference toolkit for this particular problem, or at least, in the worst scenario, can point us to the general direction of the parameter-space for further sophisticated inference.

%%%%%%%%%%%%%%%%%%%%%%%%%%%%%%%%%
\section{Methodology}\label{sec:methodology}
\subsection{The Classifier Network}\label{sec:network}
%%%%%%%%%%%%%%%%%%%%%%%%%%%%%%%%%
As mentioned earlier, we have used SNN as the classifier/model-selection algorithm in this analysis.  
These were created to overcome a shortcoming of fully connected neural-nets (FNN), in that it is extremely difficult to train an FNN more than a few layers deep. Even then, their performances are not even at par with traditional machine-learning algorithms such as Random Forest, Support Vector Machine etc. The reason for this difficulty in training was of either vanishing \cite{gradVanish} or exploding gradients. The traditional activation functions, like hyperbolic tangent or logistic sigmoids, used in FNNs, are bound to give outputs in a small region. Thus, for some problems, during stochastic gradient descent and back-propagation, some gradients in a `deep' FNN network become vanishingly small and prevent the weights from updating. Newer and more widely used rectifier alternatives like Rectified linear units (RELU) suffer less from this problem, as they saturate only in one direction \cite{reluVanish}. A different problem occurs for these, where the activation outputs are unbounded on one side. Here, in some cases, the changes in weights are so large that the activation functions get stuck on the positive side and thus become `dead'. This is the exploding gradient problem. Ad-hoc procedures like gradient clipping or $L2$ regularization are used to solve this problem.

%%%%%%%%%%%%%%%%%%%%%%%%%%%%%%%%%%%
\begin{figure*}[!t]
	\small
	\centering
	\subfloat[]{\includegraphics[width=0.32\textwidth]{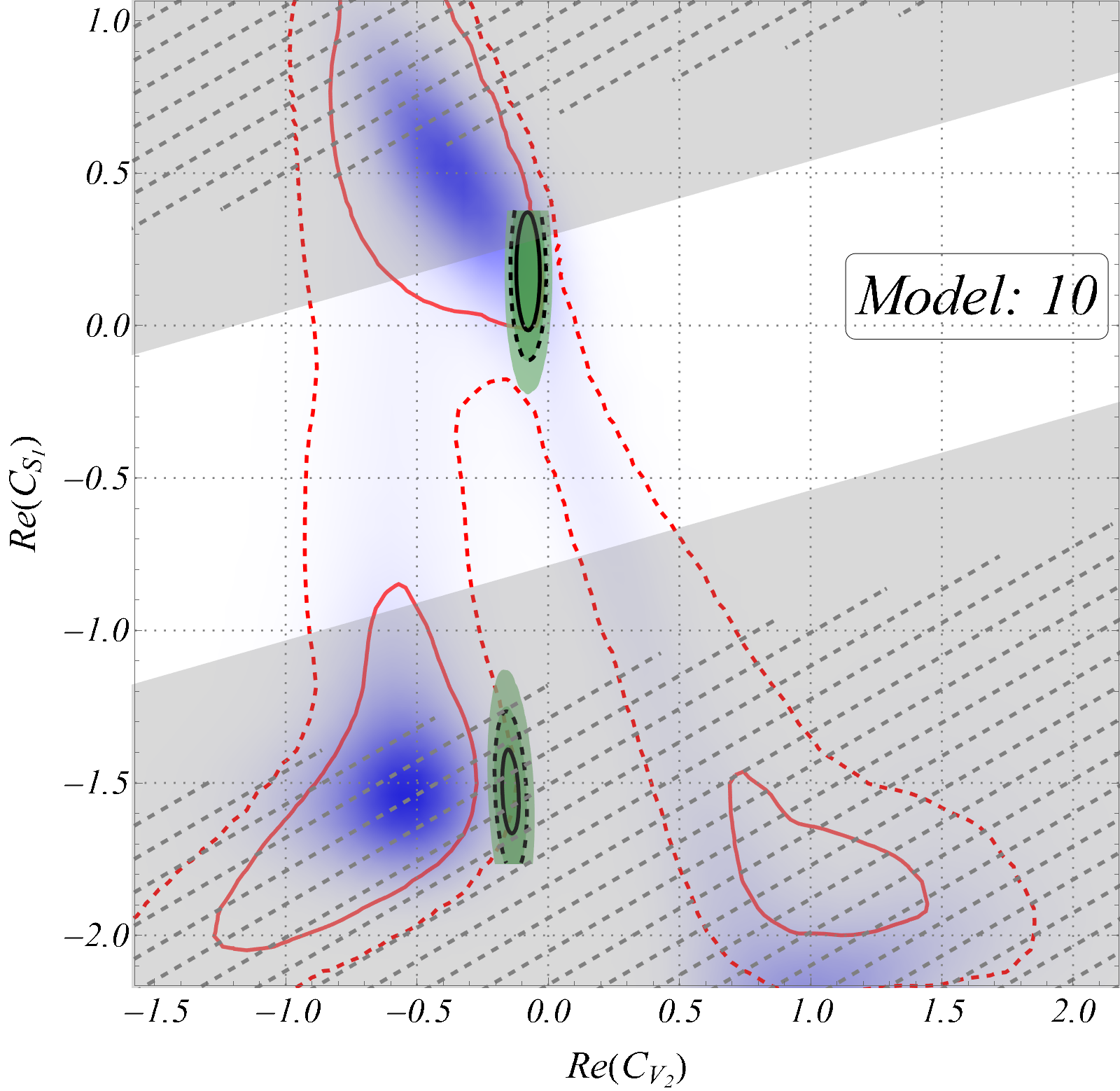}\label{fig:parspace10}}~
	\subfloat[]{\includegraphics[width=0.32\textwidth]{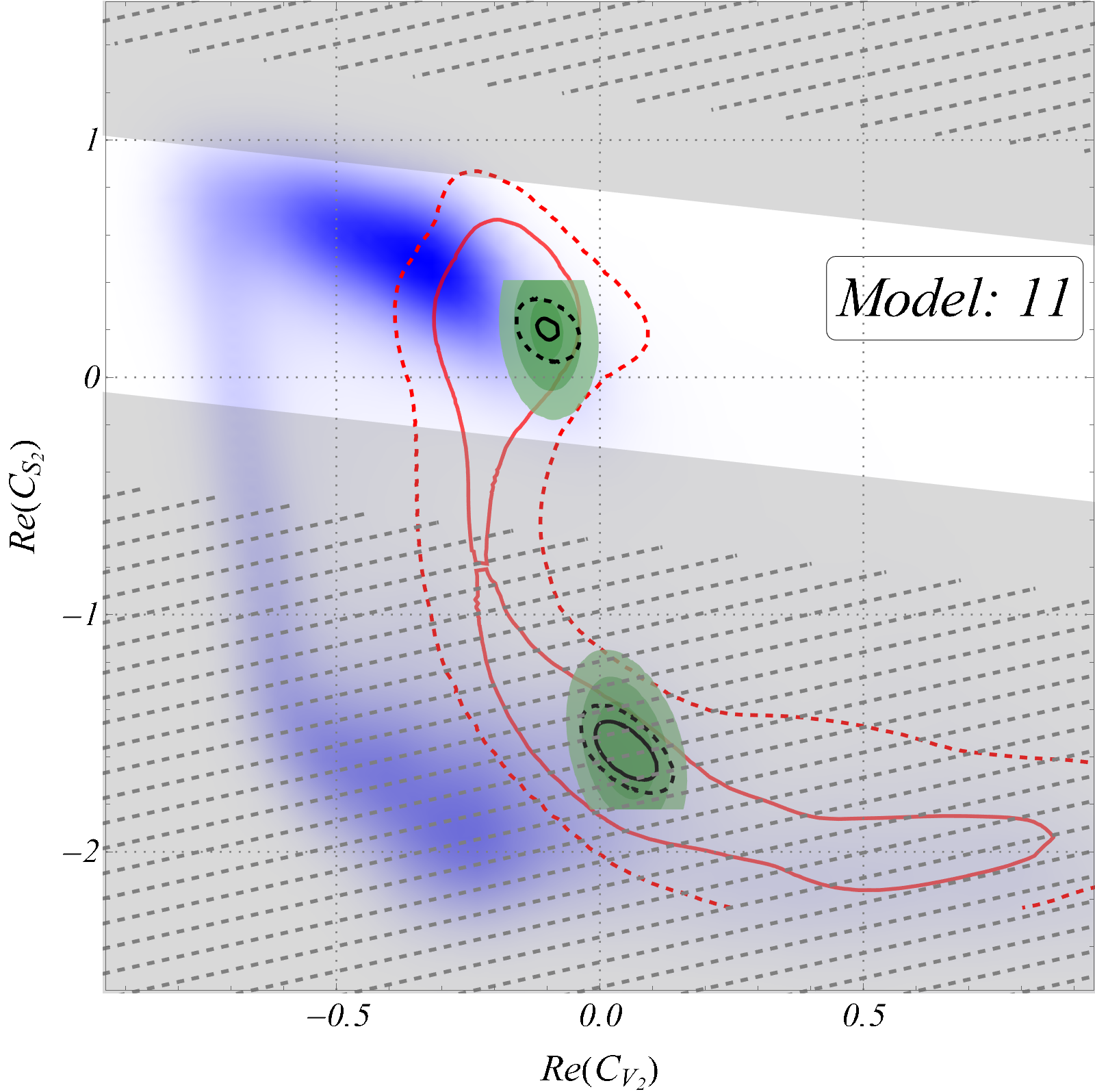}\label{fig:parspace11}}~
	\subfloat[]{\includegraphics[width=0.32\textwidth]{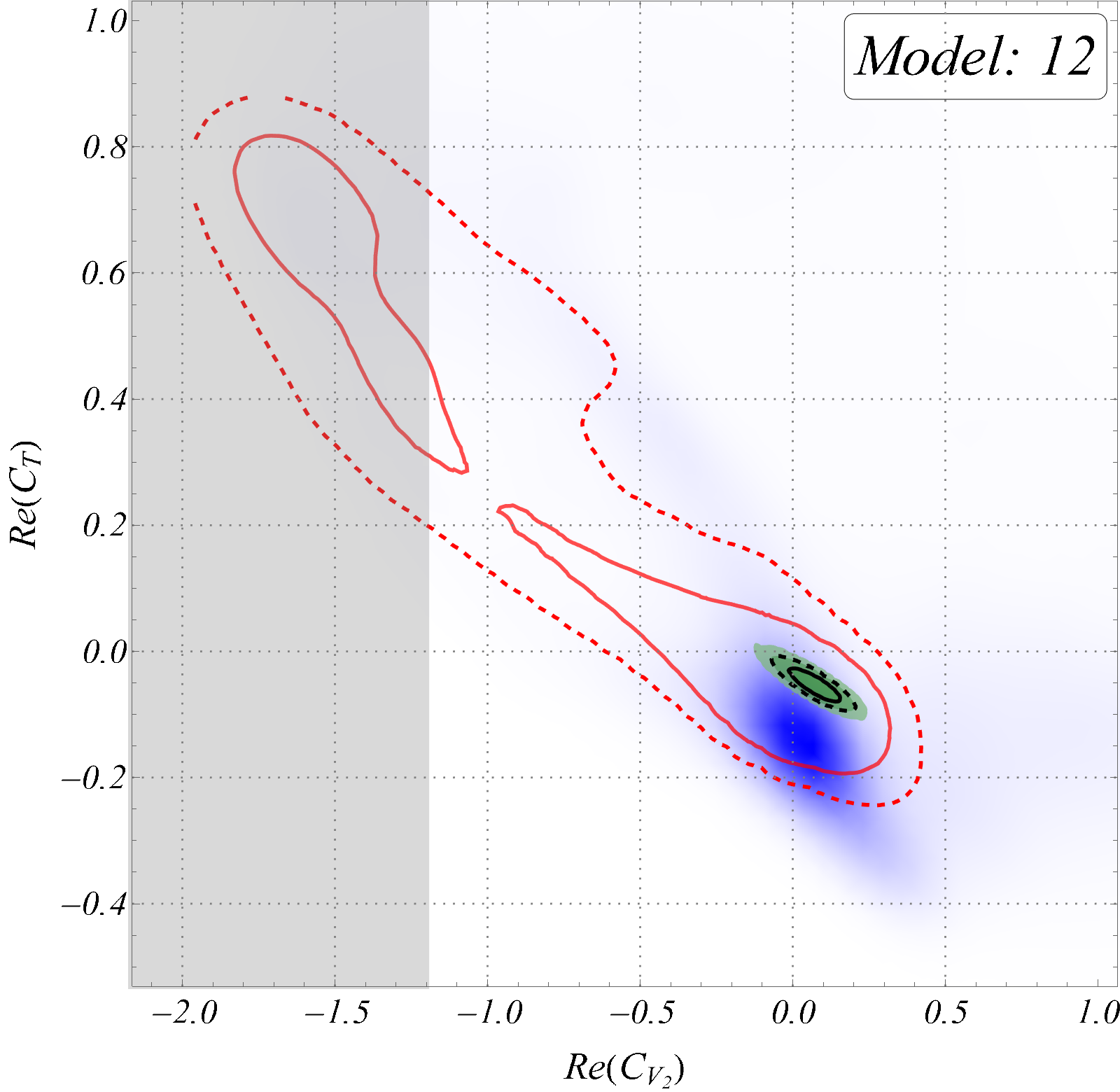}\label{fig:parspace12}}\\
	\subfloat[]{\includegraphics[width=0.32\textwidth]{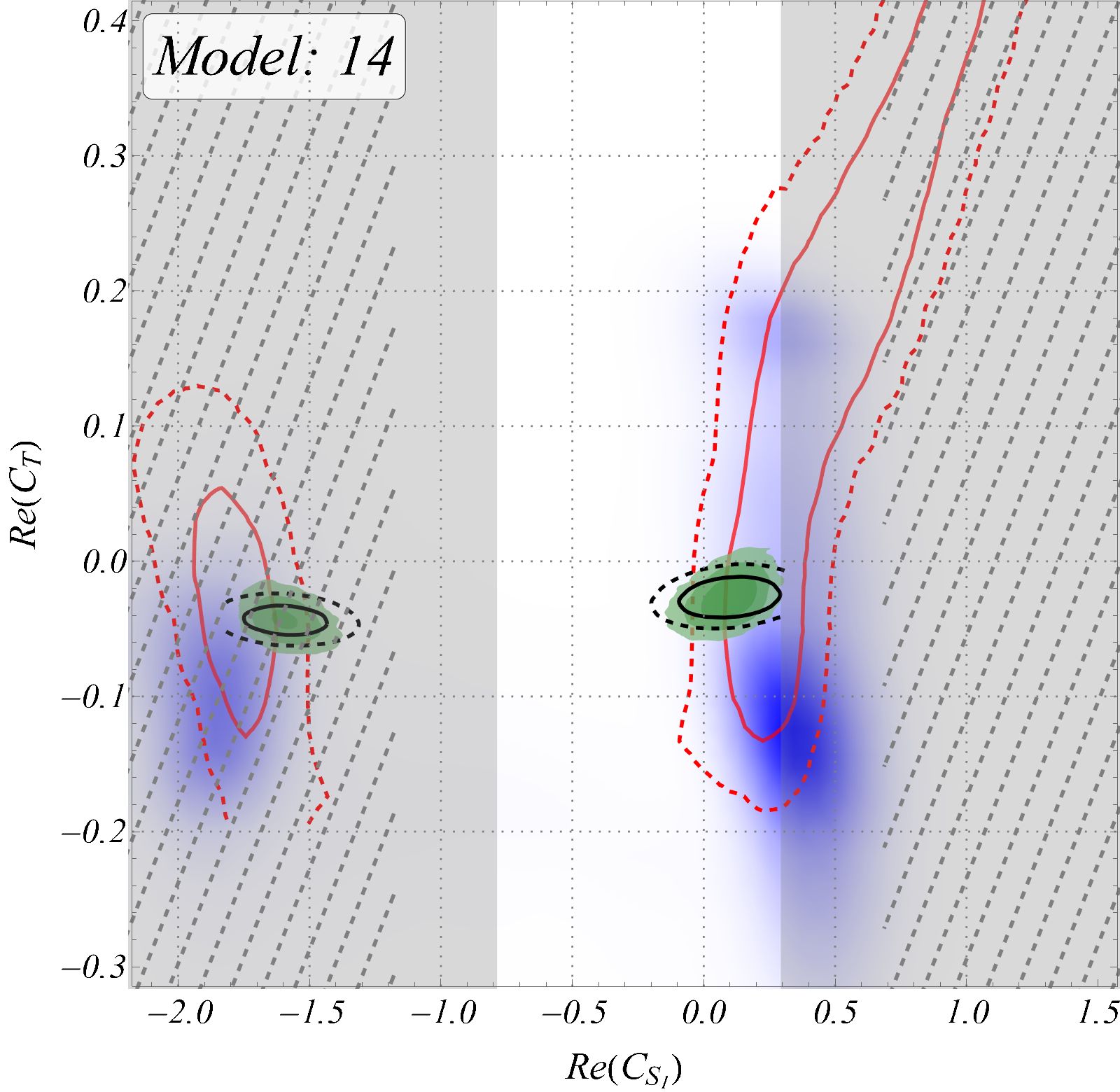}\label{fig:parspace14}}~
	\subfloat[]{\includegraphics[width=0.32\textwidth]{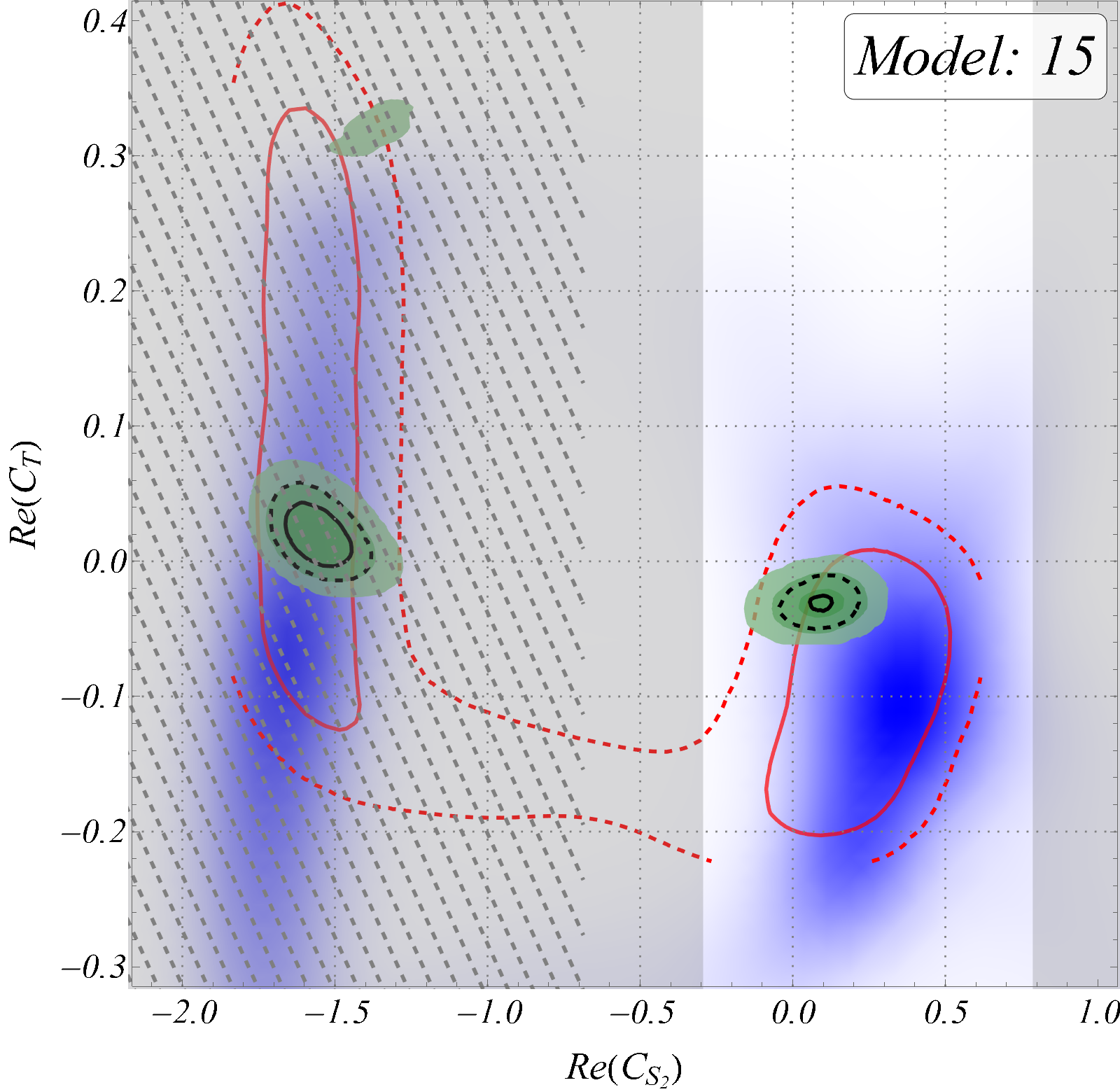}\label{fig:parspace15}}~~~
	\subfloat[]{\includegraphics[width=0.32\textwidth]{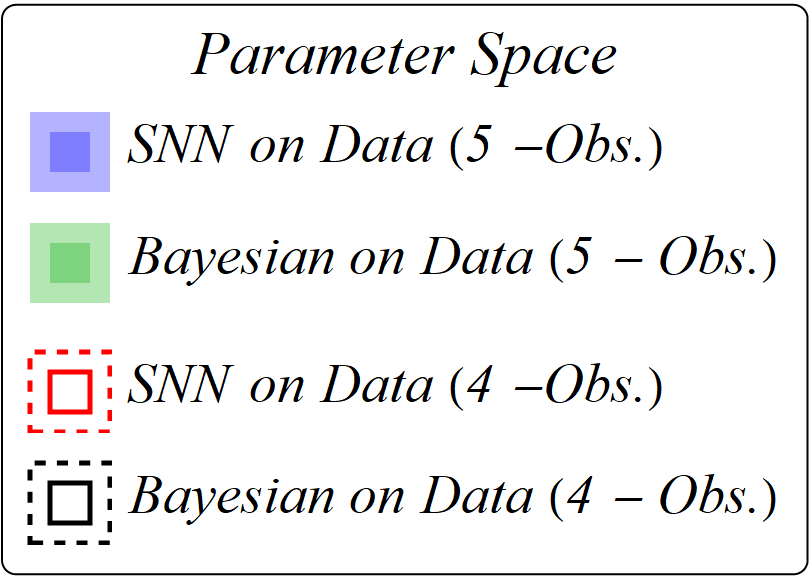}\label{fig:legComp2D}}
	\caption{2-D parameter spaces for the best models. The green (filled) and black (empty) contours show Bayesian fits for 5 and 4-observable cases respectively. Three contours correspond to $68.27\%$, $95.45\%$, and $99.73\%$ credible regions. Red contours show 2 credible regions with the predictor SNNs applied on 4 observables. Corresponding 5-observable cases are blue density histograms. The gray-shaded and the diagonally hatched regions are discarded by $10\%$ and $30\%$ limits on $Br(B_c\to\tau\nu_{\tau})$, respectively.}
	\label{fig:parspace2D}
\end{figure*}
%%%%%%%%%%%%%%%%%%%%%%%%%%%%%%%%%%%
%%%%%%%%%%%%%%%%%%%%%%%%%%%%%%%%%%%
\begin{figure*}[htbp]
	\centering
	\subfloat[]{\includegraphics[width=0.32\textwidth]{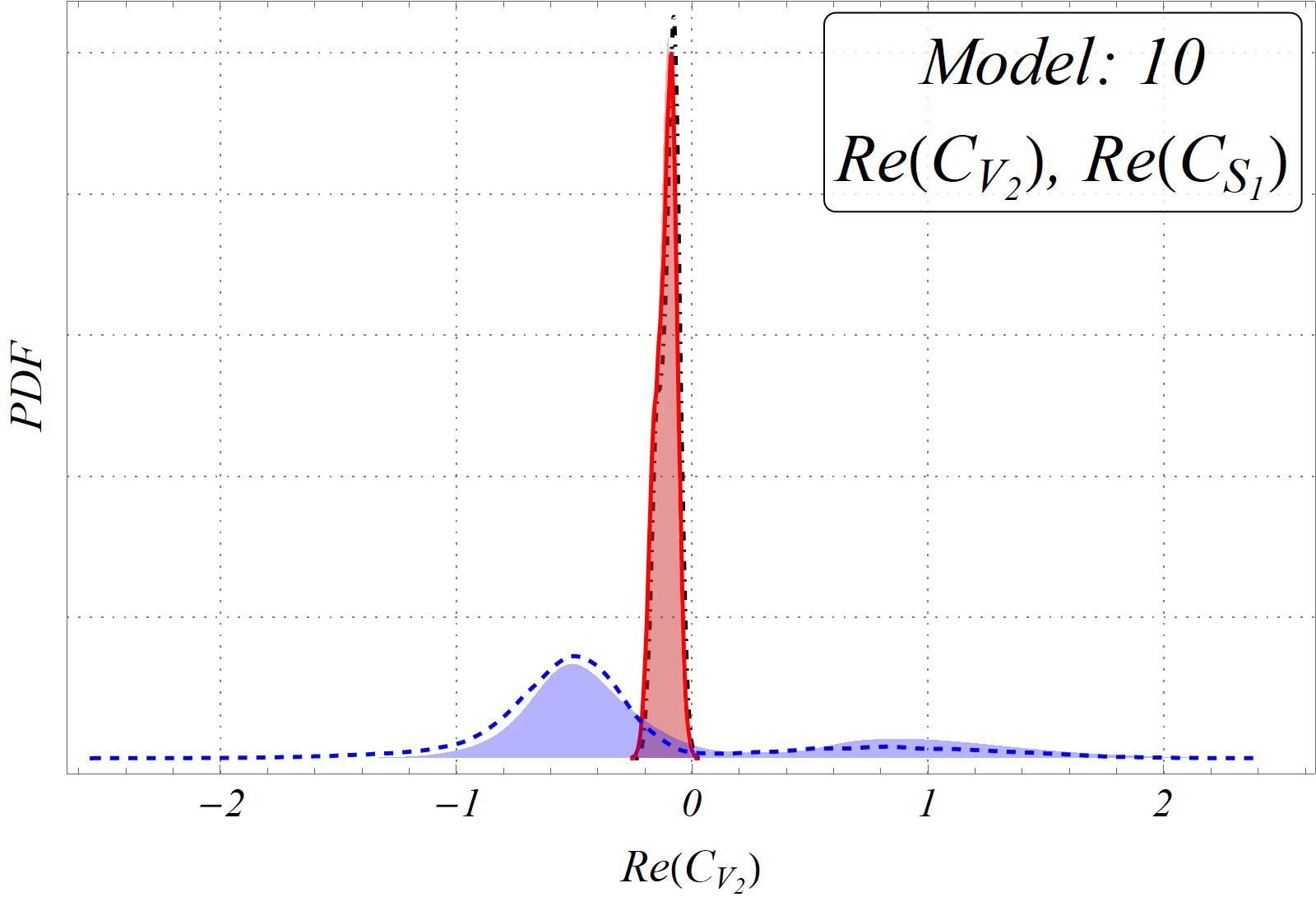}\label{fig:parspace101}}~
	\subfloat[]{\includegraphics[width=0.32\textwidth]{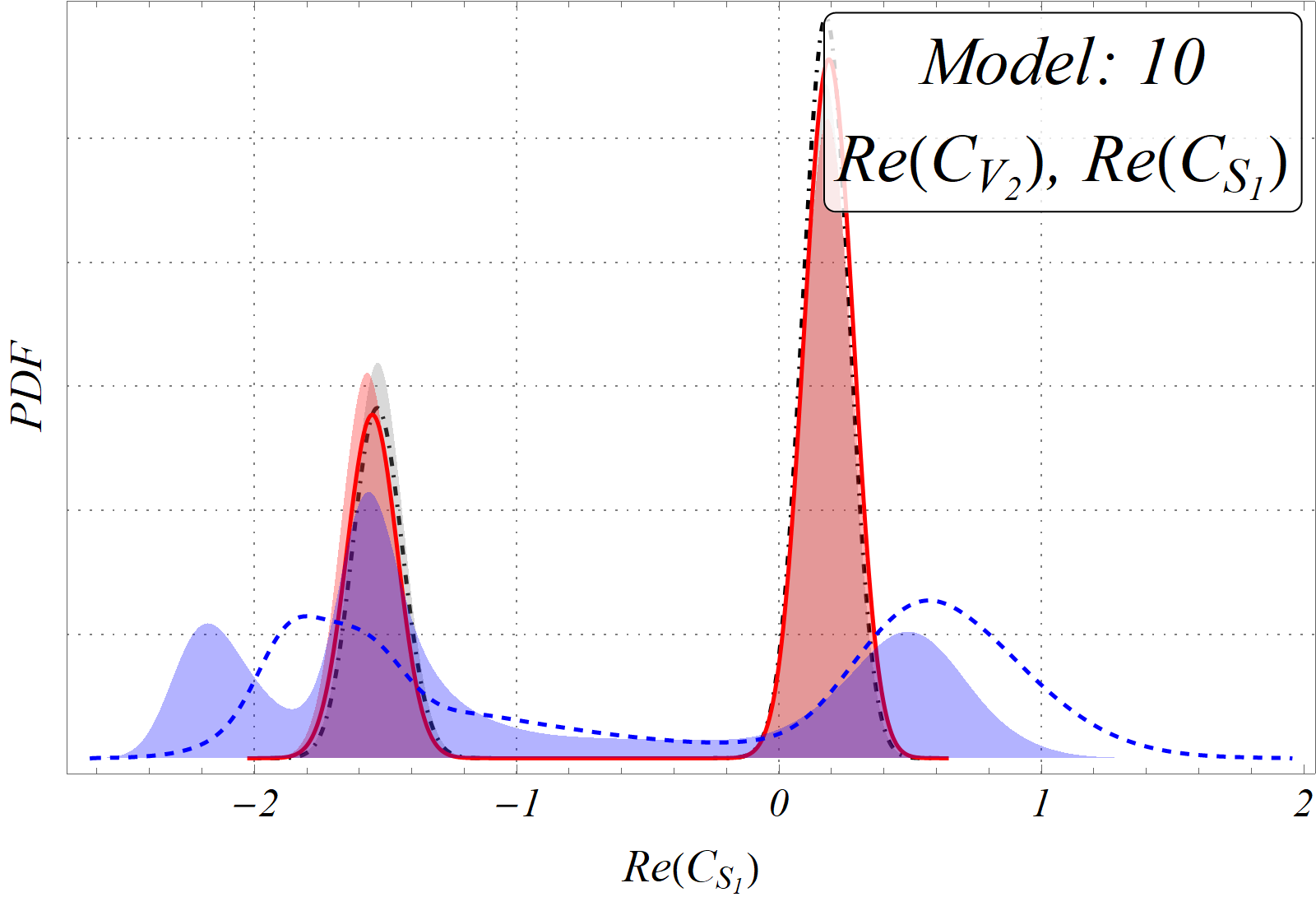}\label{fig:parspace102}}~
	\subfloat[]{\includegraphics[width=0.32\textwidth]{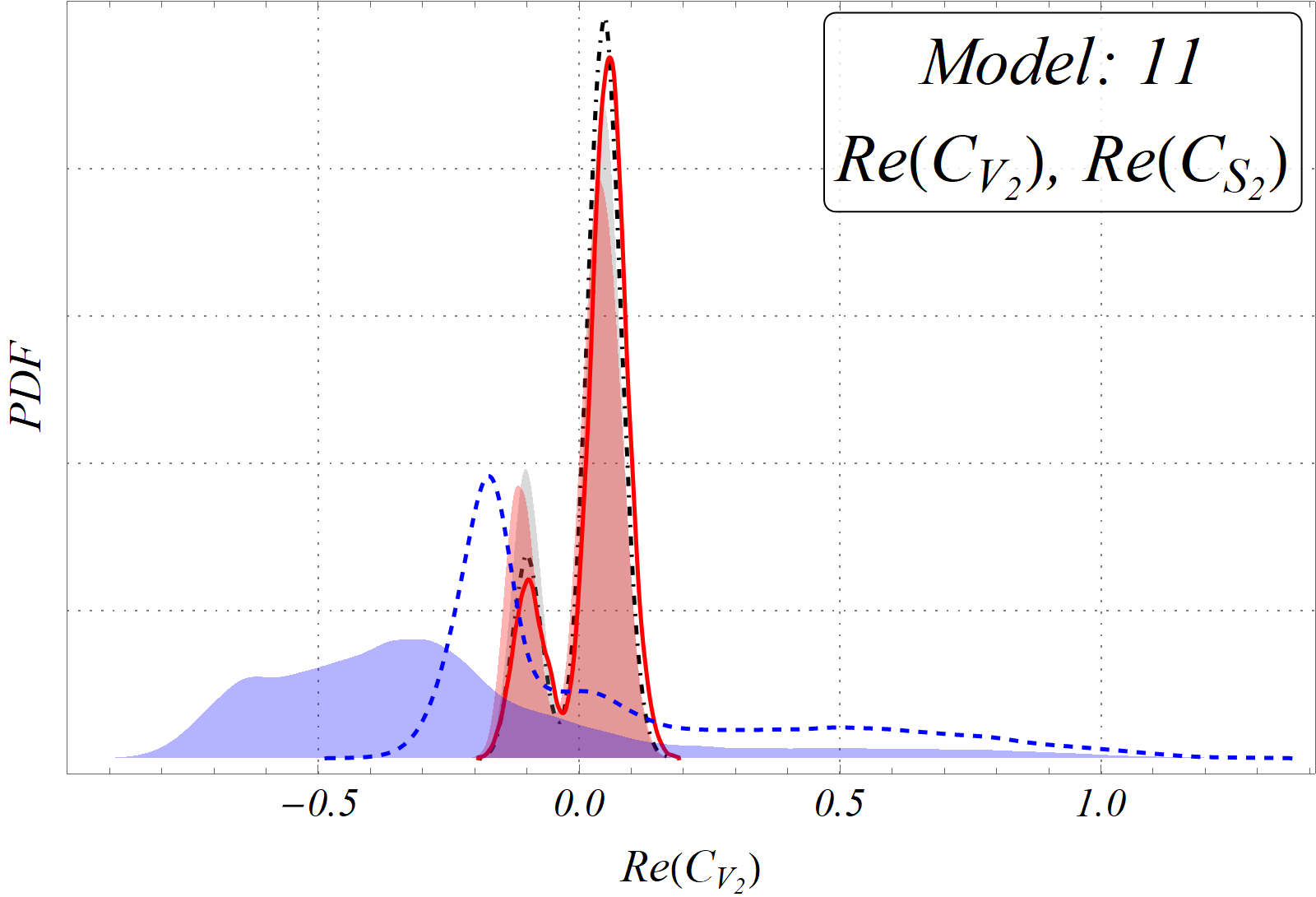}\label{fig:parspace111}}\\
	\subfloat[]{\includegraphics[width=0.32\textwidth]{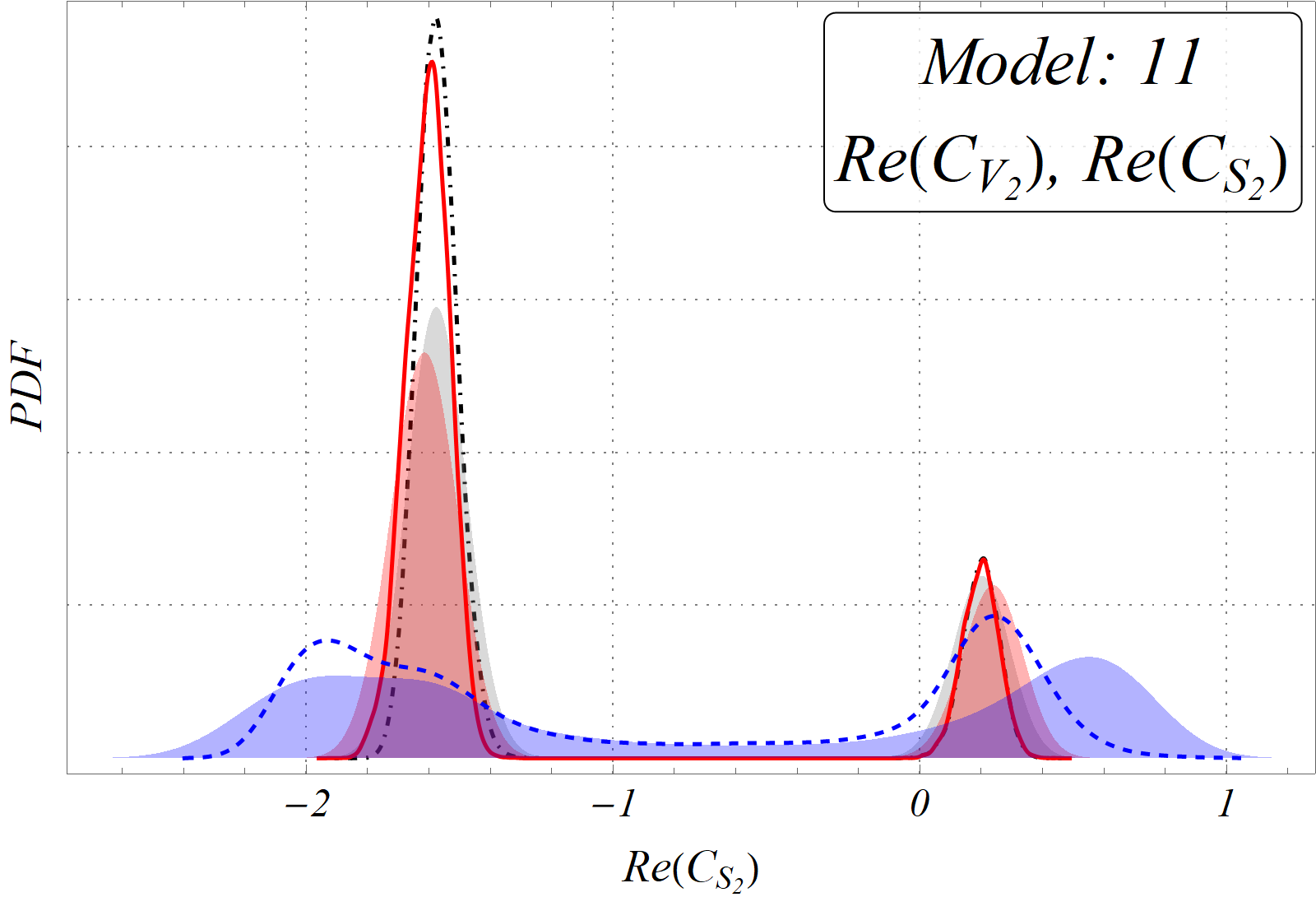}\label{fig:parspace112}}~
	\subfloat[]{\includegraphics[width=0.32\textwidth]{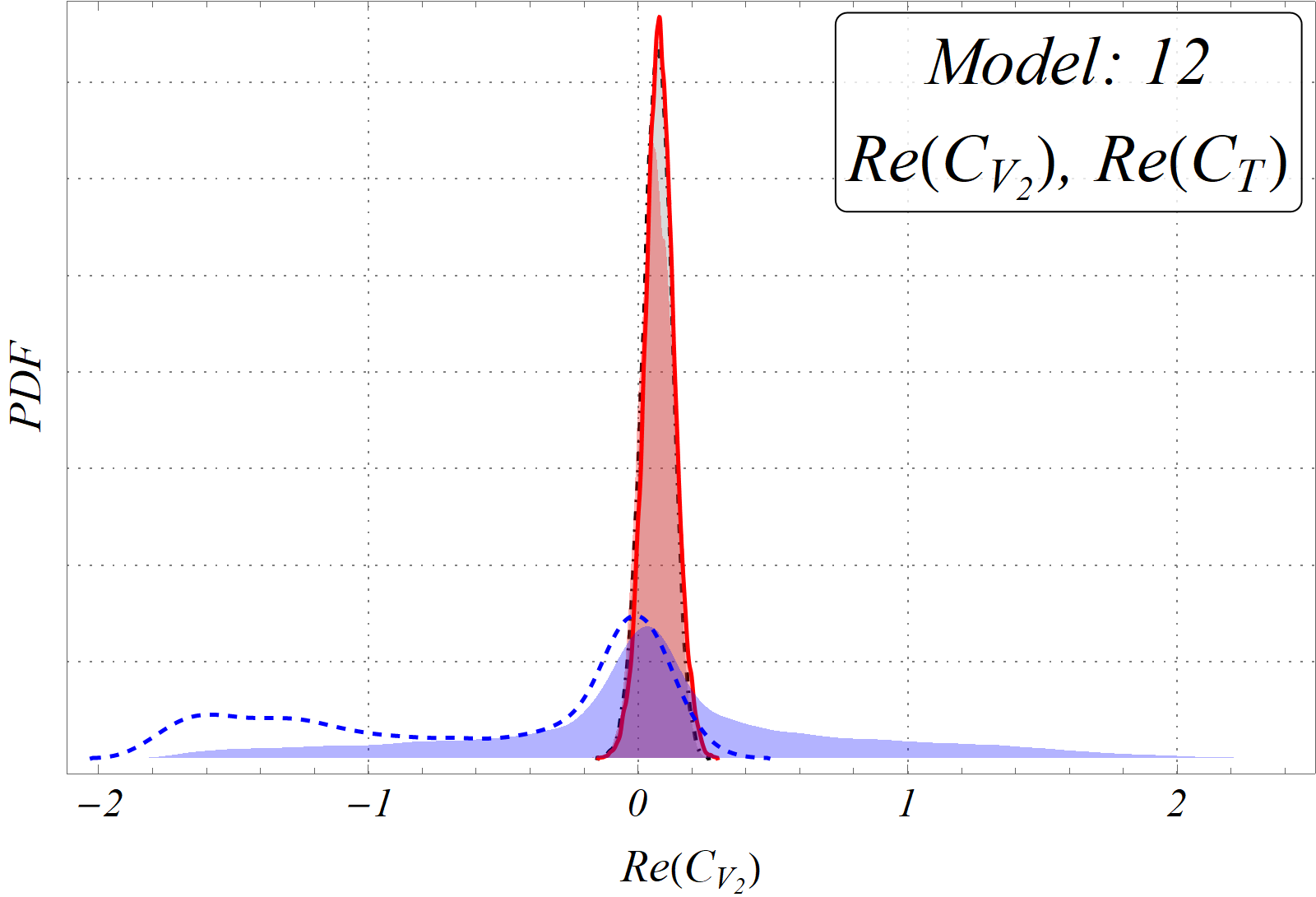}\label{fig:parspace121}}~
	\subfloat[]{\includegraphics[width=0.32\textwidth]{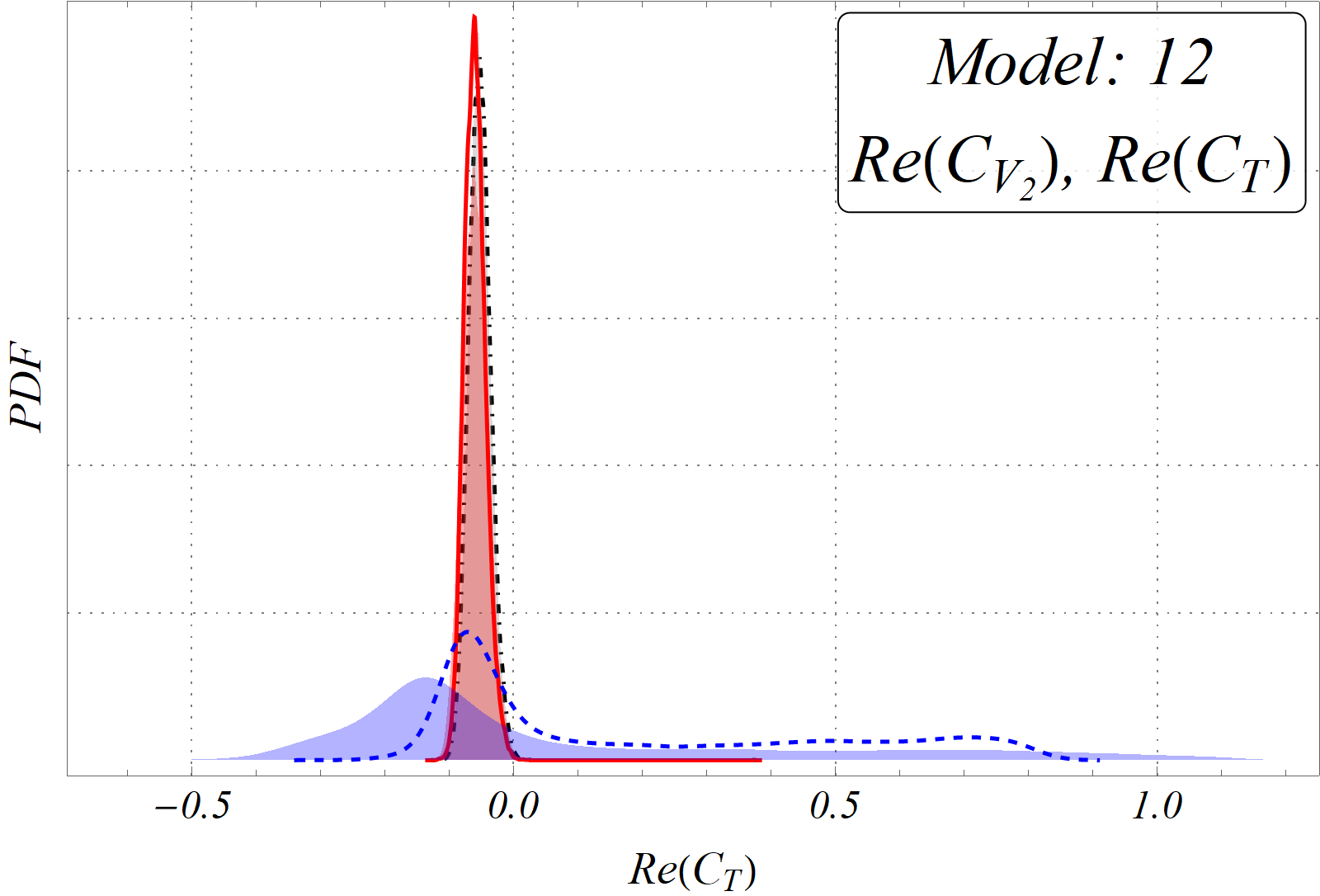}\label{fig:parspace122}}\\
	\subfloat[]{\includegraphics[width=0.32\textwidth]{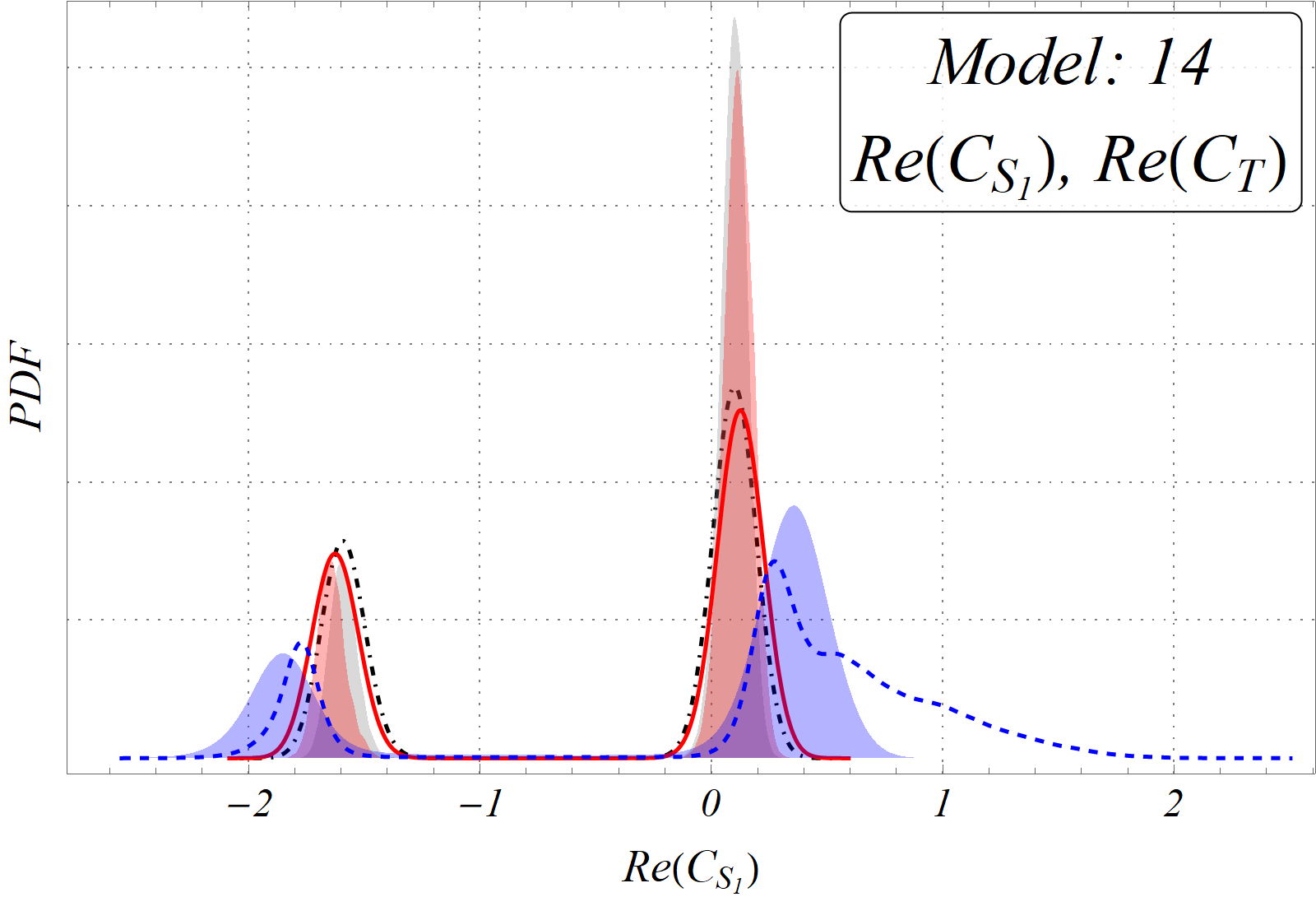}\label{fig:parspace141}}~
	\subfloat[]{\includegraphics[width=0.32\textwidth]{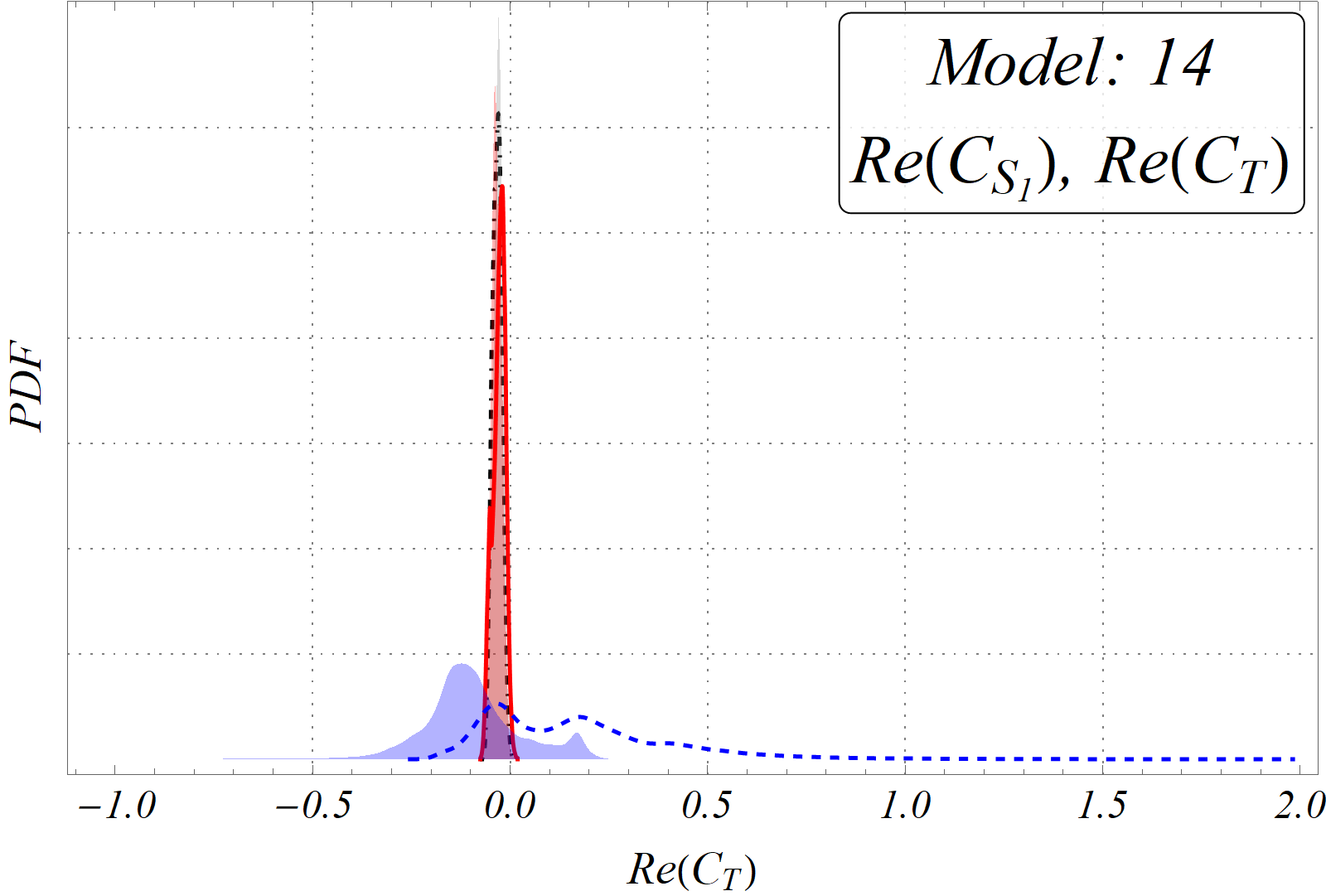}\label{fig:parspace142}}~
	\subfloat[]{\includegraphics[width=0.32\textwidth]{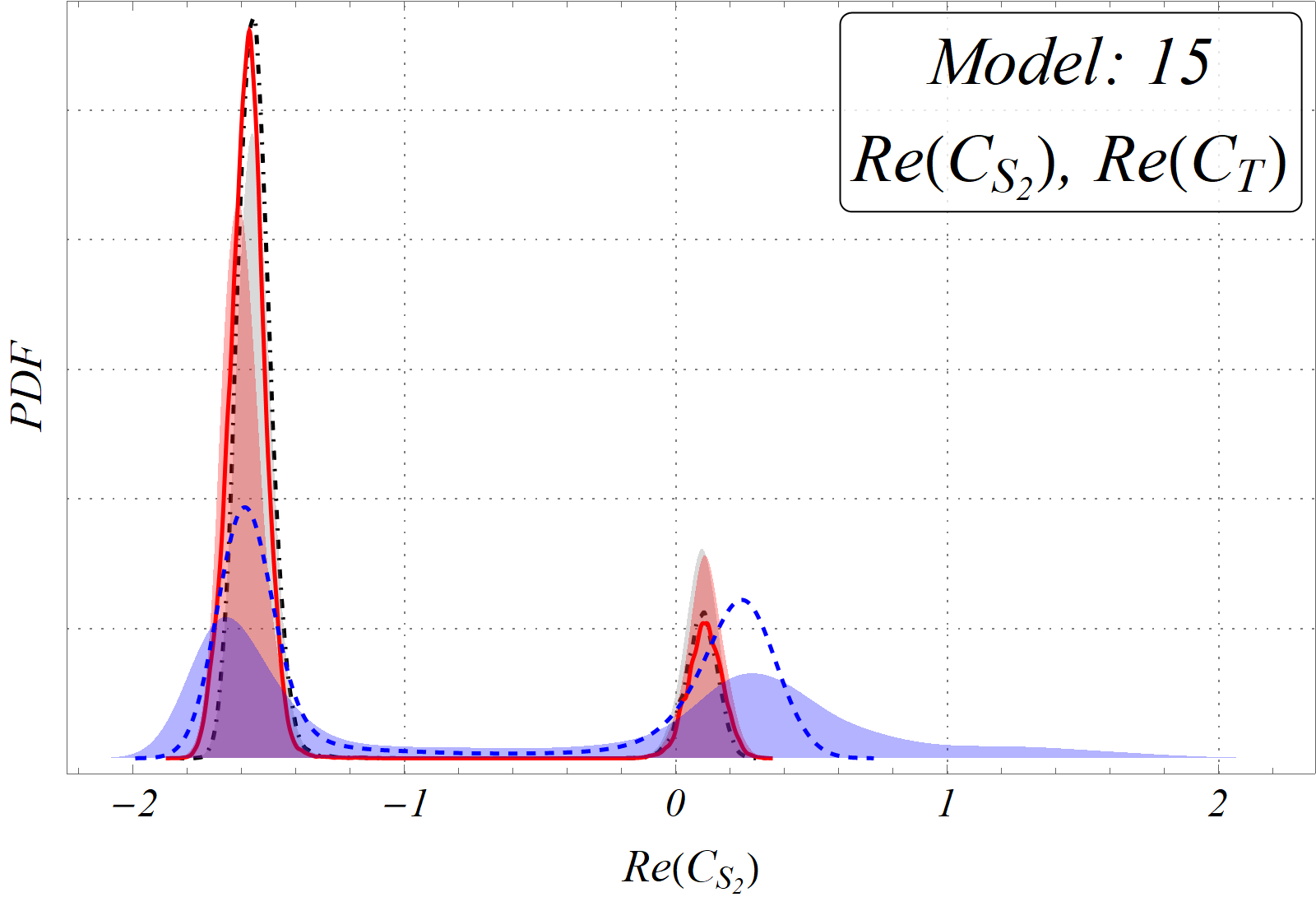}\label{fig:parspace151}}\\
	\subfloat[]{\includegraphics[width=0.3\textwidth]{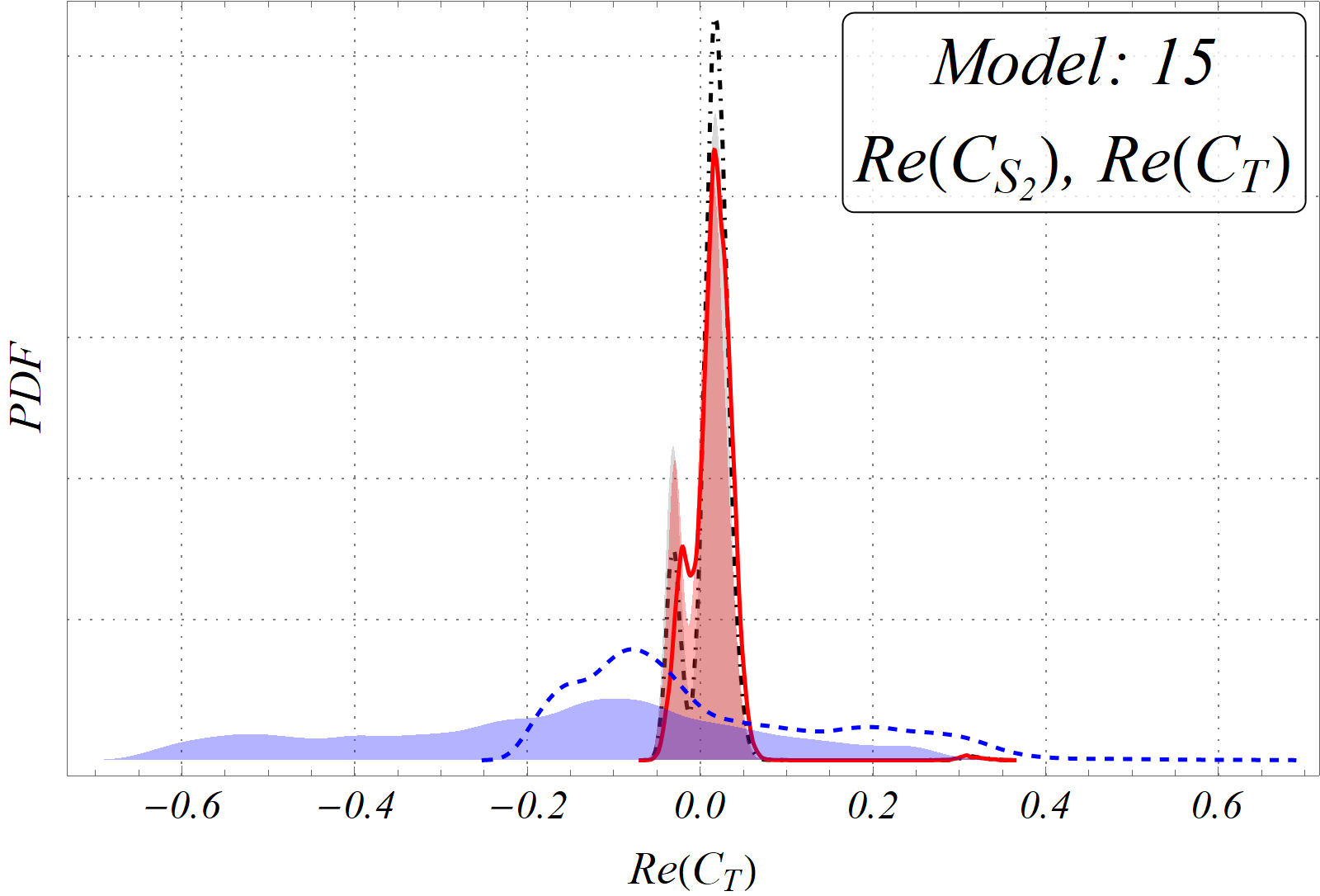}\label{fig:parspace152}}~
	\subfloat[]{\includegraphics[width=0.32\textwidth]{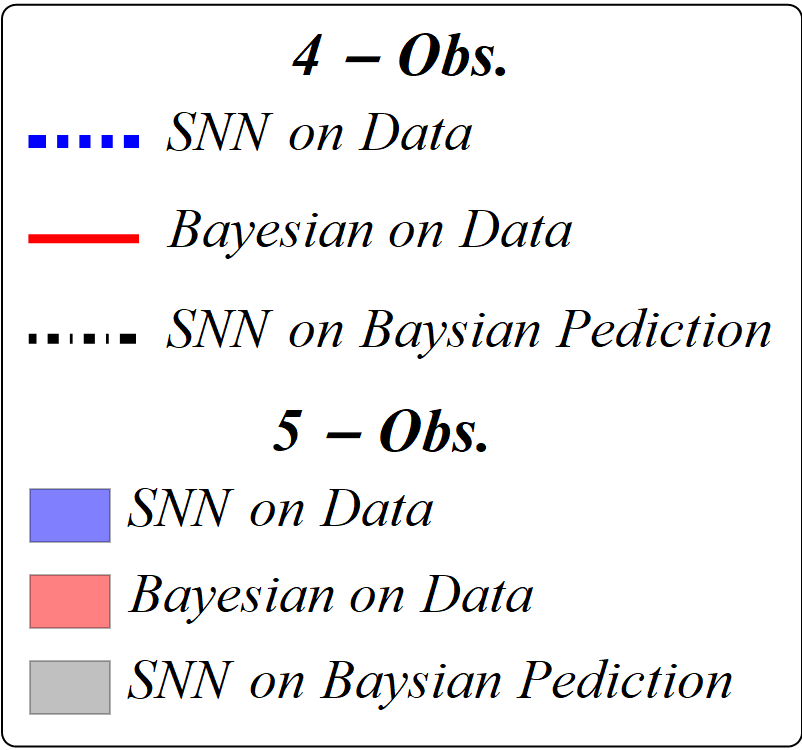}\label{fig:legComp1D}}
	\caption{1-D parameter spaces for the selected best models.}
	\label{fig:parspace1D}
\end{figure*}
%%%%%%%%%%%%%%%%%%%%%%%%%%%%%%%%%%
Though other types of weight-sharing networks, such as convolutional (CNN) or recurrent type (RNN) are extremely successful at perception tasks (e.g. computer vision or natural language processing), and the use of batch normalization layers and skip-connections (ResNets) in these networks make them stable and faster to train, a deep FNN alternative of traditional machine-learning methods for non-perception tasks, such as classification and regression of numerical or nominal data, was not present even a few years back. 

Proposed in 2017, SNNs are demonstrated to have better accuracy than all FNN alternatives in all 121 UCI tasks \cite{Dua:2019} while being very deep at the same time. Extensive theoretical proofs are also provided to show that if input data is standard-normalized (distributed in an unit normal), the activation outputs converge to a unit normal as well - that too, throughout the depth of the network. To achieve this, these networks use a special type of activation function, released in the same paper, called scaled exponential linear units (SELU). Like the unique activation function, there are some other specialties regarding SNNs. If dropout layers are to be used in the network, they have to be of a special type, called Alpha-Dropout, proposed in the same paper. Also, at the beginning of training, the initial choice of weights should be such that variance of arrays is preserved when propagated through layers and for this network, they need to be distributed normally.

In our work, we have used a seven-module SNN, where each module comprises of a fully connected/linear layer of 50 artificial neurons, a SELU activation and an Alpha-Dropout layer with dropout-probability of $1\%$. At the end of the network-chain, another linear layer changes the dimension of the output to the number of classes/models considered and a Soft-max layer changes those outputs to probabilities. During training, the cross-entropy between the outputs and the targets is minimized using the stochastic {\em ADAM} optimizer \cite{Kingma2015AdamAM}. We set the initial learning-rate $\ell = 0.001$ and apply no $L2$-regularization. Nets are initialized with the method proposed in ref. \cite{HeInit}. This standardized net-model was curated and stored by Wolfram organization in their Neural-Net Repository \cite{SNNWolfram}. Fig. \ref{fig:netgraph} shows a schematic graph of a representative network which takes 4 dimensional vectors as inputs and classifies it among 20 classes.
%%%%%%%%%%%%%%%%%%%%%%%%%%%%%%%%%%%%%%%%%
\begin{table*}[hbt]
	\begin{ruledtabular}
		\renewcommand*{\arraystretch}{1.2}
		\begin{tabular}{ccllrcc}
			Dataset & Model Index & \multicolumn{2}{c}{$D_{KL}$} & \multicolumn{3}{c}{Net Performance (\%)} \\
			\cline{3-7}
			& & (Net) & (Bayes) & $\sigma$ & $R^2$ & MSE \\
			\hline
			& $15$  &  $3.75$    &  $31.78$ 			&  $1.80$   &  $99.93$  &  $0.03$  \\
			& $14$  &  $10.43$   &  $1.15\times 10^5$  &  $2.03$ 	&  $99.91$  &  $0.04$  \\
			4-Obs. & $10$  &  $14.53$   &  $1.21\times 10^6$ 	&  $3.08$  	&  $99.84$  &  $0.10$  \\
			& $12$  &  $18.20$   &  $6.53\times 10^4$  &  $1.30$   &  $99.93$  &  $0.02$  \\
			& $11$  &  $29.67$   &  $5.49\times 10^5$  &  $2.34$  	&  $99.91$  &  $0.05$  \\
			%			 & $9$   &  $449.30$  &  $6.72\times 10^6$ 	&  $54.05$ 	&  $45.88$  &  $29.21$  \\
			\hline
			& $12$  &  $95.60$   &  $6.29\times 10^6$ 	&  $1.58$  	&  $99.90$ 	&  $0.02$  \\
			%			 & $9$   &  $925.09$  &  $3.42\times 10^8$ 	&  $54.05$ 	&  $45.87$  &  $29.21$  \\
			& $15$  &  $1146.50$ &  $1.92\times 10^6$ 	&  $2.19$  	&  $99.89$  &  $0.05$  \\
			5-Obs. & $14$  &  $1631.02$ &  $9.60\times 10^7$ 	&  $2.33$  	&  $99.88$  &  $0.05$  \\
			& $10$  &  $3232.05$ &  $5.20\times 10^6$ 	&  $3.28$  	&  $99.82$  &  $0.11$  \\
			& $11$  &  $3905.8$  &  $2.41\times 10^6$ 	&  $2.97$  	&  $99.86$  &  $0.09$  \\
		\end{tabular}
	\end{ruledtabular}
	\caption{Main results for regression with the predictor SNNs for 4 and 5-obs. cases. 5 best models, $D_{KL}$ scores of the SNN-parameter-space and of the Bayesian fits are in $2^{nd}$ - $4^{th}$ columns. Last three columns list performances of SNNs from test data-set. See section \ref{sec:resregression} for details.}
	\label{tab:klcompare}
\end{table*} 
%%%%%%%%%%%%%%%%%%%%%%%%%%%%%%%%%%%%%%%%%%%%
%%%%%%%%%%%%%%%%%%%%%%%%%%%%%%%%%%%%%%%%%%%%
\subsection{Simulated Data and Training}\label{sec:simdata}
\subsubsection{Classification}\label{sec:methodclass}
%%%%%%%%%%%%%%%%%%%%%%%%%%%%%%%%%%%%%%%%%%%%
The workflow is in three parts. First, we check the model-selection capabilities of the 4 observables from $B\to D^{(*)}\tau\nu$ channel with experimental measurements. As the number of observables is small, we restrict our choice of models with up to two parameters, including real and imaginary parts of the possible WCs. This gives us 20 possible models/classes. We generate 100,000 points in each class, calculate the observables for each parameter-vector and ensure that they are not too far away from the corresponding experimental results (e.g. $\mathcal{R}_{D^{(*)}}$ values are generated within $(0,0.6)$); the class labels are saved as the training-target for each point. Next we train the aforementioned SNN with this training data-set and save the trained network with all information about training in a database-structure.

To take care of over-fitting, we use early stopping and validation. The data-set used for training is randomly broken into two parts, with $85\%$ of points as `training-data' and the remaining $15\%$ used as `validation-data'. Training is then done in batches, where a large batch-size of 10,000 is used to minimize training time. When all of training-data are used, a `round' of training is done. After one round, the trained net is applied on the validation-data to obtain the performance-metrics of the net, e.g., `validation-loss' and `validation-error'. The whole process is then repeated. The training automatically stops when the validation-loss does not decrease after 25 rounds of getting the previous lowest value. Fig \ref{fig:errloss} demonstrates how loss and error-rate (for both training and validation) change over rounds. The orange curves, denoting the training metrics, are obtained by collecting info after each batch, whereas the blue validation curves are created with information obtained after each round. The best round (for which the validation-loss is minimum) is also depicted in each plot.

If we train identical copies of an SNN over the same data-set, the slight differences in the initial random seeds and the general training procedure will render the final trained networks equivalent, but ever so slightly different. Consequently, their accuracy measures will also be similar but slightly different. We can then create an ensemble of such SNNs acting on a data simultaneously, with the output as the mean of the individual outputs of each net. It has been seen that the accuracy of the ensemble-net exceeds that of each individual net \cite{Goodfellow-et-al-2016,BishopBook}. We thus follow this procedure using an ensemble of 10 such SNNs and after taking all possible combinations of the individual nets, find the ensemble which gives us quite a considerable increase in accuracy. During this process, we found that the accuracy of an $n$-net ensemble first increases with increasing $n$, but then starts decreasing after a certain value of $n$. Taking all possible combinations of the individual nets, we plot the maximum accuracy of any $n$-ensemble with $n$ in figure \ref{fig:accucompare}, for all data-sets. For our final analysis, we use the ensemble corresponding to the maximum accuracy for a specific data-set.

In the next part, we repeat the whole training and testing procedure detailed above for all the five observables with experimental results (including $\mathcal{R}_{J/\Psi}$) and then again with all 13 observables listed in section \ref{sec:thobs}. For the 13-obs. data-set, as the dimensionality of the data-set almost doubles, we expect the model-discriminating capability of the SNN to increase a lot. We keep all models with up to 4 parameters for this case, making the class-size 45. Table \ref{tab:modlst} lists the model-indices and the corresponding parameters of these 45 models. The model indices in parentheses are the corresponding serials of those models in the 20-model list used in the 4 and 5-observable classification. Table \ref{tab:netdetails} showcase the details of the training of a single SNN, part of the final classifier ensemble, for the 4-obs. data-set.

To have a clear idea about the performance of our trained networks, we have applied them on unseen test data-sets created for all the different observable cases. The size of the test data-set is $10\%$ of the corresponding training data-set. The results of various performance measures for each ensemble nets trained on the three observable-sets are listed in table \ref{tab:netmeasuresClass} and the confusion matrices of the resulting classifications are shown in fig. \ref{fig:confusion}. These provide us with a visual estimate of misclassification among all available classes. The deeper the color, the larger are the number of data in each box. The $i$-$j$th box contains all data actually coming from class $j$, identified as class $i$ by the classifier. In an ideal classifier, all test data will be correctly classified and thus only diagonal boxes will be populated. Unsurprisingly, the model-discriminating performance of the net trained on all observables is better throughout all measures than those with 4 or 5 observables, in spite of the considerable increase in the number of available models (from 20 to 45) \footnote{A note here: in realistic cases, the effective accuracy of the nets are even higher. For those cases where one of the parameters (WCs) in a model has significantly smaller value (especially the imaginary parts), the net identifies it as the simpler model with only real WC. This is why there are more misclassified examples in the lower half of the matrices than the upper ones. This, in real cases, is consistent with parsimony and will be identified as a correct classification for all intents and purposes.}.

To know how the SNN performs compared to some other classical machine-learning techniques, we created small data-sets (using the 5-obs. data-set) of 10,000 and 1500 points for each model as respectively training and validation data, and performed classification with various different shallow-learning techniques, as well as our ensemble-SNN. We then applied the resultant trained classifiers from each method on a test data-set of 2000 points and compared their accuracy. Table \ref{tab:classcompare} contains the results of those. The ensemble-net clearly has a considerably larger accuracy than even the best of the shallow-learning processes, which in this case is `Random Forest'\footnote{In a recent work using deep neural networks to encode likelihood functions \cite{Coccaro:2019lgs}, the authors also found SNNs, specifically SELU activation to perform much better than RELUs or ELUs.}.

To compare the classification results with a standard statistical model selection procedure, we perform both frequentist and Bayesian fits for each model, using the present global averages of the 4 and 5-observable data-sets. All fits and subsequent analyses are done using a \emph{Mathematica\textsuperscript \textregistered} package \cite{sunando_patra_2019_3404311}. Calculating AIC$_c$ with the frequentist MLE estimates, we calculate the $w^{\Delta \text{AIC}_c}$ (defined in section \ref{sec:infocrit}) to get the probabilistic estimates for the data to come from each model. For the Bayesian fit results, we apply a slightly different tactic of using the $D_{KL}$. Using the obtained multivariate posterior density of the parameters, we obtain the predicted distribution in the observable-space. The goal here is to check which model predicts a observable-distribution with the minimum $D_{KL}$ with respect to the original data-distribution, i.e., which predicted observable distribution is the closest to that of the original data. Finally, we compare these results with the prediction of the trained ensemble SNN, applied on a large random set sampled from the data-distribution. The obtained probability estimates for each sample-point are aggregated and the final vector is normalized to get a probability distribution. This is repeated for each model. 

This procedure can now only be done for the 4 and 5-observable cases, but with more measurements in the future, model-selection capabilities of the global trained network can also be tested.

\subsubsection{Regression}\label{sec:methodregress}
As mentioned earlier, we saved each parameter-vector while generating training data-set. This enables us to train another variant of SNN, able to perform regression on the data, given a specific model. For this purpose, we have increased the number of points in the training-set to 1,000,000, for each model and trained a separate network indexed for that model. This procedure is repeated for all the 4, 5 and 13 observable cases. The idea is to first select the best candidate-models for a particular real data-set, then apply the predictor SNN for each of those models to a large set of points sampled from the data-distribution, to obtain the parameter space of the said model.

The predictor SNN differs from the classifier SNN only in the fact that the last Soft-Max layer is absent in the regression SNN. While training, we set the ``coefficient of determination" ( `$R^2$' \cite{draper1998applied}) metric as the training-stopping-criterion, i.e., training stops if `$R^2$' does not increase after 30 rounds. In all other aspects, the training and validation processes are more-or-less identical to the classification one.

To validate the results of the regression SNN, we use the Bayesian parameter spaces for the 4 and 5-observable case. For a selected model, we take the large samples generated for the posterior distribution of the parameters in the Bayesian analysis and create the corresponding predicted observable-distribution. If the regression SNN is working correctly, we should exactly get back the Bayesian posterior by applying it on this predicted observable-distribution. Next, we apply the net to the sampled distribution of the real-world 4 and 5-observable global averages to compare the outcome with that of the Bayesian results. This process is repeated for all the best-selected models.  
%%%%%%%%%%%%%%%%%%%%%%%%%%%%%%%%%%%%%%%%%
\begin{table}[hbt]
	\begin{ruledtabular}
		\renewcommand*{\arraystretch}{1.2}
		\begin{tabular}{ccl}
			Luminosity & Model Index & $D_{KL}$ (Net)  \\
			\hline
			& $12$  &  $35.04$    \\
			$5 ab^{-1}$ & $14$  &  $231.44$   \\
			& $15$  &  $306.41$   \\
			\hline
			& $12$  &  $239.15$   \\
			$50 ab^{-1}$ & $15$  &  $454.78$ \\
			& $14$  &  $2790.54$ \\
		\end{tabular}
	\end{ruledtabular}
	\caption{SNNs predictions for synthetic data of the 4-obs. case with same central values as present and predicted uncertainties at future luminosities of Belle II, with corresponding $D_{KL}$ scores. See section \ref{sec:future} for details.}
	\label{tab:klcomparefuture}
\end{table} 
%%%%%%%%%%%%%%%%%%%%%%%%%%%%%%%%%%%%%%%%%%%%

%%%%%%%%%%%%%%%%%%%%%%%%%%%%%%%%%
\section{Results}\label{sec:results}
\subsubsection{Classification}\label{sec:resclass}
%%%%%%%%%%%%%%%%%%%%%%%%%%%%%%%%%
As discussed in the previous section, we use the present world average of the 4 and 5-observable data-sets to perform both frequentist and Bayesian fits to all the 20 models used in the classification training. This enables us to do multiple things:
\begin{itemize}
	\item We perform a classical model-selection using AIC$_c$. Comparing the results of these with the more sophisticated techniques tells us how biased AIC$_c$ is in selecting less complex models for a very small data-set such as the present one.
	\item The posterior parameter-distributions obtained in the Bayesian analysis helps us in validating both classification and regression results. During classification, we use them to generate the predicted observable-distribution for each model. The better the model is, the closer the predicted distribution would be to the original data-distribution. We use $D_{KL}$ as the measure for dissimilarity between these two distributions. Then the models are sorted according to this $D_{KL}$. \textit{In absence of any other regression result}, this is the most faithful model-selection possible.
	\item In regression, the same predicted observable-distribution lets us check whether our regression network can faithfully reconstruct the (source) posterior parameter-distributions.
\end{itemize} 

%%%%%%%%%%%%%%%%%%%%%%%%%%%%%%%%%%%
\begin{figure*}[hbt]
	\small
	\centering
	\subfloat[]{\includegraphics[width=0.45\textwidth]{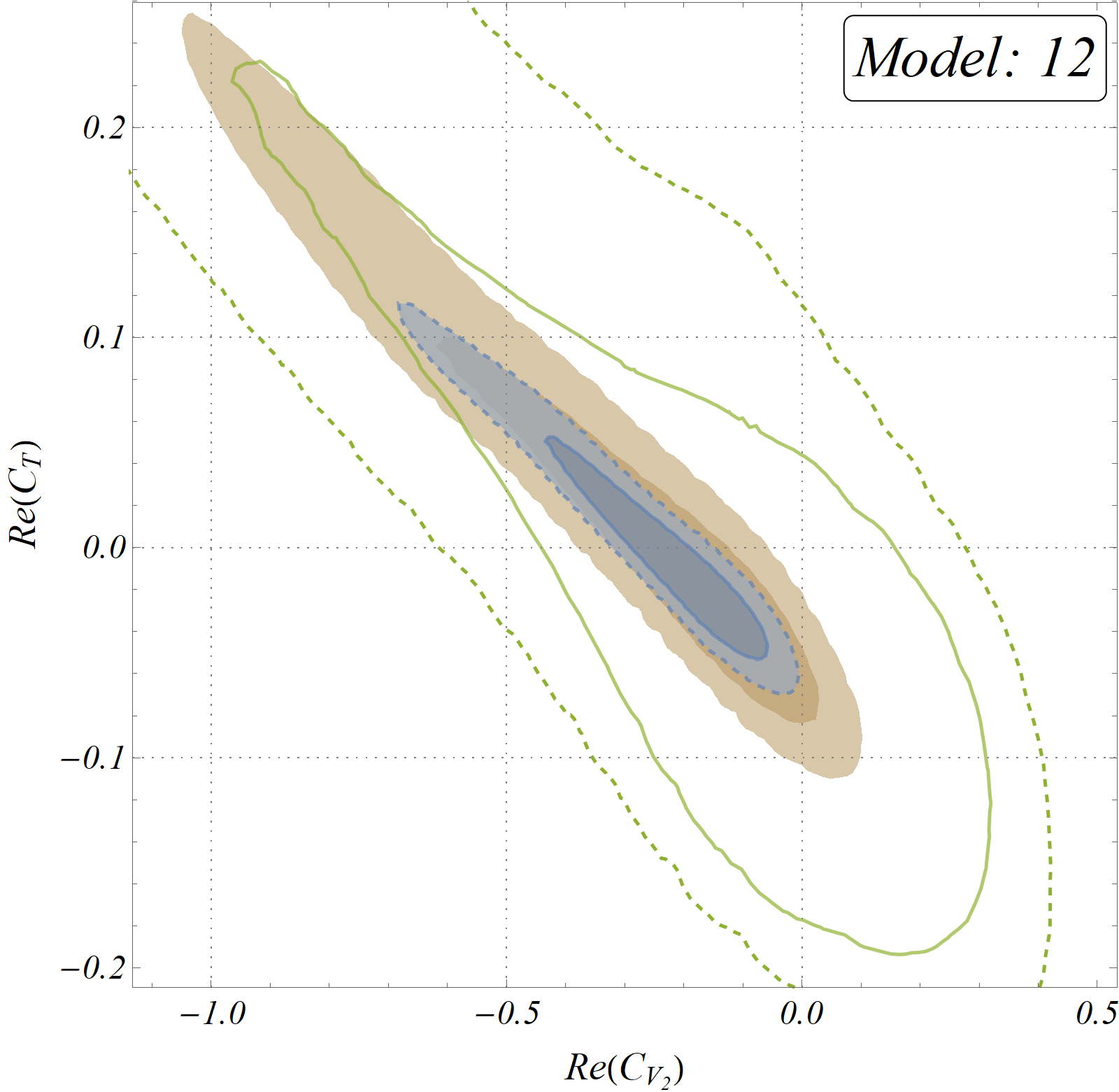}\label{fig:parspace12fu}}~
	\subfloat[]{\includegraphics[width=0.45\textwidth]{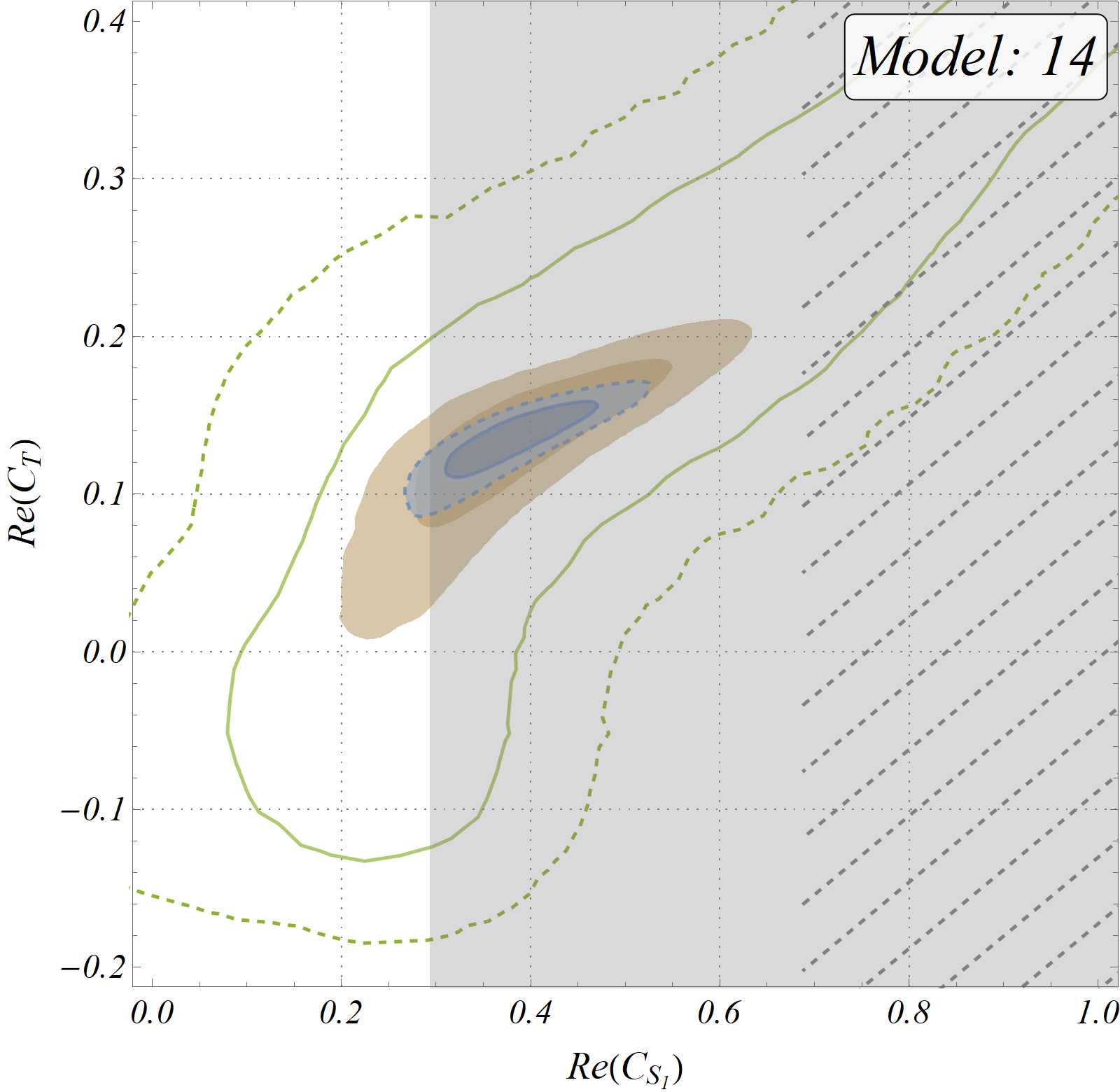}\label{fig:parspace14fu}}\\
	\subfloat[]{\includegraphics[width=0.45\textwidth]{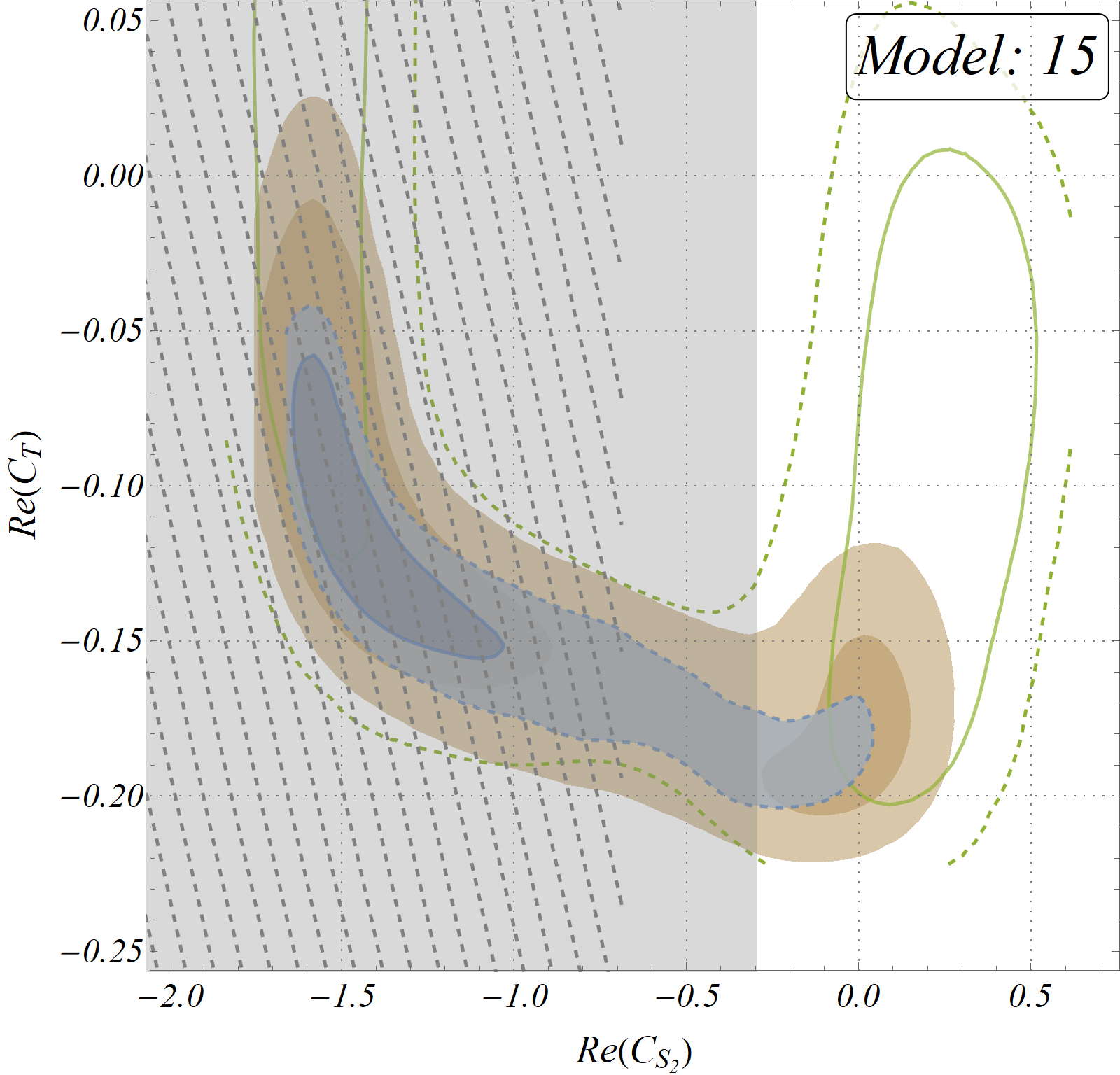}\label{fig:parspace15fu}}~~~
	\subfloat[]{\includegraphics[width=0.4\textwidth]{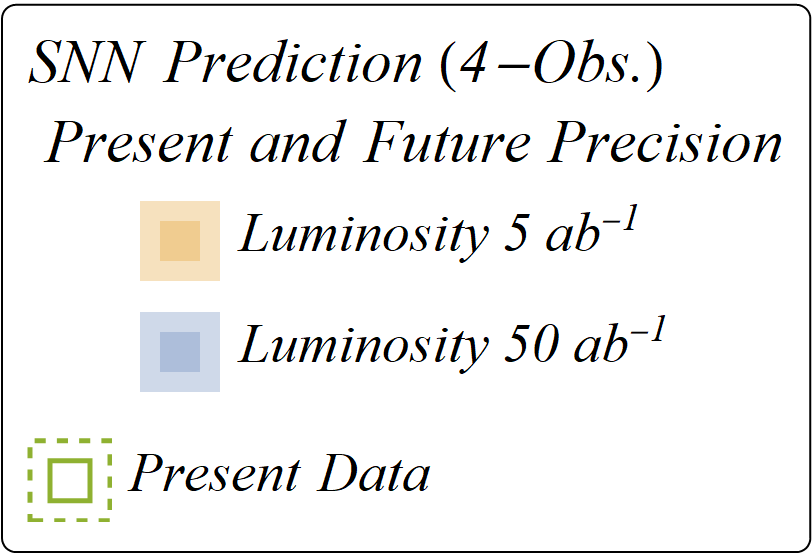}\label{fig:legFuture2D}}
	\caption{2-D parameter spaces for the best models. The green (filled) and black (empty) contours show Bayesian fits for 5 and 4-observable cases respectively. Three contours correspond to $68.27\%$, $95.45\%$, and $99.73\%$ credible regions. Red contours show the first two credible regions with the predictor SNNs applied on 4 observables. Corresponding 5-observable cases are blue density histograms. The gray-shaded and the diagonally hatched regions are discarded by $10\%$ and $30\%$ limits on $Br(B_c\to\tau\nu_{\tau})$, respectively.}
	\label{fig:parspace2Dfuture}
\end{figure*}
%%%%%%%%%%%%%%%%%%%%%%%%%%%%%%%%%%
Table \ref{tab:classnetrescompare} encapsulates the main results of model selection using SNN classifier for the 4 and 5-observable cases. The second column lists the serials of the best models picked by the classification SNN with probability $>1\%$.  The model-parameters and the corresponding probabilities are in the next two columns, respectively. The fifth column lists the positions of those models in the list of models sorted by $D_{KL}$. Similarly, the next two columns list the positions of those models in the list sorted by model-probabilities after applying the SNN on just the central value of the data-set and $\Delta\text{AIC}_c$, respectively. The corresponding $w^{\Delta\text{AIC}_c}$ values are listed in the last column. Note that the current data select a couple of two-operator scenarios with real WCs. A few of them which include a tensor type operator are highly probable.     

Taking the $D_{KL}$ measure as our optimum model-selection criterion, it becomes clear that in this small sample-sized data-set, $\Delta\text{AIC}_c$ completely misses the best models and rejects them in preference for less complex ones. Actually, the only models picked up within $0\leq\Delta\text{AIC}_c\leq 4$ are one-parameter models with model indices $1,~4$ and $5$  - none of which appear in the list.

On the other hand, 5 of the best candidate models, according to $D_{KL}$ are there in the list picked up by the SNN, for both data-sets. The corresponding model-indices are in bold in table \ref{tab:classnetrescompare}. The appearance of some less probable models, selected by the SNN, can be attributed to the large uncertainty in the observable $P_{\tau}(D^*)$. Note that their respective $D_{KL}$ measures select some of these models. As was previously explained in section \ref{sec:methodclass}, these model probabilities (fourth column in table) are not obtained by just applying the SNN on the central values. To take the full-data-distribution into account, we have aggregated the probabilities of each class for a large sample of points from this distribution. As the uncertainty of $P_{\tau}(D^*)$ is huge ($\sim 200\%$), different model-sets are selected in different regions of the sample. This enhances the final aggregated probabilities of those models, and though they are not viable candidates near the central value, they still appear in the list. This variation of model prediction can be easily checked by sorting the results of the SNN on just the central value of the data-set and comparing the serials of the aggregate-probability-selected models in that sorted list. The sixth column of table \ref{tab:classnetrescompare} contains the positions of the selected models in that list. 

Armed with the knowledge of the classification capability of the SNNs, we then have trained ensemble classifier SNNs for the 13-observable case. The performance of the chosen ensemble, compared to those of the 4 and 5-observable ones, is listed in table \ref{tab:netmeasuresClass}. Using the higher dimensionality of the data-sets, these can not only discriminate between a larger number of models, but also does that with far better accuracy. 
%%%%%%%%%%%%%%%%%%%%%%%%%%%%%%%%%%%
\begin{figure*}[hbpt]
	\small
	\centering
	\subfloat[]{\includegraphics[width=0.31\textwidth]{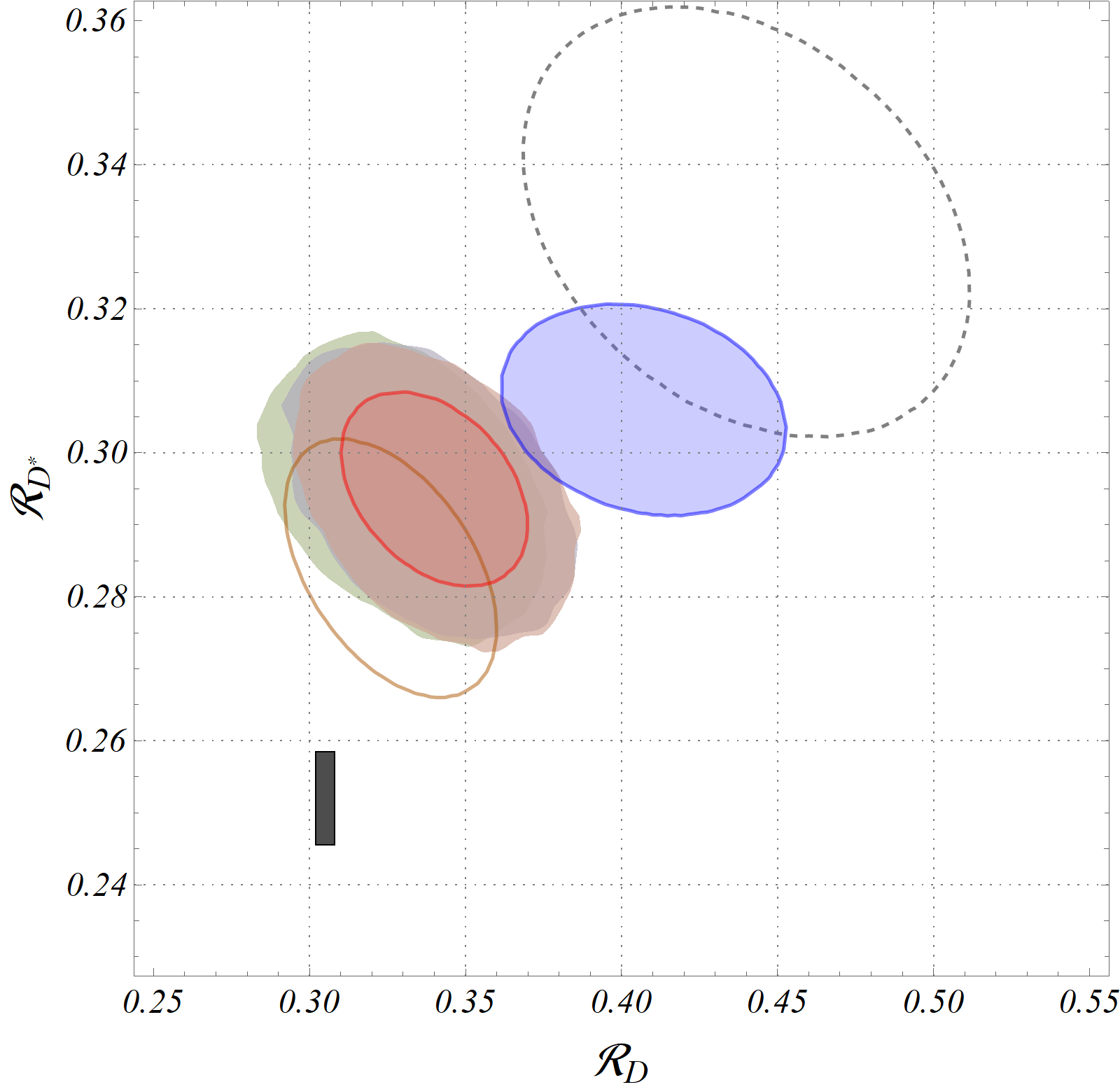}\label{fig:obsPlotBayesRDRDst}}~
	\subfloat[]{\includegraphics[width=0.31\textwidth]{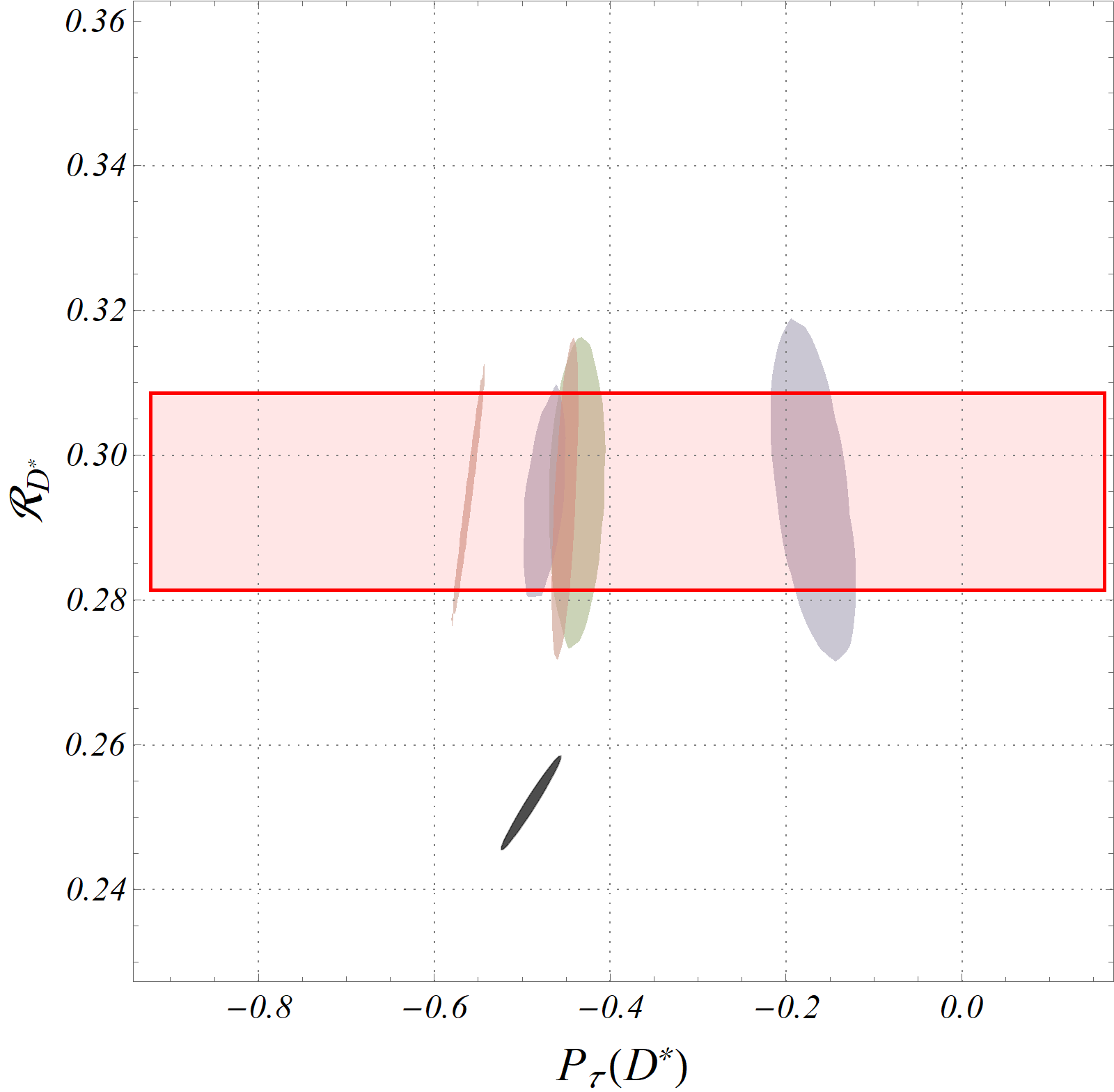}\label{fig:obsPlotBayesPtauRDst}}~
	\subfloat[]{\includegraphics[width=0.31\textwidth]{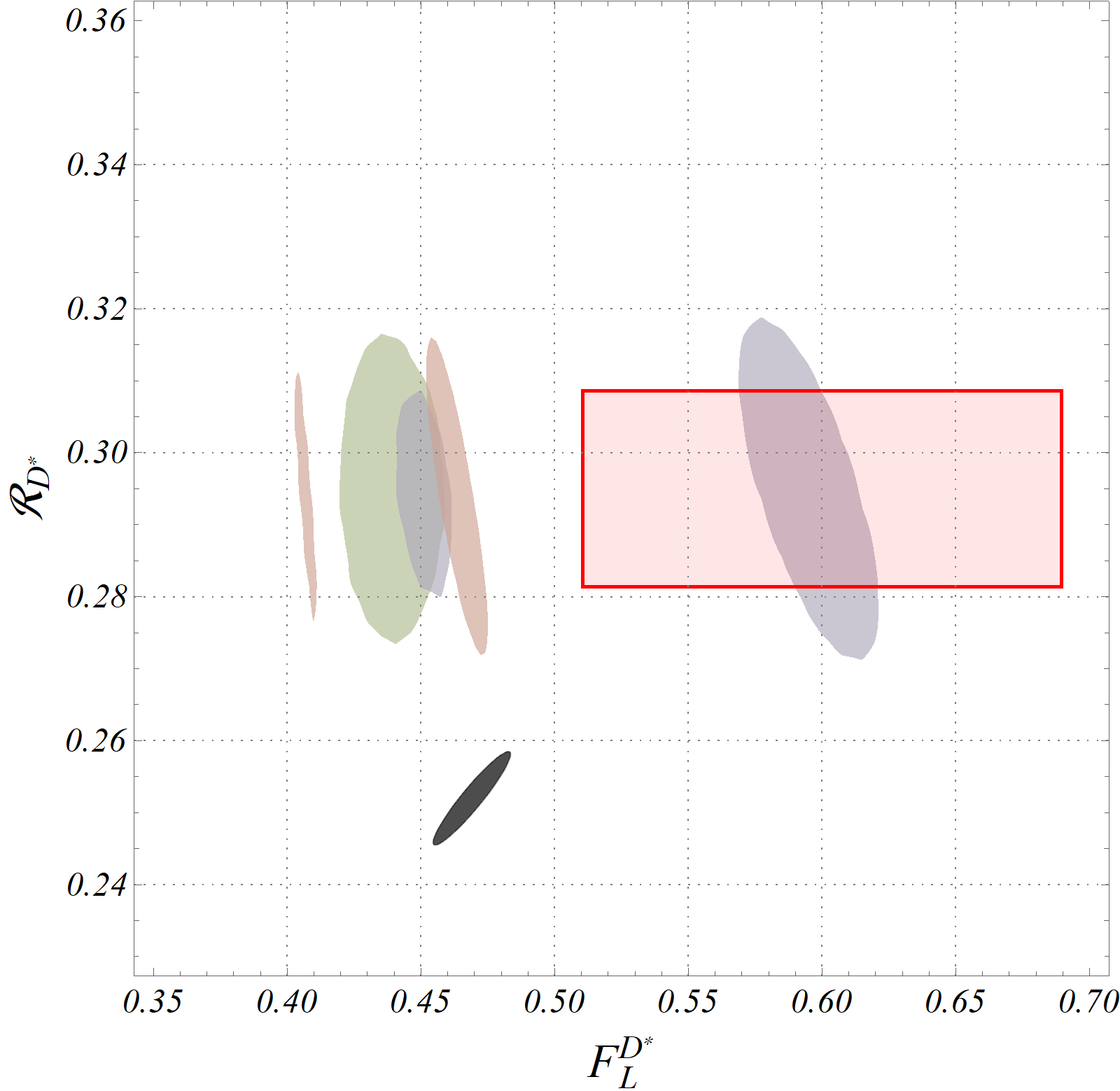}\label{fig:obsPlotBayesFLDstRDst}}\\
	\subfloat[]{\includegraphics[width=0.31\textwidth]{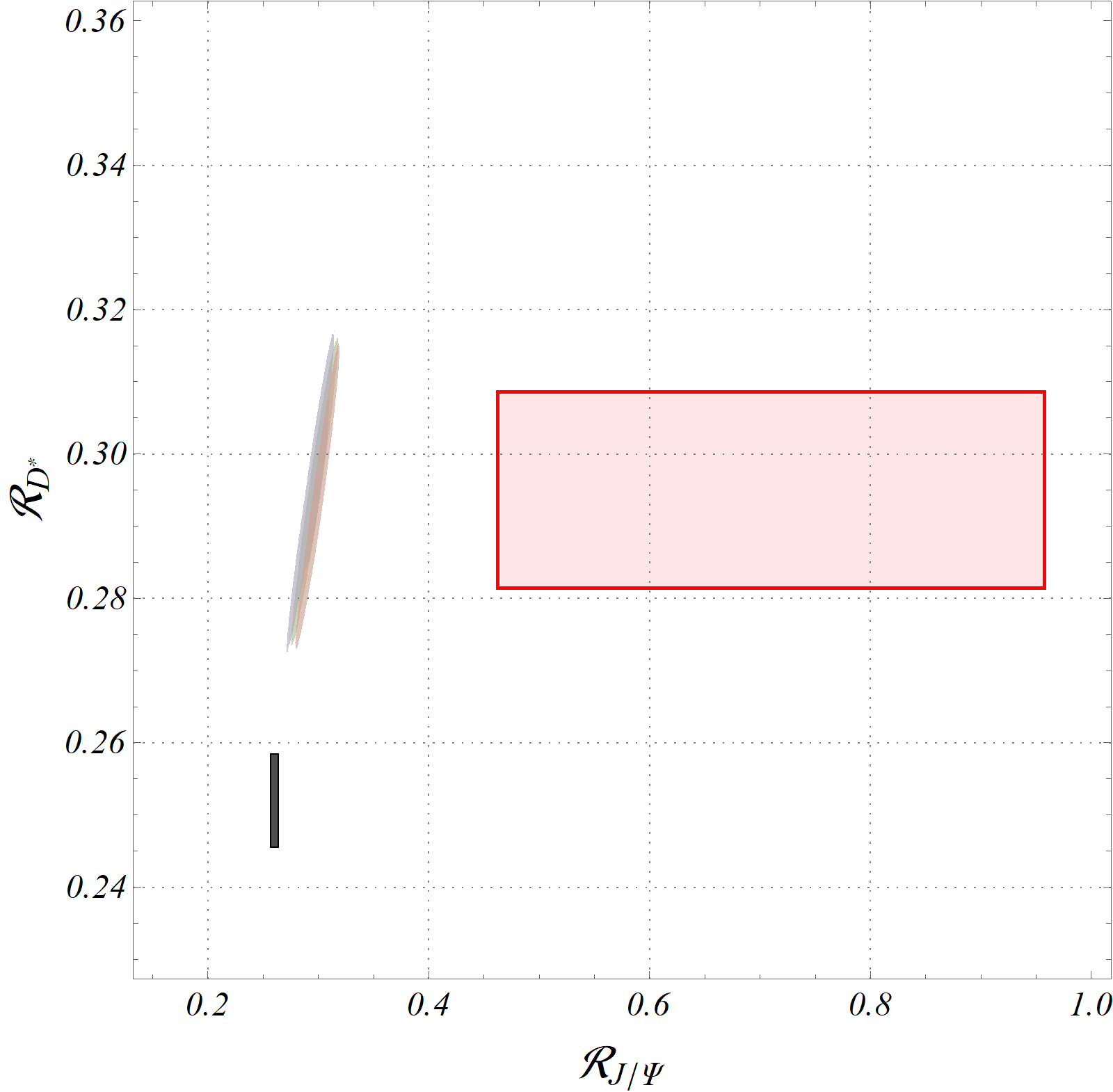}\label{fig:obsPlotBayesRJRDst}}~
	\subfloat[]{\includegraphics[width=0.31\textwidth]{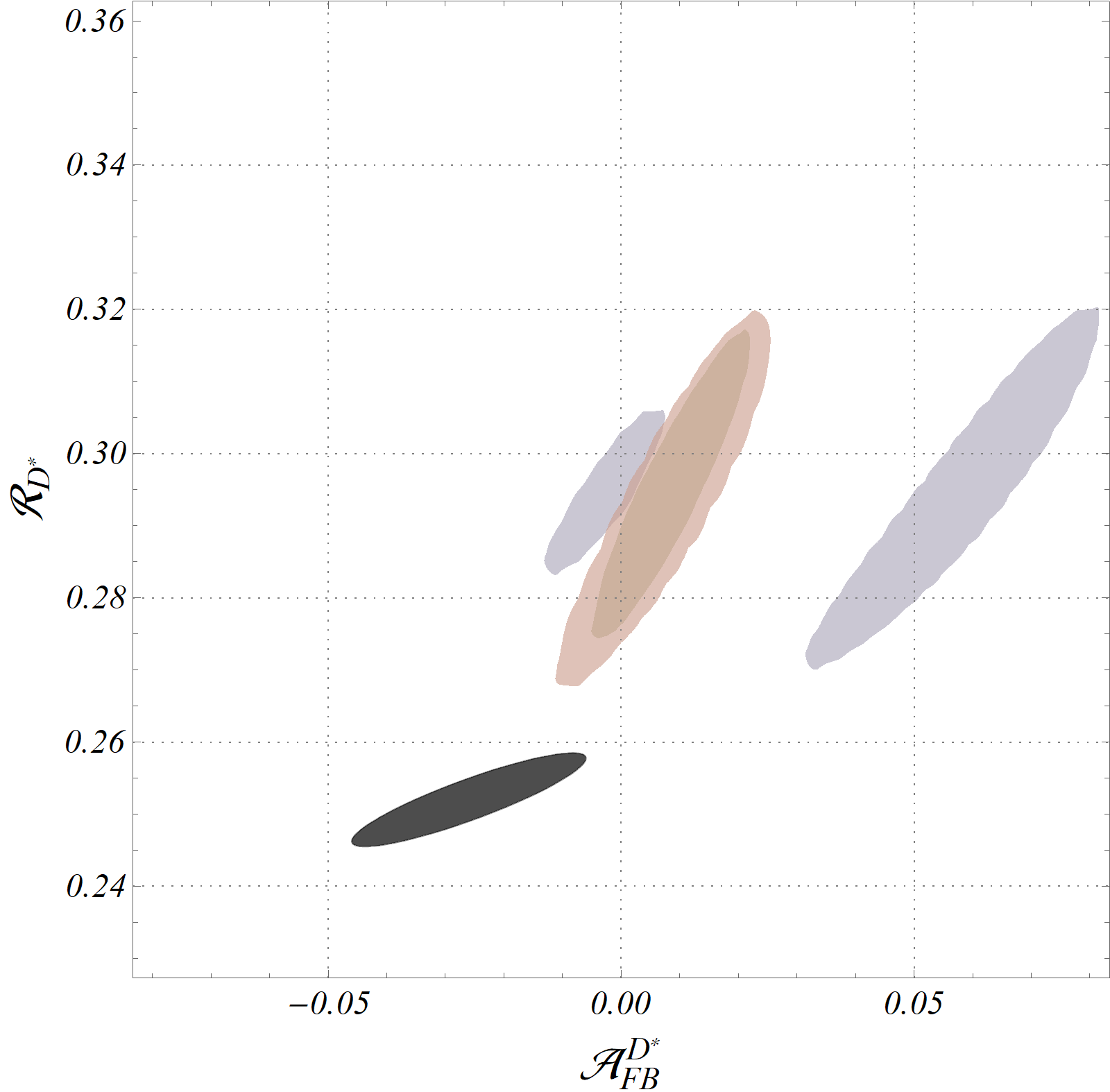}\label{fig:obsPlotBayesAFBDstRDst}}~
	\subfloat[]{\includegraphics[width=0.31\textwidth]{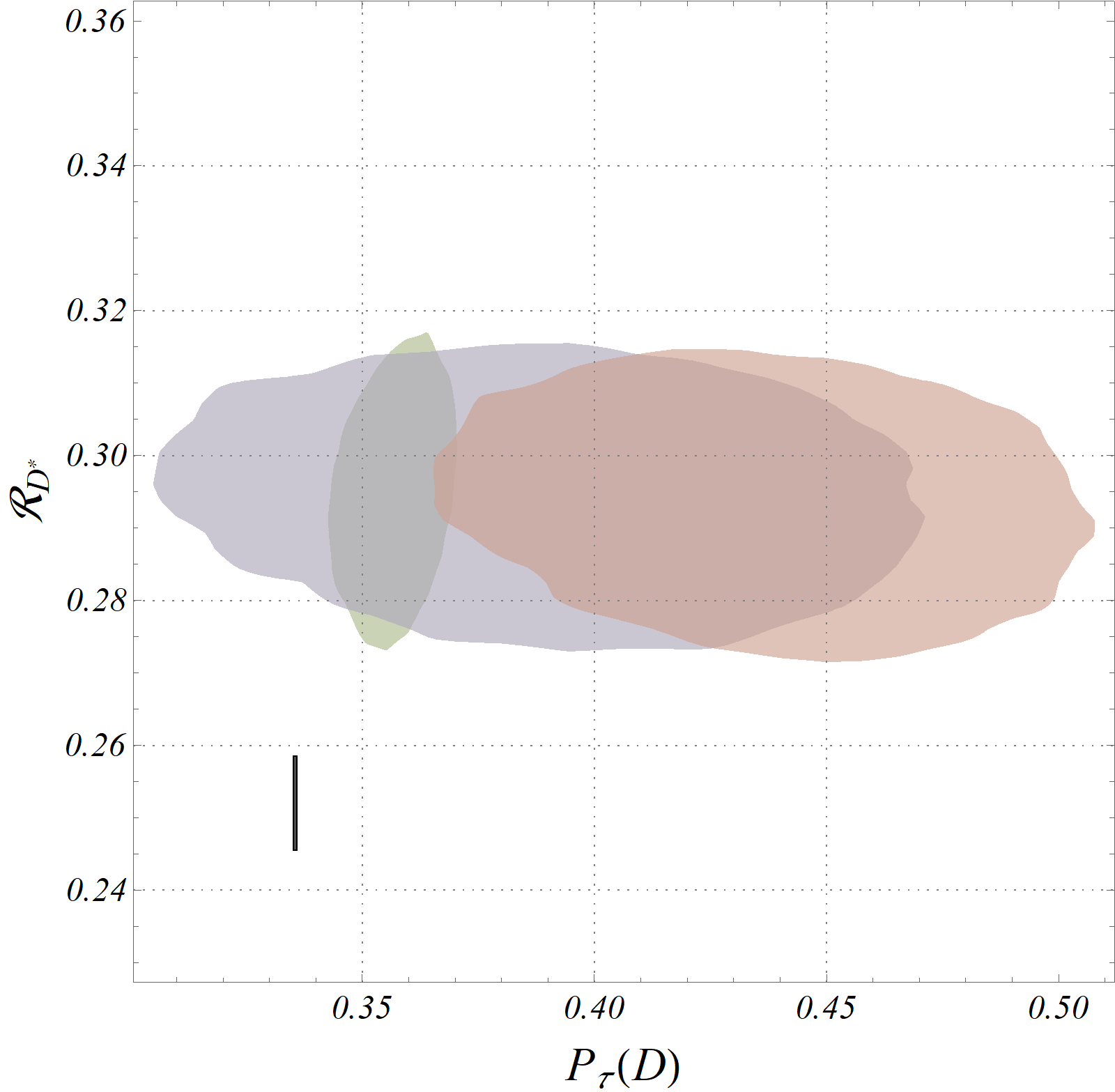}\label{fig:obsPlotBayesPtauDRDst}}\\
	\subfloat[]{\includegraphics[width=0.31\textwidth]{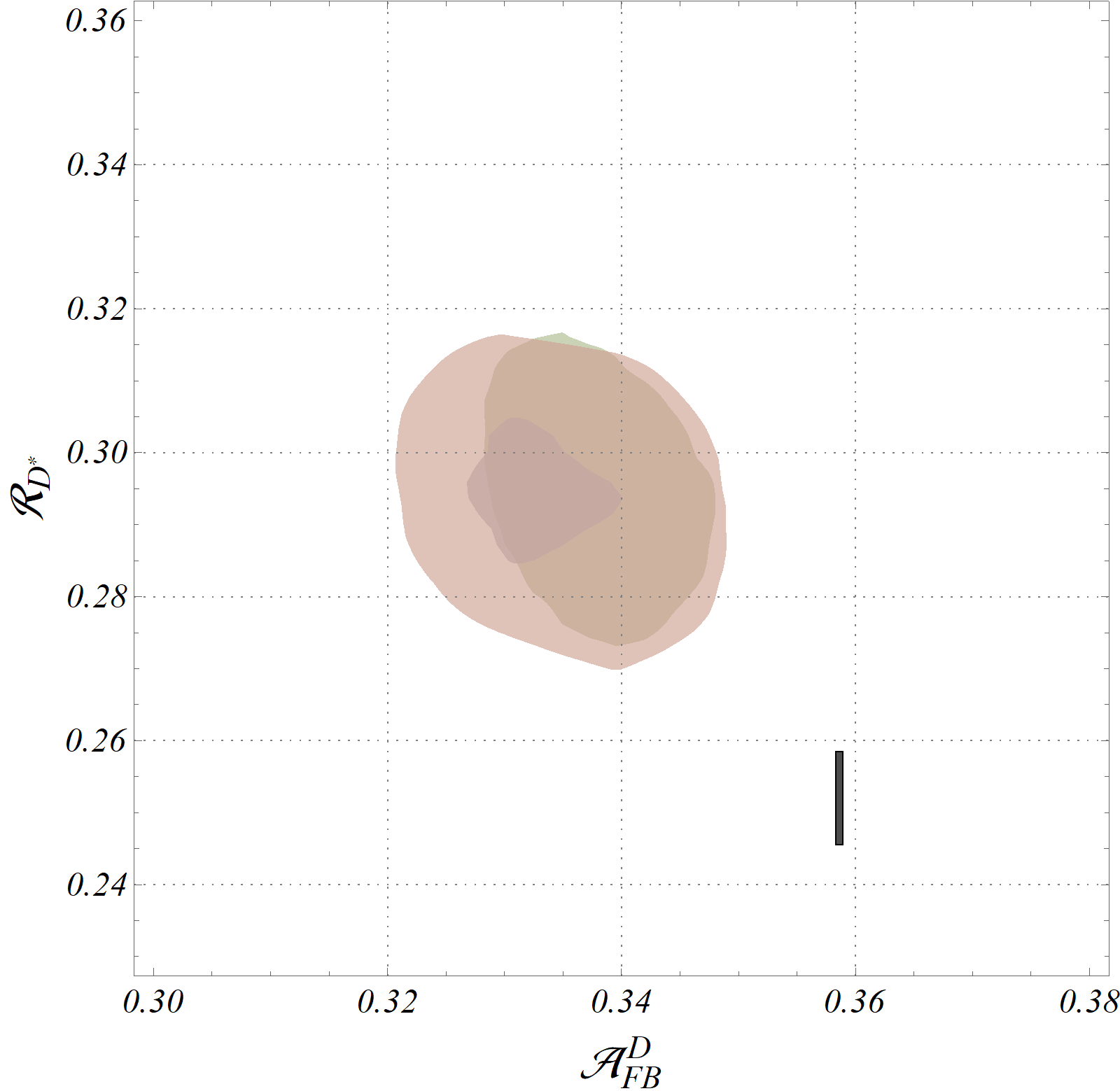}\label{fig:obsPlotBayesAFBDRDst}}~
	\subfloat[]{\includegraphics[width=0.31\textwidth]{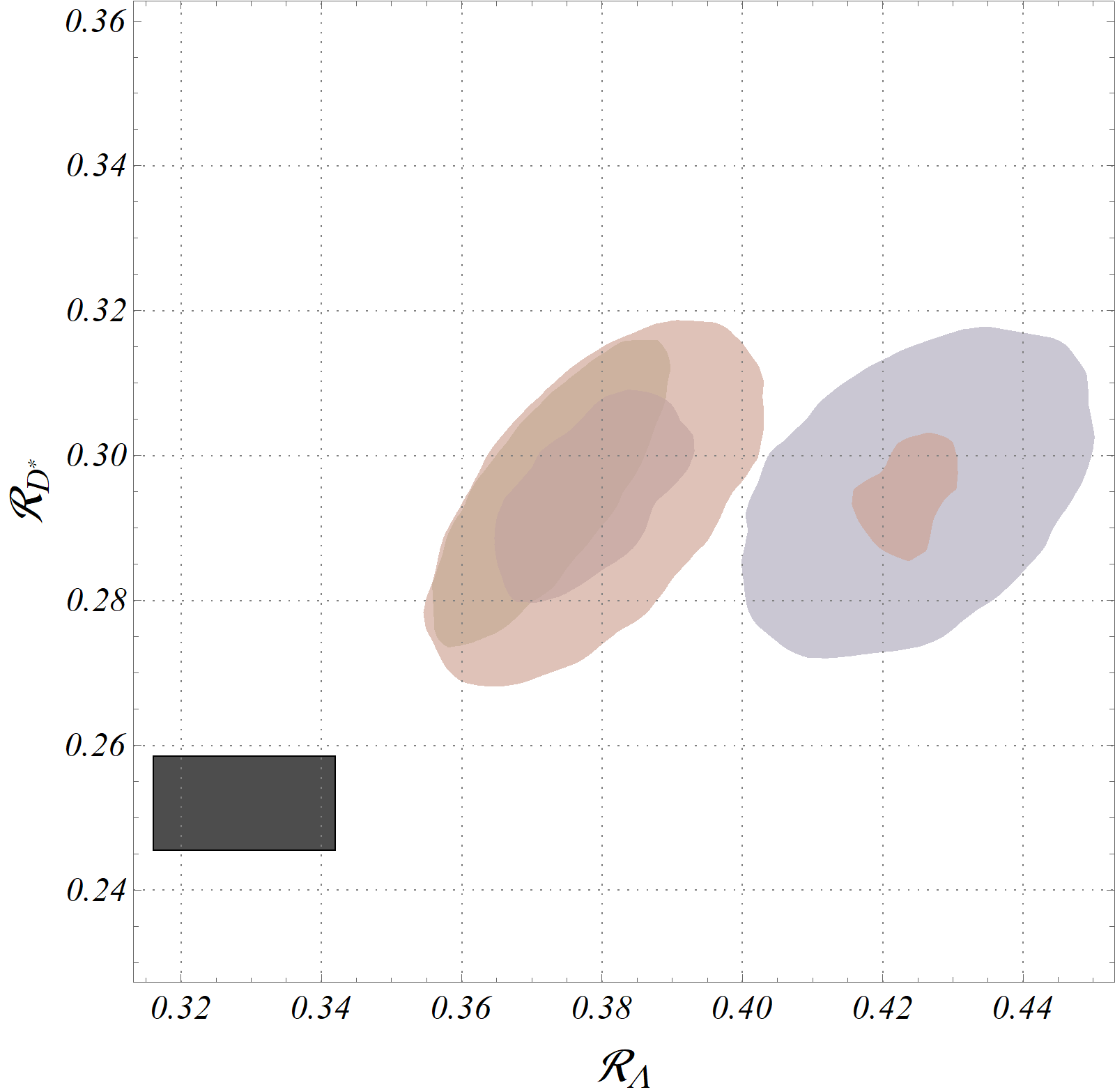}\label{fig:obsPlotBayesRLambdaRDst}}~
	\subfloat[]{\includegraphics[width=0.31\textwidth]{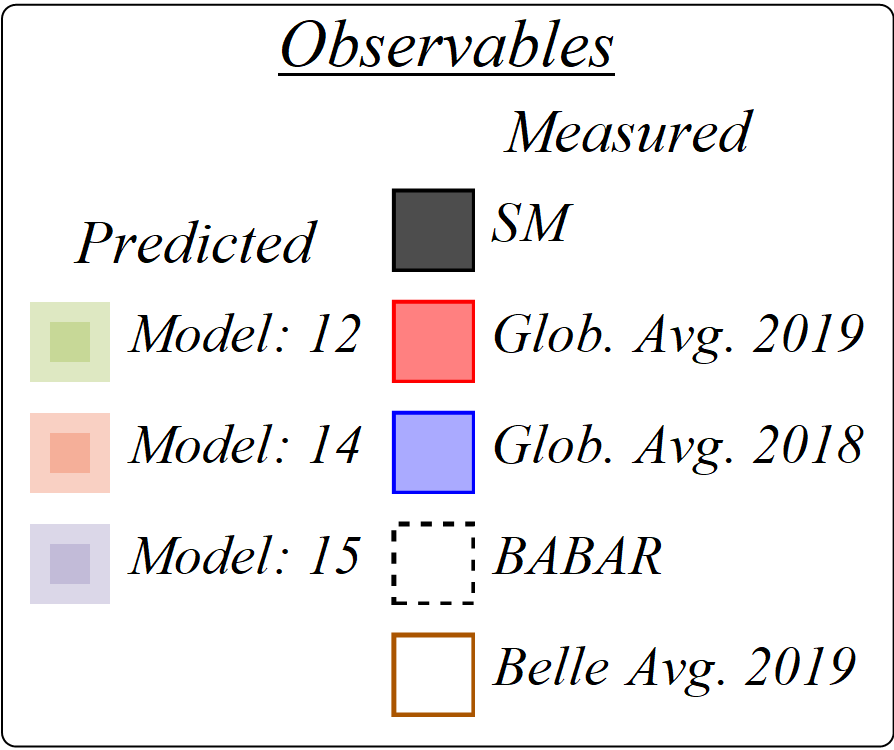}\label{fig:legObs}}\\
	\caption{Comparison of measured, predicted (with the Bayesian results for the 5-Obs. case of the best models), and SM observable spaces. Only $68.27\%$ C.Is are shown. All SM predictions for observables for the $D^*$ mode are taken from ref. \cite{Jaiswal:2020wer}. Uncorrelated measurements are shown as rectangles, while the correlated ones are ellipses, with rotation of semi-major axis of the ellipse denoting the correlation. Experimental measurements have boundaries, while the predictions from models have none.}
	\label{fig:obspace2D}
\end{figure*}
%%%%%%%%%%%%%%%%%%%%%%%%%%%%%%%%%%
\subsubsection{Regression}\label{sec:resregression}
Following the procedure described in section \ref{sec:methodregress}, we have trained regression SNNs for each model for all of the 4, 5, and 13-observable cases. The 4 and 5-observable networks help us to validate the regression capability of the SNNs by comparing them with the results of the Bayesian fits. Figure \ref{fig:parspace2D} shows this comparison for five of the best selected models (indices: 10, 11, 12, 14, and 15). Though model 13 is selected by the network as well, it does not represent the data-distribution faithfully, as seen from the $D_{KL}$ score and we refrain from using it.

The green (filled) and black (empty) contours show the parameter spaces after Bayesian fits for 5 and 4-observable cases respectively. For these cases, three credible probability regions ($68.27\%$, $95.45\%$, and $99.73\%$) are shown. With the exception of model 12, all posterior distributions are multi-modal, with a stray $>99.73\%$ credible region near $Re(C_T) \approx 0.3$ for model 15. We have used these parameter spaces to create Bayesian-predicted observable distributions and then applied SNNs on them. The predictor SNNs reproduce the original parameter spaces quite faithfully. This consistency can be checked from the 1-dimensional parameter spaces depicted in figure \ref{fig:parspace1D}. The faithful reproduction of the Bayesian parameter-spaces ensures the validity of the regression nets while using them on the actual data-distribution.

The red contours in figure \ref{fig:parspace2D} show the $68.27\%$ and $95.45\%$ credible regions for the parameter-space obtained by directly applying the corresponding regression SNNs on the data-distribution of the 4 observables. Similar regions for the 5-observable case are shown as blue density histograms. Across all models, these distributions are much flatter, i.e.,though the maximum posterior (MAP) estimates are almost always overlapping, the equivalent credible regions are much larger than the Bayesian results. The locations of the modes match, but not the spreads. The gray-shaded and the diagonally hatched regions show the parameter-spaces discarded by the tighter $10\%$ and moderate $30\%$ limits on $Br(B_c\to\tau\nu_{\tau})$, respectively. Figure \ref{fig:parspace2D} shows that for all models, there are at least parts of the $68.27\%$ credible probability regions allowed by the tightest constraints. Note that in all the selected scenarios, the absolute values of the WCs are  $< 1.0$.   

The results make sense, as we already know about the huge spread of $P_{\tau}(D^*)$. Still, to validate that the predictive observable-distributions generated by the SNN parameter spaces represent the data-distribution more faithfully than those for the Bayesian results, we have compared the $D_{KL}$ scores between these with respect to the original data-distribution for all models. Table \ref{tab:klcompare} lists the 5 best models (same for both data-sets, but with different order) in terms of the $D_{KL}$ scores of the SNN-parameter-space. As can be checked from the fourth column of that table, corresponding $D_{KL}$ scores of the Bayesian fits are orders of magnitude larger. These first 5 best models are the ones whose parameter-spaces are shown in fig. \ref{fig:parspace2D}. All other models have orders of magnitude larger $D_{KL}$ scores than these. Each row of the last three columns of the same table lists the performance of the predictor SNN for the corresponding model, evaluated on the independent test data-set. As expected, the standard deviation (fifth column) and the mean squared error (last column) are very low for the selected models, and the coefficient of determination (penultimate column) is really high.

Relatively larger values of the $D_{KL}$ for predictor SNNs for the 5-observable case compared to the 4-observable ones point to the fact that addition of $\mathcal{R}_{J/\Psi}$ actually worsens both the net prediction and the Bayesian fit results. This finding is consistent with our earlier analyses \cite{Bhattacharya:2018kig,Biswas:2018jun} and the possible reason is the large deviation of the central value of the experimental result of $\mathcal{R}_{J/\Psi}$ ($\sim 2.5$ times the SM). As the underlying quark structure of both $B\to D^{(*)}\tau\nu$ and $B\to J/\Psi \tau\nu$ are same, this enhancement should have been of similar order as that of $\mathcal{R}_{D^{(*)}}$ \cite{Bhattacharya:2018kig,Biswas:2018jun,Alok:2017qsi}. This worsens the predictions over all models.

This way of first using the regression SNNs on the data-distribution for each model and then using $D_{KL}$ to find the best ones among them constitutes a lengthier and computationally costlier version of model-selection. However, this is the most robust and theoretically most stable technique of all. 

\subsubsection{Probing Future Precision}\label{sec:future}
Belle II has already started taking data and is expected to take 50 times the present Belle data sample by 2025. The unprecedented increased luminosity and the resultant statistics will make the statistical uncertainty so small, that the ultimate relative uncertainty is expected to be dominated by systematics, which are also expected to decrease with the increased sample size of the dominant background channels \cite{Kou:2018nap}. To test the effectiveness of the trained SNNs with more precise data in the future, we have created a synthetic data-set of our 4-observables case, following the predicted relative uncertainties of these 4 observables at $5$ and $50~ab^{-1}$ luminosity, from ref. \cite{Adamczyk:2019wyt}, and keeping the central values fixed at their present value. 

As the central values are kept fixed and the uncertainties are decreased, we expect to pick a subset of the selected models obtained using the rigorous technique described in the previous section and listed in table \ref{tab:klcompare}. Indeed, we find that with increasing precision, only some of the models are favored as shown in table \ref{tab:klcomparefuture}. Applying the trained classifier ensemble-SNN on these data also yield similar results. Model 12 ($Re(C_{V_2}),~Re(C_{T})$) has aggregate probabilities of $68.87\%$ and $76.49\%$ for data corresponding to $5$ and $50~ab^{-1}$ luminosities, respectively, whereas the same values for model 15 are $15.48\%$ and $9.97\%$. Comparing these values with the corresponding ones in table \ref{tab:classnetrescompare} shows that with increasing precision, model selection becomes increasingly decisive.

The effect of increased precision is reflected in the model parameter-space as well. In figure \ref{fig:parspace2Dfuture}, we showcase the decrease in parameter-uncertainty for the best models and compare them with the present parameter space. Along with being more precise, this points to the fact that if the central values of these 4 observables remain unchanged in the future, all of the $68\%$ and most of the $96\%$ credible regions of the parameter spaces of models 14 ($Re(C_{S_1}),~Re(C_{T})$) and 15 ($Re(C_{S_2}),~Re(C_{T})$) will be discarded by the $10\%$ upper limit on $\mathcal{B}\left(B_c\to\tau\nu\right)$. The two-operator scenario with a right-handed vector and a tensor type quark current will be only possible solution.

In figure \ref{fig:obspace2D}, we have compared the measured values and SM predictions of some of the observables, with their predicted regions obtained from the Bayesian results (for the 5-Obs. case) of the best three models. As was expected, though different models have different predictions for $P_\tau(D^*)$, they are all equivalently consistent with different parts of the $1 \sigma$ C.I. of the experimental result, due to its very large uncertainty. The mentioned tension between SM and measured values of $\mathcal{R}_{J/\Psi}$ is also evident from the corresponding plot. None of the models are able to explain the experimental result within $1\sigma$. Only one model-prediction (Model-15; with operators $\mathcal{O}_{S_2}$ and $\mathcal{O}_{T}$) is consistent with the experimental $1 \sigma$ of $F_L^{D^*}$. We have also provided the predicted distributions of some other observables without any measurement till date. All predictions are consistent with the experimental results within $2 \sigma$.

\paragraph*{$\underline{\text{Supplied additional Material:}}$} We have provided the trained classifier SNNs for all data-sets and the corresponding predictor SNNs for all models in MxNet(\texttt{JSON}) and Wolfram Language (\texttt{WLNET}) formats. These, along with some simulated data examples and requisite example codes in a \emph{Mathematica\textsuperscript \textregistered} notebook are in a GitHub repository \cite{GitRepo}. Further updates, with future Wolfram Language resources will be notified and archived in the our group web-page \cite{GroupPage}.
%%%%%%%%%%%%%%%%%%%%%%%%%%%%%%%%%%%
	\section{Summary}
%%%%%%%%%%%%%%%%%%%%%%%%%%%%%%%%%%%
In this article, we have looked at the opportunities of solving the `inverse problem', in other words, model selection, in $b\to c \tau \nu_{\tau}$ decays. With a small number of available experimental results, this sector is challenging in this respect, as standard statistical and information-theoretic techniques like cross-validation and AIC$_c$ become unstable for a proper model selection. Noting that the training of a multi-class classifier machine learning model by optimizing a cross-entropy loss function is equivalent to minimizing the information-loss as measured by K-L divergence, we attempt to solve this problem using a deep self-normalizing neural network architecture, through supervised training over simulated data-sets.

By generating the pool of `models' with different combinations of dimension-six operators, with the corresponding Wilson coefficients considered complex in general, we first perform a standard frequentist analysis for model selection with AIC$_c$ and a Bayesian one as well, for inferring the parameter spaces. All of this is repeated for three separate data-sets, two with observables already measured by some experimental collaboration, and one with all significant NP-sensitive observables, both measured and unmeasured. The recent lattice update of the SM prediction of ratio $\mathcal{R}_{J/\Psi}$ enables us to use it in our data-sets as a NP-sensitive observable.

To increase the accuracy of the SNN classifiers, they are combined in ensembles and we also study the variation of accuracy with the depth of the nets. The model-selection-prowess of these classifiers are tested by comparing there predictions with both AIC$_c$ results and the K-L divergence between the data and predicted distributions of the Bayesian results for the corresponding models. Performance of the ensemble classifier is tallied with a collection of various shallow machine learning algorithms and it is found to be exceeding that of the best of them by a large margin. We find that due to the small size of data, AIC$_c$ unerringly picks simpler models, neglecting the ones which have predicted distributions most consistent with the real data-distribution.

In the next step, we use a variant of the SNNs as predictors of parameter space for each model. We see that though the MAP estimates of these are equivalent to the Bayesian results in most cases, the obtained parameter distributions are much flatter. Ordering these models in increasing K-L divergence between the data and predicted distributions (predictions with SNN-obtained parameter space) is the most robust possible way of doing model selection at present, though computationally cumbersome. After doing that we see that best selected models in this way are consistent with results of the classifier SNNs.

The trained predictor networks provide us a way of probing the reach of future experiments with higher luminosity with the model predictions. Using the SNN results, we have qualitatively shown how the uncertainties in the parameter spaces decrease with increasing luminosity. We have also provided observable-predictions of the selected best three models in our analysis, with their Bayesian fit results. Models with either a simultaneous presence of scalar and tensor interactions, or right-handed vector and tensor interactions are picked as best models all through the analysis process. If the central values of the measurements remain unchanged, only one model, with simultaneous scalar and tensor interactions will survive the constraints from $\mathcal{B}(B_c\to\tau\nu)$ with increased luminosity and precision of future measurements. 

With the upcoming experiments like Belle-II and high-luminosity LHCb, we expect to see a hoard of new measurements. In view of that, the supplied trained networks can be useful to the community for both model selection and at least crude regression of the selected models, without the need for running a sophisticated statistical analysis every time. This has the potential to help model builders to come up with suitable models to explain the data at that time.

\begin{acknowledgments}
	S.N.is supported by the Science and Engineering Research Board, Govt. of India, under the grant CRG/2018/001260.
\end{acknowledgments}

\bibliography{MLref1}

\end{document}